\documentclass[a4paper]{article}
\pdfoutput=1 

\usepackage{jcappub} 
\usepackage{float}

\usepackage[T1]{fontenc} 

\usepackage{dcolumn}
    \newcolumntype{d}[1]{D{.}{\cdot}{#1}}
    \newcolumntype{.}{D{.}{.}{-2}}
    \newcolumntype{,}{D{,}{,}{-3}}

\usepackage{booktabs}

\title{Unraveling the CMB lack-of-correlation anomaly with the cosmological gravitational wave background}

\author[a,b,c,1]{Giacomo Galloni,\note{Corresponding author.}}
\author[d,e,f]{Mario Ballardini,}
\author[g,h,i]{Nicola Bartolo,}
\author[d,f,j]{Alessandro Gruppuso,}
\author[d,e,k]{Luca Pagano}
\author[g,l]{and Angelo Ricciardone}

\affiliation[a]{Dipartimento di Fisica, Universit\`a di Roma Tor Vergata,\\Via della Ricerca Scientifica 1, 00133, Roma, Italy}
\affiliation[b]{Dipartimento di Fisica, Universit\`a La Sapienza,\\P.le A. Moro 2, 00185, Roma, Italy}
\affiliation[c]{Istituto Nazionale di Fisica Nucleare, Sezione di Roma 2,\\Via della Ricerca Scientifica 1, 00133 Roma, Italy}
\affiliation[d]{Dipartimento di Fisica e Scienze della Terra, Universit\`a degli Studi di Ferrara,\\Via Giuseppe Saragat 1, 44122, Ferrara, Italy}
\affiliation[e]{Istituto Nazionale di Fisica Nucleare, Sezione di Ferrara,\\Via Giuseppe Saragat 1, 44122, Ferrara, Italy}
\affiliation[f]{INAF - Osservatorio di Astrofisica e Scienza dello Spazio di Bologna,\\Via Piero Gobetti 101, 40129, Bologna, Italy}
\affiliation[g]{Dipartimento di Fisica e Astronomia ``G. Galilei'', Universit\`a degli Studi di Padova,\\Via Marzolo 8, I-35131, Padova, Italy}
\affiliation[h]{Istituto Nazionale di Fisica Nucleare, Sezione di Padova,\\Via Marzolo 8, I-35131, Padova, Italy}
\affiliation[i]{INAF - Osservatorio Astronomico di Padova,\\Vicolo dell’Osservatorio 5, I-35122, Padova, Italy}
\affiliation[j]{Istituto Nazionale di Fisica Nucleare Sezione di Bologna, Viale C. Berti Pichat 6/2, 40127, Bologna, Italy}
\affiliation[k]{Université Paris-Saclay, CNRS, Institut d’Astrophysique Spatiale, 91405, Orsay, France}
\affiliation[l]{Dipartimento di Fisica “E. Fermi”, Universit\`a di Pisa, I-56127, Pisa, Italy}

\emailAdd{giacomo.galloni@roma2.infn.it}
\emailAdd{mario.ballardini@unife.it}
\emailAdd{nicola.bartolo@pd.infn.it}
\emailAdd{alessandro.gruppuso@inaf.it}
\emailAdd{luca.pagano@unife.it}
\emailAdd{angelo.ricciardone@unipi.it}

\abstract{
Since the very first observations, the Cosmic Microwave Background (CMB) has revealed on large-scales unexpected features known as anomalies, which challenge the standard $\Lambda$ cold dark matter ($\Lambda$CDM) cosmological model. One such anomaly is the ``lack-of-correlation'', where the measured two-point angular correlation function of CMB temperature anisotropies is compatible with zero, differently from the predictions of the standard model. This anomaly could indicate a deviation from the standard model, unknown systematics, or simply a rare realization of the model itself. In this study, we explore the possibility that the lack-of-correlation anomaly is a consequence of living in a rare realization of the standard model, by leveraging the potential information provided by the cosmological gravitational wave background (CGWB) detectable by future gravitational wave (GW) interferometers. We analyze both constrained and unconstrained realizations of the CGWB to investigate the extent of information that GWs can offer. To quantify the impact of the CGWB on the lack-of-correlation anomaly, we employ established estimators and introduce a new estimator that addresses the ``look-elsewhere'' effect. Additionally, we consider three different maximum multipoles, denoted as $\ell_{\rm max}$, to account for the anticipated capabilities of future GW detectors ($\ell_{\rm max} = 4, 6, 10$). Summarizing our findings for the case of $\ell_{\rm max} = 4$, we identify the angular range $\qty[63^\circ - 180^\circ]$ as the region where future observations of the CGWB maximize the probability of rejecting the standard model. Furthermore, we calculate the expected significance of this observation, demonstrating that 98.81\% (81.67\%) of the constrained GW realizations enhance the current significance of the anomaly when considering the full-sky (masked) \textit{Planck} SMICA map as our CMB sky.}


\begin{document}

\maketitle

\flushbottom

\section{Introduction}\label{sec:introduction}

The \textit{Planck} satellite is widely recognized as a major milestone in modern cosmology \cite{planckcollaboration2020Planck2018ResultsVI, planckcollaboration2020Planck2018ResultsConstraints, planckcollaboration2019Planck2018ResultsIXa, planckcollaboration2020Planck2018ResultsVIIa}. Its unprecedented precision measurement of the temperature and polarization anisotropies of the Cosmic Microwave Background (CMB) allowed us to severely constrain the $\Lambda$ cold dark matter ($\Lambda$CDM) model, which today is considered the standard description of our Universe.

Despite the success of a spatially flat $\Lambda$CDM, there are intriguing anomalies in the large angular scales of the CMB sky that we observe. For example, the alignment of low multipole moments \cite{tegmark2003HighResolutionForegroundcleaned, deoliveira-costa2004SignificanceLargestScaleCMB, copiMultipoleVectorsNew2004}, the hemispherical power asymmetry \cite{eriksenAsymmetriesCosmicMicrowave2004}, the low CMB variance \cite{monteserin2008LowCMBVarianceWMAP}, the parity asymmetry \cite{land2005UniverseOdd, kim2010AnomalousParityAsymmetryWilkinson, gruppuso2011NewConstraintsParitySymmetry}, and the cold spot \cite{cruz2004DetectionNonGaussianSpotWMAP, vielva2004DetectionNonGaussianityWMAP1year} (see also \cite{copi2010LargeangleAnomaliesCMB, schwarzCMBAnomaliesPlanck2016, muir2018CovarianceCMBAnomaliesa}). Their statistical significance lies between 2-3$\sigma$ depending on which estimator is used and which anomaly is considered. Another interesting anomalous feature of the CMB is its topology \cite{gott1986SpongelikeTopologyLargeScaleStructure, gott1990TopologyMicrowaveBackgroundfluctuations, colley1996TopologyCOBEMicrowaveBackground}. It can be studied employing a zoology of different tests, such Minkowski functionals \cite{mecke1994RobustMorphologicalMeasureslargescale, schmalzing1998MinkowskiFunctionalsUsedmorphological, winitzki1998MinkowskiFunctionalDescriptionmicrowave} or the skeleton length \cite{novikov2006SkeletonProbeCosmicweb}, and have also shown a relatively high significance \cite{eriksen2004TestingNonGaussianityWilkinsonMicrowave, hansen2004AsymmetriesLocalCurvatureWilkinson, park2004NonGaussianSignaturesTemperatureFluctuation, pranav2019UnexpectedTopologyTemperatureFluctuations, pranav2022AnomaliesTopologyTemperaturefluctuations, akrami2022SearchTopologyUniverseHasa}. In this work, we focus on the anomaly that historically was noticed first. Unlike what was expected, we observed a strange feature in the so-called two-point angular correlation of the CMB temperature anisotropies $C(\theta)$: it is almost zero when evaluated on large angular scales (see section~\ref{sec: ang_corr} below). For this reason, it was named ``lack-of-correlation'' anomaly \cite{hinshaw19962PointCorrelationsCOBEDMR, spergel2003FirstYearWilkinsonMicrowave, copi2007UncorrelatedUniverseStatisticalAnisotropy, copi2009NoLargeangleCorrelationsnonGalactic, bennettNineyearWilkinsonMicrowave2013, copi2015LackLargeangleTTcorrelationsa}. This characteristic of $C(\theta)$ was able to survive the test of time, since it was first observed by COBE \cite{boggessCOBEMissionIts1992, bennett19964YearCOBEDMRCosmic, hinshaw19962PointCorrelationsCOBEDMR}, then reassessed with WMAP \cite{cruz2011AnomalousVarianceWMAPdata, gruppuso2014TwopointCorrelationFunctionWMAP, monteserin2008LowCMBVarianceWMAP}, and further confirmed by \textit{Planck} \cite{planckcollaboration2014Planck2013ResultsXXIII, planckcollaboration2016Planck2015ResultsXVI, planckcollaboration2020Planck2018ResultsVIIa}, suggesting that it is not the consequence of some unknown systematic effect. 

In order to assess the significance of these anomalies, a number of different techniques have been employed. These can be based on some two-point statistic (such as the angular power spectrum), or on peak statistics, and N-point correlation functions (see \cite{planckcollaboration2020Planck2018ResultsVIIa} and the references therein). Independently from the analysis, the fundamental question one tries to answer is: are these features the consequence of a rare realization of the standard $\Lambda$CDM model, or do we need to abandon it in favor of a more complex one? One of the obstacles that this question poses is where to search for new information on the anomalies. In fact, CMB temperature has been observed in a cosmic variance-limited fashion on low and intermediate multipole scales; thus, we cannot unveil any new information from the temperature alone. To determine whether these are the consequence of a physical phenomenon, we must exploit other observables, correlated with the CMB temperature, such as the CMB polarization 
\cite{copi2013LargeAngleCMBSuppressionPolarizationa, billi2019PolarisationTracerCMBanomalies, yoho2015MicrowaveBackgroundPolarizationProbe, chiocchetta2021LackofcorrelationAnomalyCMBlargea, shi2023TestingCMBAnomaliesEmode, dvorkinTestablePolarizationPredictions2008}.

In the same spirit, we want to explore the capabilities of a cosmological gravitational wave background (CGWB) to shed light on the lack-of-correlation anomaly. Indeed, we know that a fundamental prediction of any inflationary model is a stochastic background of GWs \cite{guzzettiGravitationalWavesInflation2016}. This CGWB can be explored indirectly through CMB thanks to the relic polarization it sources. Indeed, we know that primordial GWs produce the divergenless component of the CMB polarization, the B-modes \cite{kamionkowskiQuestModesInflationary2016}. Many ongoing and upcoming experiments are targeting this feature, such as BICEP/Keck Array (BK) \cite{adeDetectionBModePolarization2014}, Simons array \cite{suzukiPolarbear2SimonsArray2016}, Simons Observatory \cite{ade2019SimonsObservatorySciencegoals}, Stage-IV \cite{abazajianCMBS4ScienceBook2016} and the Light satellite for the study of B-mode polarization and Inflation from cosmic microwave background Radiation Detection (LiteBIRD) \cite{hazumiLiteBIRDSatelliteStudies2019, litebirdcollaboration2022ProbingCosmicInflationLiteBIRDa} (see \cite{galloniUpdatedConstraintsAmplitude2022} for the updated constraints on the primordial GWs spectrum). In addition, there are well motivate models of inflation that can produce a CGWB that can be the target of future GW detectors like LISA \cite{amaro-seoaneLaserInterferometerSpace2017, barausseProspectsFundamentalPhysics2020, bartolo2016ScienceSpacebasedInterferometerLISA, capriniReconstructingSpectralShape2019, pieroniForegroundCleaningTemplatefree2020} and the Einstein Telescope \cite{sathyaprakashScientificPotentialEinstein2012, maggioreScienceCaseEinstein2020,branchesi2023ScienceEinsteinTelescopecomparison}, becoming a window on high-energy phenomena that are not accessible in any other way. Such a background will be characterized by its frequency dependence \cite{capriniReconstructingSpectralShape2019, flaugerImprovedReconstructionStochastic2021}, and by other peculiar observables, e.g. spatial anisotropies of their energy density \cite{albaPrimordialGravityWave2016, contaldiAnisotropiesGravitationalWave2017, geller2018PrimordialAnisotropiesGravitationalWave, bartoloCharacterizingCosmologicalGravitational2020, bartoloAnisotropiesNonGaussianityCosmological2019, valbusadallarmiImprintRelativisticParticles2021}. These are generated both at the time of the GW production and during the propagation in a very similar way as CMB photons. Future GW interferometers, both on the ground and in space, will have limited angular sensitivity ($\ell\sim 15$) \cite{auclair2022CosmologyLaserInterferometerSpace}, however, as we will see, this does not limit our findings.

Recently in \cite{galloni2022TestStatisticalIsotropyUniversea}, CGWB has also been used to explore a dipolar modulation model trying to describe the hemispherical power asymmetry \cite{hansen2009PowerAsymmetryCosmicMicrowave, dvorkinTestablePolarizationPredictions2008}. This has shown that the great degree of correlation between the CGWB and the temperature of the CMB \cite{ricciardone2021CrosscorrelatingAstrophysicalCosmologicalGravitational} is key to enhancing our ability to probe the physical origin of these anomalies. Despite a lower degree of correlation, also the cross-correlation between the CMB and the astrophysical GWB can be used to probe early universe initial conditions \cite{perna2023NonGaussianityCrosscorrelationAstrophysicalGravitational}. We will show that in the lack-of-correlation context, this correlation is even more crucial.

Another nontrivial limitation in assessing the physical nature of the CMB anomalies is the fact that one typically uses ``a posteriori'' statistics. Indeed, estimators are often designed to maximize the significance of a certain anomaly under some a posteriori assumption on the data. Thus, one has to face the following antithesis: neglecting the assumption made on the data, is the evidence of the anomaly still significant? This is often called the ``look-elsewhere effect'' \cite{planckcollaboration2020Planck2018ResultsVIIa}. In this work, we will take care of this aspect of the lack-of-correlation anomaly.


To address these issues, we investigate the fluke hypothesis, exploring the possibility that we inhabit a rare realization of $\Lambda$CDM, considering both CMB temperature and CGWB anisotropies. We develop an estimator to determine if a specific value of the lack-of-correlation anomaly excludes the fluke hypothesis. Our analysis includes constrained and unconstrained realizations of the CGWB \cite{bertschingerPathIntegralMethods1987, hoffmanConstrainedRealizationsGaussian1991, bucherFillingCMBMap2012, kimHarmonicInpaintingCMB2012, manzottiMappingIntegratedSachsWolfe2014,ricciardone2021CrosscorrelatingAstrophysicalCosmologicalGravitational}, utilizing \textit{Planck}'s SMICA for temperature observations. Additionally, we provide a forecast of the expected improvement in significance from GWs, accounting for the look-elsewhere effect\footnote{All the code to reproduce our analysis is available at \url{https://github.com/ggalloni/lack_of_correlation_with_GWs}.}.

The structure of this paper is as follows: section~\ref{sec: methodology} outlines our methodology, including dataset definitions, computation of CMB temperature and CGWB realizations, and quantification of the lack-of-correlation anomaly using information from the CGWB. We also introduce the aforementioned new estimator. In section~\ref{sec: results}, we present the results of our analysis, focusing on the very realistic (in some sense pessimistic) case of $\ell_{\rm max}=4$ and providing additional cases with $\ell_{\rm max}=6$ and $\ell_{\rm max}=10$ in appendix~\ref{app: sensitivity}. Section~\ref{sec:conclusions} contains our conclusions.


\section{Datasets and Methodology} \label{sec: methodology}
As mentioned before, in this work we want to study and build an estimator which is able to reject the fluke hypothesis of the lack of correlation anomaly. Therefore, we investigate the idea that what we observe in the CMB sky is just a rare realization of the standard $\Lambda$CDM model. In particular, we want to use the CGWB and try to forecast its ability to shed some light on the physical origin of the anomaly.

Before presenting the results, we describe the datasets used for our analysis (see section~\ref{sec: dataset}). In addition, we provide more information on our methodology, with a focus on specific key aspects. This includes developing a procedure to simulate the CGWB sky (as outlined in section~\ref{sec: constrained}), gaining a deeper understanding of the crucial quantity that results in the lack of correlation in our data (as described in section~\ref{sec: ang_corr}), defining an estimator that measures the anomaly (outlined in section~\ref{sec: quantify}), and specifying the analysis that we performed in this study (as presented in sections~\ref{sec: metho_optimal} and~\ref{sec: signi}). Indeed, the final goal of this analysis is to be able to associate a value with the lack of correlation anomaly; for each value, we want to be able to conclude whether it is compatible or not with the fluke hypothesis. 

\subsection{Datasets} \label{sec: dataset}
We use the \textit{Planck} SMICA temperature map as our observation for CMB \cite{planckcollaboration2020Planck2018ResultsIV}\footnote{\url{http://pla.esac.esa.int/pla/\#maps}.}. It comes at a resolution of $\approx 5$ arcmin, which corresponds to a pixelization of the sky in $\approx5\times10^{7}$ equal-area pixels\footnote{This is usually expressed in terms of the $N_{\rm side}$ parameter of \texttt{Healpy}, which define the sky partition. In this case, $N_{\rm side}=2048$.}. Furthermore, masking the galactic plane enhances the discrepancy between the data and the $\Lambda$CDM predictions \cite{gruppuso2013LowVarianceLargescales, gruppuso2014TwopointCorrelationFunctionWMAP, nataleLackPowerAnomaly2019, planckcollaboration2020Planck2018ResultsVIIa}, therefore, it is important to treat the cut-sky case. To capture this feature, we will consider the full-sky SMICA map and a masked version, where we use the \textit{Planck} common mask for intensity\footnote{\url{http://pla.esac.esa.int/pla/\#maps}.}. In the latter case (and whenever a mask is involved in the computations), we use the pseudo-$C_\ell$ formalism to recover the unbiased angular power spectra (\texttt{NaMaster} \cite{alonso2019UnifiedPseudoEllframework})\footnote{\url{https://github.com/LSSTDESC/NaMaster}.}.

Regarding the CGWB, we obtain the theoretical angular power spectrum with a modified version of {\tt CLASS} \cite{blasCosmicLinearAnisotropy2011, lesgourguesCosmicLinearAnisotropy2011}\footnote{\url{https://github.com/lesgourg/class_public}.}. Specifically, we use the expressions shown in \cite{ricciardone2021CrosscorrelatingAstrophysicalCosmologicalGravitational} and set the $\Lambda$CDM parameters to their best-fit values provided by \textit{Planck} 2018 (see table~\ref{tab: params}).
\begin{table}[t]
 \centering
\begin{tabular}{l c}
\toprule
\multicolumn{1}{c}{Parameter} & \multicolumn{1}{c}{Best-fit value} \\ \midrule
$A_{\rm s} \times 10^9$      & 2.100549  \\ 
$n_{\rm s}$      & 0.9660499  \\ 
$\Omega_{\rm b} h^2$      & 0.0223828  \\ 
$\Omega_{\rm cdm} h^2$      & 0.1201075  \\ 
$\tau_{\rm reio}$      & 0.05430842  \\ 
$H_{\rm 0}$      & 67.32117 $\qty[\rm km\ s^{-1}\ Mpc^{-1}]$ \\ \midrule
$Y_{\rm He}$      & 0.2454006  \\ 
$T_{\rm 0}$      & 2.7255 $\qty[\rm K]$  \\ 
$\sum m_\nu$          & 0.06 $\qty[\rm eV]$ \\
\bottomrule
\end{tabular}
\caption{Assumed values of the 6 $\Lambda$CDM parameters and other important ones. The variables $A_{\rm s}$ and $n_{\rm s}$ represent the amplitude and tilt of the primordial scalar perturbations. The energy densities of baryons and cold dark matter are denoted by $\Omega_{\rm b}$ and $\Omega_{\rm cdm}$, respectively. $H_{\rm 0}$ is the Hubble constant expressed in $\rm km\ s^{-1}\ Mpc^{-1}$, which is divided by 100 to obtain $h \equiv H_{\rm 0}/100$. $\tau_{\rm reio}$ represents the optical depth of reionization. Then, $Y_{\rm He}$ is the fraction of helium, $T_{\rm 0}$ is the average temperature of the CMB in Kelvin, and $\sum m_\nu$ is the mass of neutrinos, assuming that 1 is massive and the other 2 are massless.}
  \label{tab: params}
\end{table}
We consider only the scalar contribution to the anisotropies, neglecting any tensorial contribution \cite{bartoloAnisotropiesNonGaussianityCosmological2019, bartoloGravitationalWaveAnisotropies2020}. 

Soon, detectors like LISA \cite{amaro-seoaneLaserInterferometerSpace2017, barausseProspectsFundamentalPhysics2020, bartolo2016ScienceSpacebasedInterferometerLISA, capriniReconstructingSpectralShape2019, pieroniForegroundCleaningTemplatefree2020}, DECIGO \cite{kawamuraJapaneseSpaceGravitational2006}, ET \cite{sathyaprakashScientificPotentialEinstein2012, maggioreScienceCaseEinstein2020} and CE \cite{abbottExploringSensitivityNext2017} will provide the possibility to observe the CGWB, delivering fundamental knowledge on the physics of the early Universe. In fact, these could not only measure the average contribution to energy density brought by GWs, but they could also detect its fluctuations on the celestial sphere. This possibility depends on the actual monopole radiation at the frequencies of the various detectors; the higher the better. To cover the possible performance of these experiments, we will consider 3 maximum multipoles to perform our analysis: $\ell_{\rm max} = 4,6,10$. These are the multipoles one may be able to recover in a signal-dominated way in the next future, exploiting one of the experiments mentioned, or a combination of them. To be consistent with this choice, we will also assume $\ell_{\rm max} = 4, 6, 10$ for the temperature part of the analysis. Also, to reduce the computational cost of working at \textit{Planck}'s full-resolution, the CMB map is also degraded to $N_{\rm side} = 64$\footnote{Although $N_{\rm side} = 64$ allows to describe $3N_{\rm side} - 1$ multipoles (thus much more than what is used for this analysis), in our case no significant computational advantage was found in reducing $N_{\rm side}$ below 64.}. Before degrading it, we smooth it with a Gaussian beam with Full Width Half Maximum (FWHM) equal to $2^\circ$. Note that the scale corresponding to $N_{\rm side} = 64$ is $\approx 0.92^\circ$, but we follow the general principle of applying a smoothing approximately two or three times bigger than the grid scale to avoid pixelization effects \cite{gruppuso2013LowVarianceLargescales}. Regarding the mask, we also degrade it to $N_{\rm side} = 64$ to match the CMB map. The mask is then thresholded by setting to zero the pixels in which the value is less than 0.9; the others are set to unity \cite{planckcollaboration2020Planck2018ResultsVIIa}.

Furthermore, since we also want to exploit the cross-correlation of these two fields as an observable, we also produce the cross-maps from our realizations of TT and CGWB\footnote{To be consistent with what is customarily done for CMB temperature, we will indicate the auto-spectra of the CGWB with GWGW.}. Although GWs are indeed very difficult to observe, it could be possible in the next future to get a relevant signal-to-noise ratio from their cross-correlation with CMB temperature without the need to measure the GW autospectrum. In fact, cross-correlation is often used in the literature to extract information from a noise-dominated context, since it provides unbiased information on the two correlated probes (see, e.g. \cite{planckcollaboration2020Planck2018ResultsVI, piccirilli2022CrosscorrelationAnalysisCMBlensing, ricciardone2021CrosscorrelatingAstrophysicalCosmologicalGravitational, alonso2020DetectingAnisotropicAstrophysicalgravitational}).

For the sake of brevity, we will use the expression ``masked GWGW'' in the case where we apply a mask on TT (the CGWB is always evaluated full-sky). Instead, when TT is also full-sky, we will simply use GWGW. For cross-correlation we will use TGW.

In conclusion, for this work, we produce $N = 10,000$ realizations of unconstrained CGWB and TT (full sky and masked). Using both sets of full-sky and masked TT, we also compute the corresponding constrained CGWB realizations, using \textit{Planck} SMICA as our CMB observation. In addition, we also compute the cross-correlation maps between unconstrained and constrained GWGW and TT (full sky and masked). 

\subsection{Constrained realizations of masked sky} \label{sec: constrained}

Let us consider a generic example with two observables $X$ and $Y$, which share a certain degree of correlation (in our case, they will be, respectively, the CMB temperature and the CGWB).

In the full-sky case, we already know that a random Gaussian realization of $a^{XX}_{\ell m}$ of $X$ can be obtained by its angular power spectrum $C_\ell^{XX}$, which encodes the variance of the coefficients of the spherical harmonics. In a more mathematical fashion, we can write
\begin{equation}
    a_{\ell m}^{XX} = \xi_{\ell m} \sqrt{C_\ell^{XX}}\ ,
\end{equation}
where $\xi_{\ell m}$ is a random Gaussian field with null mean and unitary variance. 
Once $a_{\ell m}^{XX}$ have been measured, we can generate $a_{\ell m}^{YY}$ realizations consistent with these, namely \textit{constrained realizations}. In fact, to condition the random realizations of $Y$ on the information already available on $X$, one can use the same Gaussian seed $\xi_{\ell m}$ to write
\begin{equation}
    a_{\ell m}^{YY} = \frac{C_\ell^{XY}}{\sqrt{C_\ell^{XX}}}\xi_{\ell m} + \xi^\prime_{\ell m} \sqrt{C_\ell^{YY} - \frac{\qty(C_\ell^{XY})^2}{C_\ell^{XX}}}\ .
\end{equation}
where $C_\ell^{XY}$ is the cross-correlation spectrum of the two observables and $\xi^\prime_{\ell m}$ is another random Gaussian field ($\langle\xi_{\ell m}\xi^{\prime *}_{\ell m}\rangle=0$).
Recasting this in terms of $a_{\ell m}^{XX}$, this expression becomes
\begin{equation}
    a_{\ell m}^{YY} = \frac{C_\ell^{XY}}{C_\ell^{XX}}a_{\ell m}^{XX} + \xi^\prime_{\ell m} \sqrt{C_\ell^{YY} - \frac{\qty(C_\ell^{XY})^2}{C_\ell^{XX}}} \ .
    \label{eq: constrained}
\end{equation}
Here, the first term on the right side is extracting the Gaussian seed of $X$, i.e. $\xi_{\ell m}$, and translating it into a deterministic part of the realization $Y$. Note that eq.~\eqref{eq: constrained} depends on the underlying assumptions of Gaussianity and statistical isotropy of the coefficients of the spherical harmonics. Furthermore, the specific shape of $C_{\ell}^{XX}, C_{\ell}^{XY}, C_{\ell}^{YY}$ will depend on the assumed cosmological model, which in our case is the $\Lambda$CDM  model (e.g., the statistical isotropy is relaxed in \cite{galloni2022TestStatisticalIsotropyUniversea}).

Eq.~\eqref{eq: constrained} can be generalized to the case of masked skies. To do so, we must recall the definition of $a_{\ell m}^{XX}$ as the coefficients of the decomposition of spherical harmonics of the X field on the 2D sphere. In particular, starting from the full-sky (FS) case, we can write (we drop the apex $XX$ for the sake of notation) \cite{gorski1994DeterminingSpectrumPrimordialInhomogeneity, mortlock2002AnalysisCosmicMicrowavebackground}
\begin{equation}
    a_{\ell m}^{\rm FS} \equiv \int_{\rm full-sky} Y_{\ell m}^*(\theta, \phi) X(\theta, \phi) \dd(\cos\theta)\dd\phi\ , 
    \label{eq: alm_def}
\end{equation}
where $Y_{\ell m}$ are the spherical harmonics and $\theta, \phi$ are the angles on the celestial sphere. These $a_{\ell m}^{\rm FS}$ are predicted to be statistically isotropic Gaussian realizations with null mean and a variance defined as
\begin{equation}
    \expval{a_{\ell m}^{*,\rm FS} a_{\ell m}^{\rm FS}} = \delta_{\ell \ell^\prime} \delta_{m m^\prime} C_{\ell} \ ,
\label{eq: FS_cov}
\end{equation}
where $C_\ell$ is the angular power spectrum of $X$. From a full-sky observation, one can estimate the angular power spectrum as
\begin{equation}
    \hat{C}_\ell = \frac{1}{2\ell+1}\sum_{m=-\ell}^\ell\abs{a_{\ell m}^{\rm FS}}^2\ .
    \label{eq: alm2cl}
\end{equation}
In the presence of a mask, eq.~\eqref{eq: alm_def} gets slightly modified in order to consider that the integration is performed on the unmasked patch of the sphere. Thus, the spherical harmonics do not constitute an orthogonal basis and the $a_{\ell m}$ coefficients measured on the cut-sky (CS) have couplings between multipoles. Despite this, it is possible to recover the corresponding full-sky values knowing the geometrical couplings introduced by the mask (see appendix~\ref{app: window}) \cite{gorski1994DeterminingSpectrumPrimordialInhomogeneity, mortlock2002AnalysisCosmicMicrowavebackground}. In particular, they are related by (the sum over repeated indexes is understood)
\begin{equation}
    a_{\ell m}^{\rm CS} = a_{\ell^\prime m^\prime}^{\rm FS} \times W_{\ell m}^{\ell^\prime m^\prime}\ ,
\end{equation}
where $W_{\ell m}^{\ell^\prime m^\prime}$ is the window function of the mask considered.

Thus, eq.~\eqref{eq: constrained} must be modified to account for this since it assumes the full-sky condition. In particular, when we consider the coefficients of the cut-sky, i.e. $a_{\ell m}^{\rm CS}$, we must recast the first factor on the right-hand side as 
\begin{equation}
    \frac{C_\ell^{XY}}{C_\ell^{XX}}\times a_{\ell m}^{\rm CS} \Longrightarrow \frac{C_\ell^{XY}}{C_\ell^{XX}}\times a_{\ell' m'}^{\rm CS} \qty(W_{\ell m}^{\ell' m'})^{-1} = \frac{C_\ell^{XY}}{C_\ell^{XX}}\times \hat{a}_{\ell m}^{\rm FS}\ .
    \label{eq: new_constrained}
\end{equation}
In this way, the cut-sky coefficients are remapped to the full-sky ones and the rest of the formula can remain the same. Note that we indicated the full-sky coefficients on the right-hand side as $\hat{a}_{\ell m}^{\rm FS}$ to remark that we are not obtaining their true values, but rather an estimate of those. In other words, following this procedure, we get an estimator of the full-sky coefficients from a partial-sky observation. Note that this becomes unfeasible as the window function becomes singular for aggressive masks. For the full-sky case, the window function goes to $W_{\ell m}^{\ell^\prime m^\prime} = \delta_{\ell \ell'}\delta_{m m'}$ (see appendix~\ref{app: window}), recovering the usual expressions.
Accounting for the complete expression of this matrix allows us to correctly obtain the full-sky coefficients for $X$, which are then converted in terms of $Y$ by the factor $C_\ell^{XY}/C_\ell^{XX}$. Note that another underlying assumption of this generalized procedure is that $Y$ is full-sky; otherwise, one has to take care again of the couplings between multipoles of the $Y$ realization.

\begin{figure}[t]
    \centering
    \includegraphics[width = 0.49\textwidth]{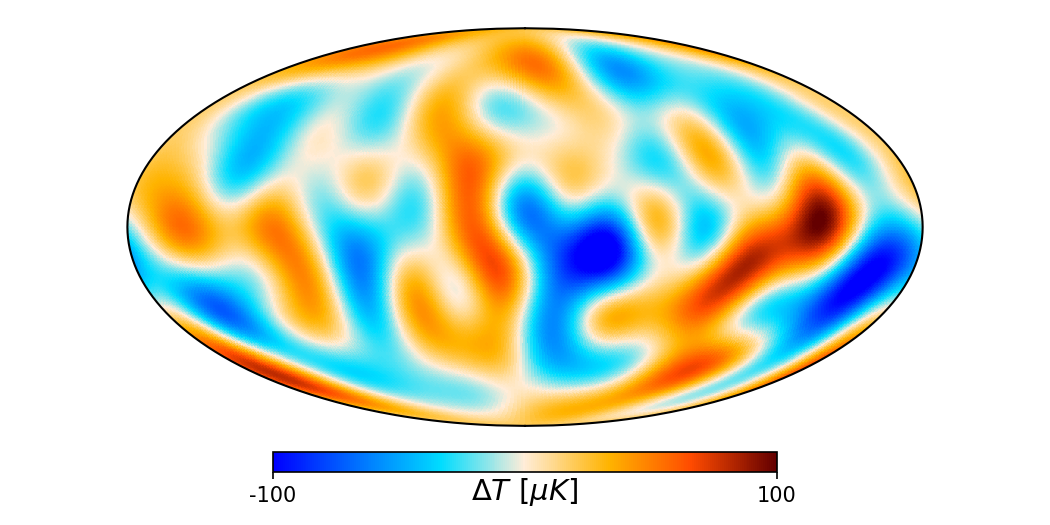}
    \includegraphics[width = 0.49\textwidth]{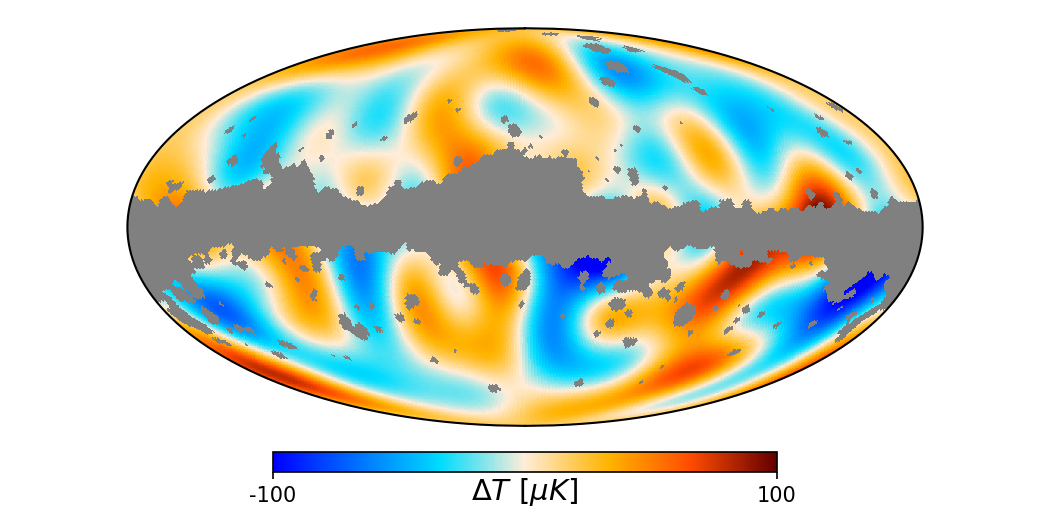}
    \includegraphics[width = 0.49\textwidth]{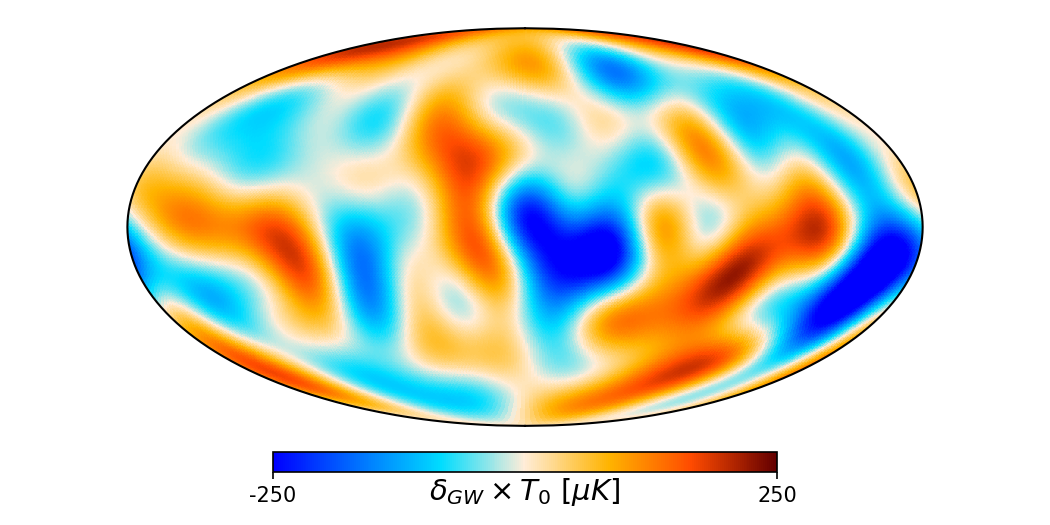}
    \includegraphics[width = 0.49\textwidth]{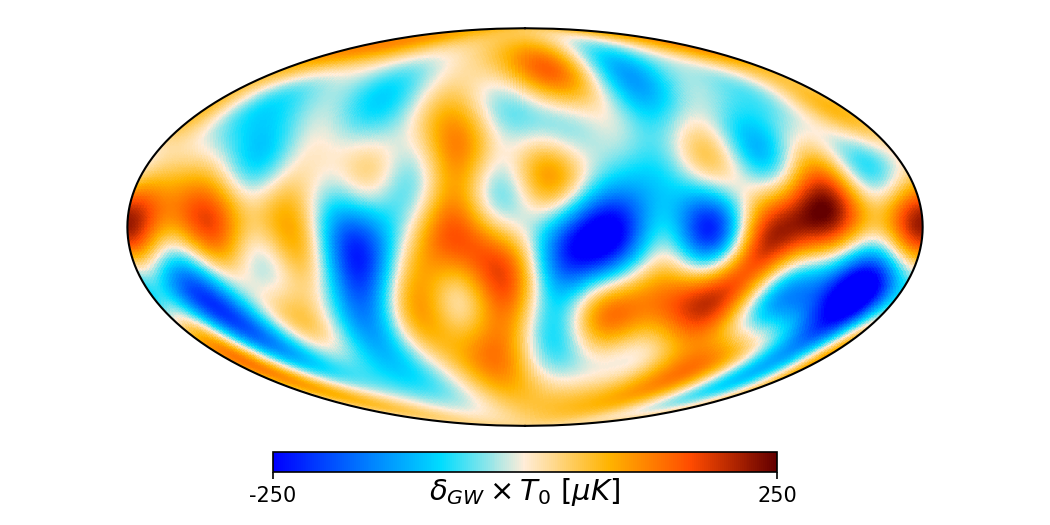}
    \caption{The two upper panels in the upper part show the SMICA map downgraded to $N_{\rm side} = 64$ and filtered to remove the multipoles above $\ell > 10$ for the full-sky and masked case. In the lower panels, we show one of the corresponding constrained realizations of the CGWB.}
    \label{fig: maps}
\end{figure}  

As an example, we show in figure~\ref{fig: maps} different sky realizations of the CGWB. In the upper part of the figure, we plot the downgraded SMICA map, for which we filter out the multipoles with $\ell>10$. Respectively, we show the full-sky and masked ones in the left and right panels. Instead, in the lower part of the figure, we show two constrained CGWB realizations based on the map above (thus full-sky or masked). Thus, the one on the left is obtained with eq.~\eqref{eq: constrained} and the other accounting for the generalization discussed in this section.

As is customarily done in the literature, we normalize the CMB anisotropies to the monopole radiation of the CMB, $T_{\rm 0} = 2.7255\times10^{6}\ \mu K$ \cite{fixsen2009TemperatureCosmicMicrowaveBackground}, so that we show the maps in units of $\mu K$. Usually in the case of the CGWB one plots the energy density contrast as defined in \cite{bartoloAnisotropiesNonGaussianityCosmological2019}. However, to be consistent with the choice for CMB and to be visually clear, we also normalize the CGWB anisotropies to the same quantity. Thus, the energy density will also be expressed in units of $\mu K$.

\subsection{Angular correlation function} \label{sec: ang_corr}

Having defined the datasets and how to obtain constrained realizations in the masked case, we briefly explore the actual quantity presenting the lack-of-correlation in current data. Indeed, we know that the fluctuations $\Delta T$ in the photon temperature $T$ that we observe are very well described by their spherical harmonic coefficients defined in eq.~\eqref{eq: alm_def}. In the case of $T$, it is recast to \cite{huWanderingBackgroundCMB1995}

\begin{equation}
    a_{\ell m} \equiv \int Y_{\ell m}^*(\theta, \phi) \frac{\Delta T}{T_{\rm 0}}(\theta, \phi) \dd(\cos\theta)\dd\phi\ , 
\end{equation}
where $T_{\rm 0}$ is the average temperature, i.e., the monopole radiation. Then, from a full-sky CMB observation, one can estimate the angular power spectrum using eq.~\eqref{eq: alm2cl}. This quantity plays a key role in almost every estimation of cosmological parameters, given that it is a very efficient summary statistic to study the properties of $a_{\ell m}$ (at least assuming Gaussianity \cite{planckcollaboration2019Planck2018ResultsIXa}). 

An alternative and equivalent way to convey the same information is the two-point angular correlation function, defined as \cite{schwarzCMBAnomaliesPlanck2016}
\begin{equation}
    C(\theta) \equiv \expval{\frac{\Delta T}{T_{\rm 0}}(\hat{n}_1)\frac{\Delta T}{T_{\rm 0}}(\hat{n}_2)}\eval{}_{\hat{n}_1 \vdot \hat{n}_2 = \cos\theta} = \sum_{\ell = 0}^{\infty}\frac{2\ell+1}{4\pi} C_{\ell} P_\ell(\cos\theta)\ ,
\end{equation}
where $P_\ell$ are the Legendre polynomials. In other words, $C(\theta)$ and $C_\ell$ are related by a series of $P_\ell$, thus the former allows us to better appreciate the large-scale behavior and the latter the small-scale one. Also, in the assumption of Gaussian fluctuations, all the available information is encoded in $C_{\ell}$, thus the same holds for $C(\theta)$.

\begin{figure}[t]
    \centering
    \includegraphics[width = 0.49\textwidth]{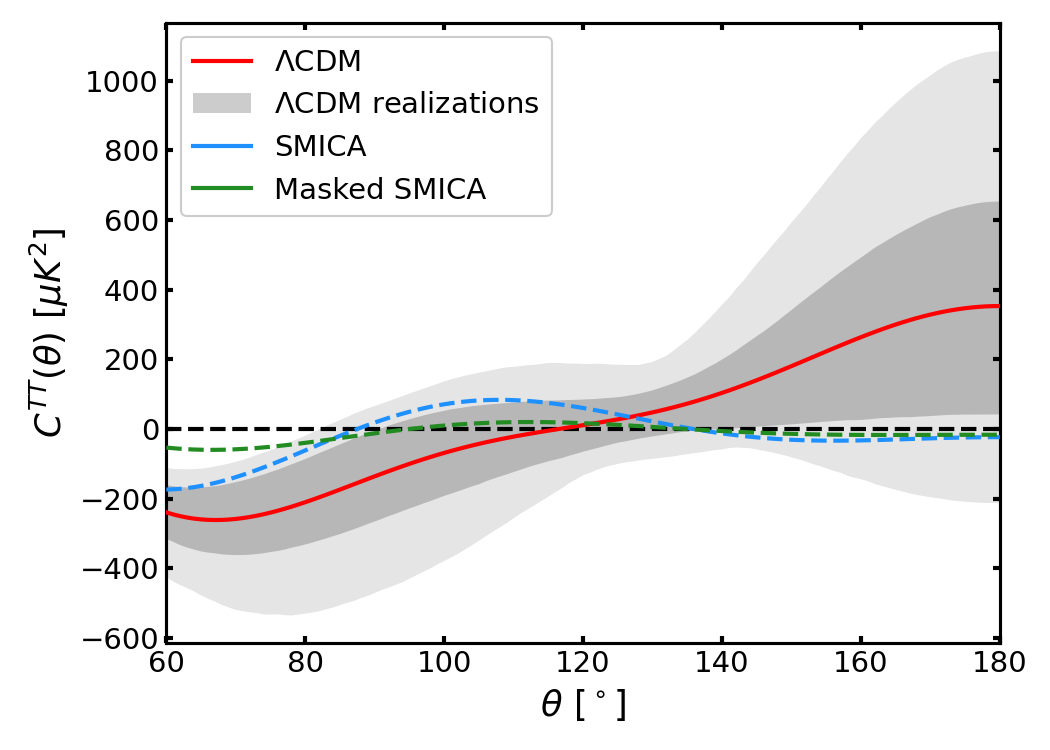}
    \caption{Two-points angular correlation function of CMB temperature. Here, we assume $\ell_{\rm max}= 4$. The red line is the mean of the $\Lambda$CDM realizations and the gray bands are the 1 and 2 $\sigma$ around that.}
    \label{fig: ang_corr_TT_lmax4}
\end{figure}

Thus, at this point, we can compute $C^{\rm TT}(\theta)$ from the SMICA map to understand the actual anomaly. Figure~\ref{fig: ang_corr_TT_lmax4} shows the result.
One can see that the SMICA map, especially the masked one, has a low correlation for scales larger than $\sim 60^\circ$. Also, note that the curves shown in figure~\ref{fig: ang_corr_TT_lmax4} are not equal to those shown in \cite{schwarzCMBAnomaliesPlanck2016}. Indeed, assuming $\ell_{\rm max} = 4$ means that figure~\ref{fig: ang_corr_TT_lmax4} shows only the angular correlation given by the first four multipoles of the expansion.

\begin{figure}[t]
    \centering
    \includegraphics[width = 0.49\textwidth]{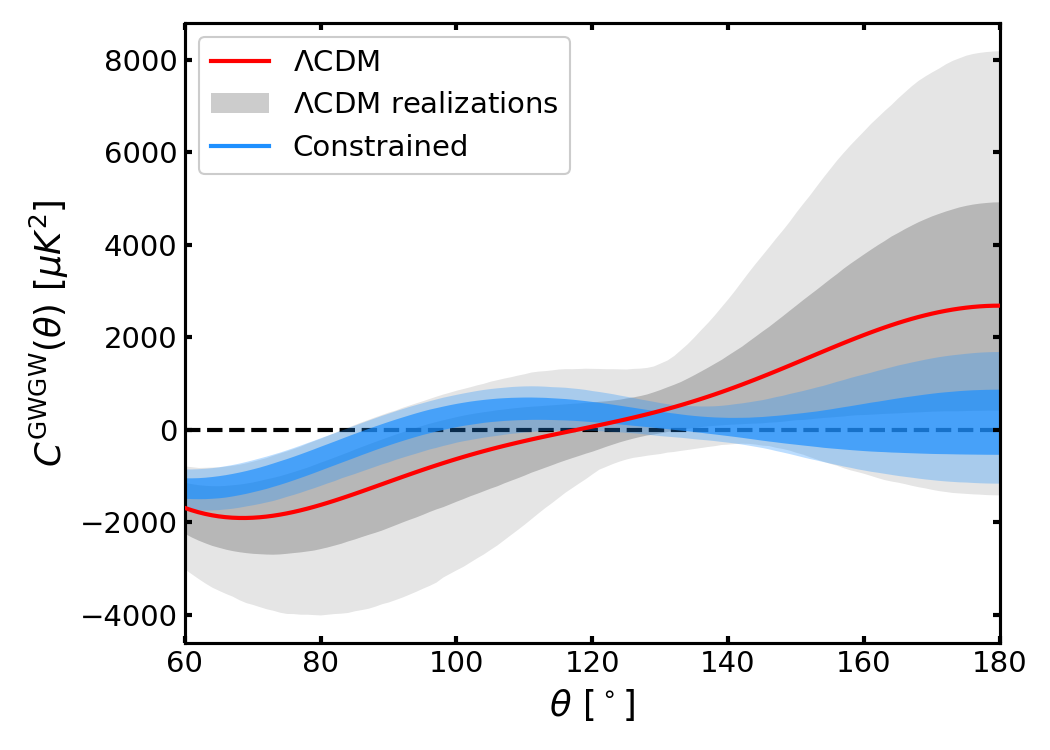}
    \includegraphics[width = 0.49\textwidth]{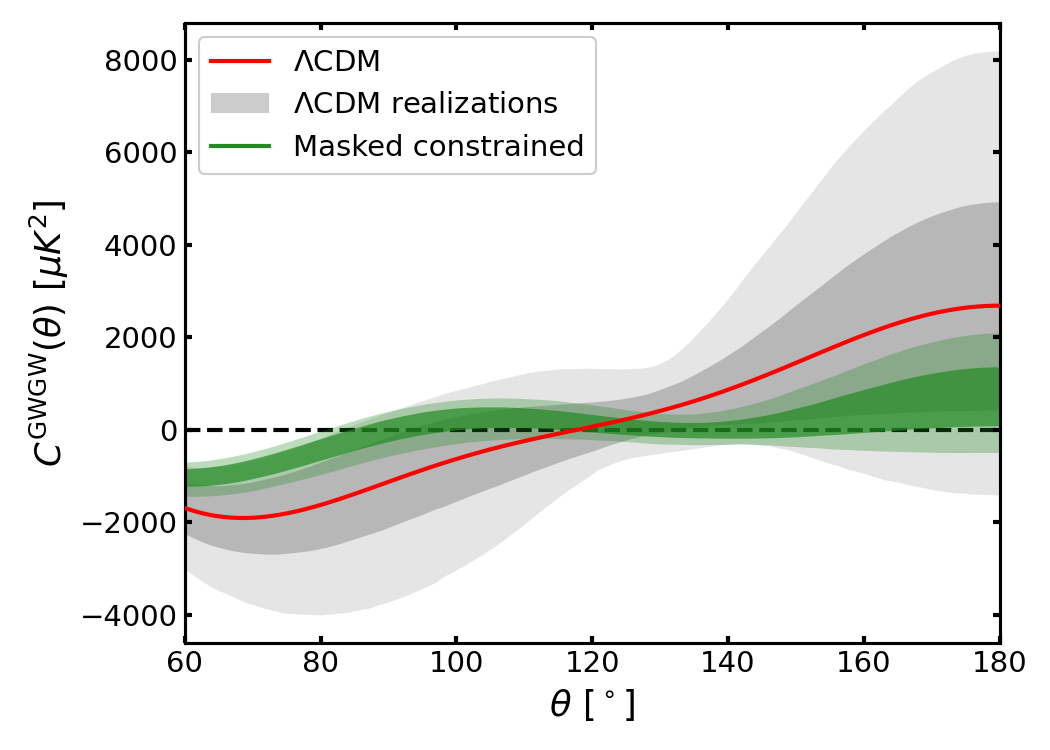}
    \caption{Two-points angular correlation function of the CGWB. The left and right panels refer respectively to the full-sky and masked cases. Here we assume $\ell_{\rm max}= 4$. The red line is the mean of the $\Lambda$CDM realizations and the gray bands are the 1 and 2 $\sigma$ around that.}
    \label{fig: ang_corr_CGWB_lmax4}
\end{figure}

Let us now look at the CGWB realizations. Figure~\ref{fig: ang_corr_CGWB_lmax4} shows both the full-sky and masked realizations.
Here, we can appreciate the fact that the very high correlation existing between TT and GWGW contributes to severely shrinking the dispersion of the CGWB realizations. Furthermore, the TT mask makes them very consistent with zero on almost all the scales considered ($>80^\circ$). These two features already show that GWs could be a pristine probe to test the fluke hypothesis.

\begin{figure}[t]
    \centering
    \includegraphics[width = 0.49\textwidth]{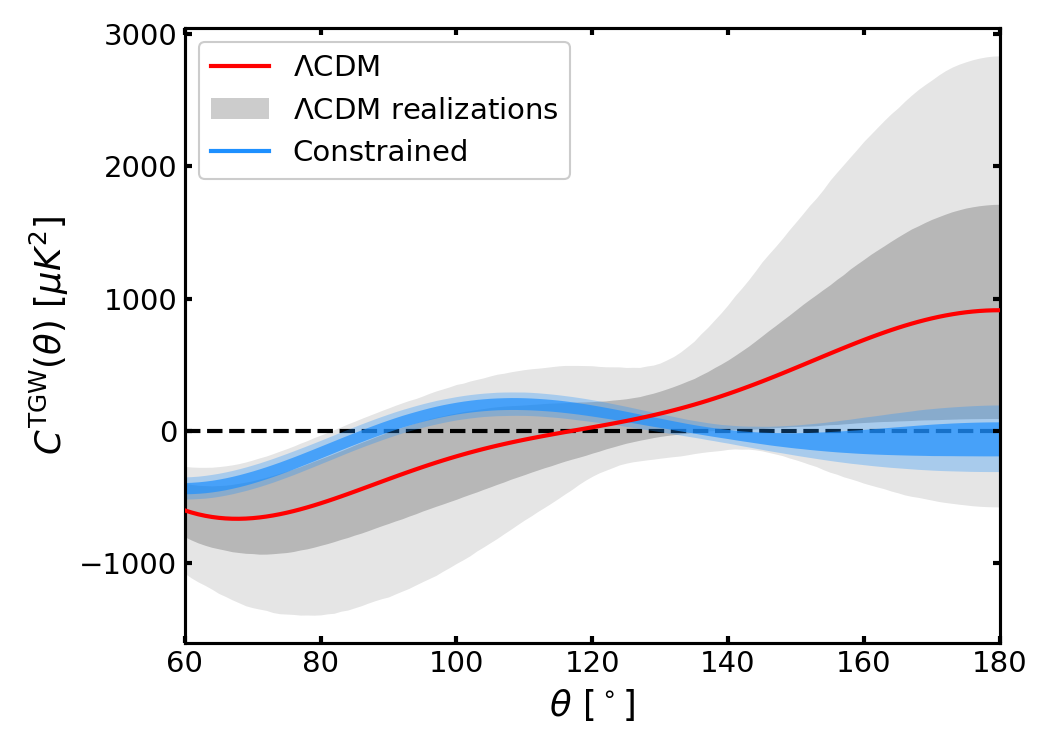}
    \includegraphics[width = 0.49\textwidth]{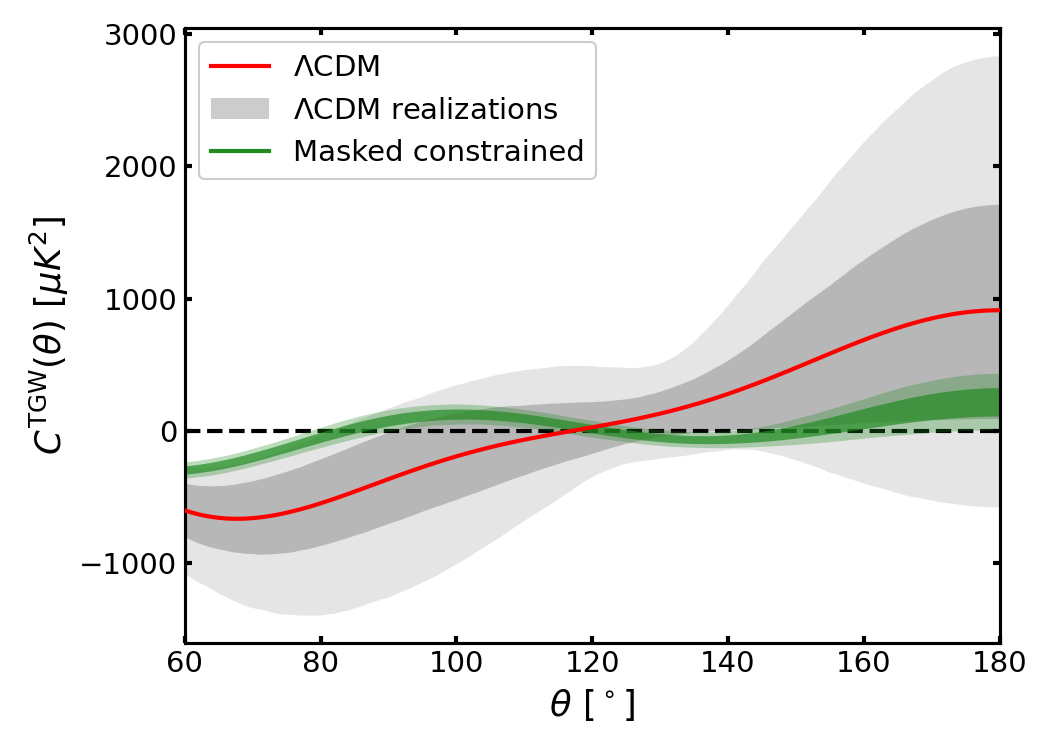}
    \caption{Two-points angular correlation function of TGW. The left and right panels refer respectively to the full-sky and masked cases. Here we assume $\ell_{\rm max}= 4$. The red line is the mean of the $\Lambda$CDM realizations and the gray bands are the 1 and 2 $\sigma$ around that.}
    \label{fig: ang_corr_X_lmax4}
\end{figure}

Finally, figure~\ref{fig: ang_corr_X_lmax4} shows the angular correlation functions of TGW. It shows that the dispersion of the constrained realizations of TGW is even smaller than that of the CGWB. This suggests that TGW could also be a very interesting probe for this analysis of the fluke hypothesis. In appendix~\ref{app: ang_corr_cgwb}, we show the angular correlation functions of GWGW and TGW assuming $\ell_{\rm max} = 2000$ to appreciate the difference w.r.t. CMB temperature, while neglecting the expected performance of future interferometers.

\subsection{S-statistic for auto- and cross-correlations} \label{sec: quantify}
To quantify this lack of temperature correlation, \citet{spergel2003FirstYearWilkinsonMicrowave} introduced the quantity
\begin{equation}
    S_{1/2} = \int_{-1}^{1/2} \qty[C(\theta)]^2 {\rm d}(\cos\theta)\ ,
\end{equation}
which integrates the squared correlation on scales larger than $60^\circ$. Clearly, this naturally captures the total distance between the angular correlation and zero in that angular range, regardless of its sign.

\citet{copi2013LargeAngleCMBSuppressionPolarizationa} introduced a way to find the optimal angular range in which an additional observable can provide most of the information. Thus, by varying the minimum and maximum angles, the $S_{1/2}$ estimator is recast to

\begin{equation}
    S_{\theta_{\rm min}, \theta_{\rm max}} = \int_{\theta_{\rm min}}^{\theta_{\rm max}} \qty[C(\theta)]^2 {\rm d}(\cos\theta) = \sum_{\ell = 2}^{\ell_{\rm max}} \sum_{\ell^\prime = 2}^{\ell_{\rm max}} \frac{(2\ell+1)}{4\pi} \frac{(2\ell^\prime+1)}{4\pi} C_\ell^{XX} I_{\ell \ell^\prime}^{\theta_{\rm min}, \theta_{\rm max}} C_{\ell^\prime}^{XX}\ ,
\label{eq: S12}
\end{equation}
with 
\begin{equation}
    I_{\ell \ell^\prime}^{\theta_{\rm min}, \theta_{\rm max}} = \int_{\theta_{\rm min}}^{\theta_{\rm max}} P_\ell(x)P_{\ell^\prime}(x)\dd x \ .
\end{equation}

Here, we have introduced $\ell_{\rm max} \neq \infty$ since we must account for the fact that a realistic observation depends on the angular resolution of the experiments considered. In fact, above a certain multipole, we know that noise will dominate the measurement. The same treatment was adopted in \cite{chiocchetta2021LackofcorrelationAnomalyCMBlargea}, where the multipole cut was made when the signal-to-noise ratio of the E-mode polarization was essentially saturated. It should be underlined that $\ell_{\rm max}$ affects the calculation of $I_{\ell \ell^\prime}^{\theta_{\rm min}, \theta_{\rm max}}$. In fact, multipoles up to $\ell_{\rm max}$ will describe scales larger than approximately $180^\circ/\ell_{\rm max}$. Instead, the behavior on smaller scales will be determined by noise or by interference of the Legendre modes considered. To be conservative on which scales we consider well described by the available multipoles, we will impose a lower bound on $\theta_{\rm min}$ and $\theta_{\rm max}$ of 
\begin{equation}
    \theta_{\rm cut} = \frac{180^\circ}{\ell_{\rm max}-1}\ ,
\end{equation}
so that $\theta_{\rm min}, \theta_{\rm max} \geq \theta_{\rm cut}$. In this way, we can also avoid ``border effects'' when approaching $180^\circ/\ell_{\rm max}$.

We can apply the estimator defined in eq.~\eqref{eq: S12} to the three angular power spectra by substituting $C_\ell^{\rm TT}$, $C_\ell^{\rm GWGW}$, or $C_\ell^{\rm TGW}$ into the above definitions. This provides us with three quantities: $S^{\rm TT}_{\theta_{\rm min}, \theta_{\rm max}}$, $S^{\rm GWGW}_{\theta_{\rm min}, \theta_{\rm max}}$, and $S^{\rm TGW}_{\theta_{\rm min}, \theta_{\rm max}}$. 

In addition to the three $S_{\theta_{\rm min}, \theta_{\rm max}}$ estimators, we study the joint estimator following the definition introduced in \citet{chiocchetta2021LackofcorrelationAnomalyCMBlargea}, which yields

\begin{equation}
    S^{\rm TT,Z}_{\theta_{\rm min}, \theta_{\rm max}} = \sqrt{ \qty(\frac{S^{\rm TT}_{\theta_{\rm min}, \theta_{\rm max}}}{\expval{S^{\rm TT}_{\theta_{\rm min}, \theta_{\rm max}}}})^2 + \qty(\frac{S^{\rm Z}_{\theta_{\rm min}, \theta_{\rm max}}}{\expval{S^{\rm Z}_{\theta_{\rm min}, \theta_{\rm max}}}})^2 }\ ,
    \label{eq: joint_S}
\end{equation}
where $\rm Z = \qty{\rm TGW, GWGW}$. We can also define the combination of all the available estimators as follows
\begin{equation}
    S^{\rm TT, TGW, GWGW}_{\theta_{\rm min}, \theta_{\rm max}} = \sqrt{ \qty(\frac{S^{\rm TT}_{\theta_{\rm min}, \theta_{\rm max}}}{\expval{S^{\rm TT}_{\theta_{\rm min}, \theta_{\rm max}}}})^2 + \qty(\frac{S^{\rm TGW}_{\theta_{\rm min}, \theta_{\rm max}}}{\expval{S^{\rm TGW}_{\theta_{\rm min}, \theta_{\rm max}}}})^2 + \qty(\frac{S^{\rm GWGW}_{\theta_{\rm min}, \theta_{\rm max}}}{\expval{S^{\rm GWGW}_{\theta_{\rm min}, \theta_{\rm max}}}})^2 }\ .
    \label{eq: super_S}
\end{equation}

When we consider constrained realizations of GWGW or TGW, $S^{\rm TT}_{\theta_{\rm min}, \theta_{\rm max}}$ will be replaced with SMICA data, that is, $S^{\rm SMICA}_{\theta_{\rm min}, \theta_{\rm max}}$. Still, both of them, GWGW and SMICA, will be normalized by their unconstrained counterparts.

Essentially, this joint estimator is a sum in quadrature of the two normalized estimators. When this estimator departs from 1, it means that both 
$TT$ and GWGW are doing so. Notice that in the constrained case, $S^{\rm TT, GWGW}_{\theta_{\rm min}, \theta_{\rm max}}$ cannot be lower than $S^{\rm SMICA}_{\theta_{\rm min}, \theta_{\rm max}}/\expval{S^{\rm TT}_{\theta_{\rm min}, \theta_{\rm max}}}$. The same applies to $S^{\rm TT, TGW}_{\theta_{\rm min}, \theta_{\rm max}}$ and $S^{\rm TT, TGW,GWGW}_{\theta_{\rm min}, \theta_{\rm max}}$.

In section~\ref{sec: constrained} we mentioned that we normalize the CMB anisotropies to $T_{\rm 0}$, so the estimator $S^{\rm TT}_{\theta_{\rm min}, \theta_{\rm max}}$ will be expressed in units of $\mu K^4$. Since we also use this normalization for the CGWB, the same is true for $S^{\rm TGW}_{\theta_{\rm min}, \theta_{\rm max}}$ and $S^{\rm GWGW}_{\theta_{\rm min}, \theta_{\rm max}}$.

\subsection{Optimal angular range} \label{sec: metho_optimal}
In order to select the optimal range of angles for the estimator $S_{\theta_{\rm min}, \theta_{\rm max}}$ (see eq.~\eqref{eq: S12}), we apply the procedure introduced in \cite{copi2013LargeAngleCMBSuppressionPolarizationa} both to the case of single estimators and to the joint case. Given an observation of the CMB temperature, the optimal range of angles is issued by $\theta_{\rm min}$ and $\theta_{\rm max}$ that maximize the displacement between the values of $S_{\theta_{\rm min}, \theta_{\rm max}}$ obtained through constrained and unconstrained realizations of the CGWB. This is somehow telling us in what angular range the CGWB is most sensitive to the signal we observe in TT, in terms of lack-of-correlation. To quantify this displacement, we proceed as follows:
\begin{enumerate}
    \item we first grid the values of $\theta_{\rm min}$ and $\theta_{\rm max}$ and at each node of the grid we compute the constrained and unconstrained estimators $S^{\rm GWGW}_{\theta_{\rm min}, \theta_{\rm max}}$;
    \item then, for each node, we compute the 99th percentile of the values given by the constrained realizations, and we count how many unconstrained realizations give a higher value of $S_{\theta_{\rm min}, \theta_{\rm max}}$. Translating this in terms of a percentage, we call this quantity Percentage Displacement (PD);
    \item finally, by studying the results on the grid, we can identify which regions give higher PDs. When searching for which specific configuration gives the optimal angular range, we have to keep in mind what our statistical error is in evaluating the PD. We can define it as $\sigma_{\rm PD} = 1/N \simeq 0.01\%$. Thus, we round the obtained PDs to the second decimal figure;
    \item then, if two or more configurations have the same PD, we privilege the one where $\Delta \theta = \theta_{\rm max} - \theta_{\rm min}$ is maximal. In this way, we are choosing the case where we integrate over the most scales.
\end{enumerate}

We repeat the same analysis for $S^{\rm TGW}_{\theta_{\rm min}, \theta_{\rm max}}$, $S^{\rm TT, GWGW}_{\theta_{\rm min}, \theta_{\rm max}}$, $S^{\rm TT, TGW}_{\theta_{\rm min}, \theta_{\rm max}}$ and $S^{\rm TT, TGW, GWGW}_{\theta_{\rm min}, \theta_{\rm max}}$. In the joint cases, we find the optimal angular range to assess the fluke hypothesis simultaneously exploiting more than one field.

\subsection{Significance accounting for the look-elsewhere effect}\label{sec: signi}

Until now, we have only mentioned the fluke hypothesis, explaining the methodology we want to use to find out whether GWs can help to reject at least some of the assumptions of the $\Lambda$CDM model. Despite this, we cannot say anything about the significance of the anomaly using the optimal angles analysis. 

Indeed, could we say that the significance of the anomaly is the distance of a certain data point (i.e. a specific value of $S_{\theta_{\rm min}, \theta_{\rm max}}$ of a realization) from the unconstrained $\Lambda$CDM simulations (in terms of sigma or p-value)? The answer is not that straightforward, and the reason is called the ``look-elsewhere'' effect \cite{planckcollaboration2020Planck2018ResultsVIIa}. Indeed, a certain realization would have a significance in each angular range, and by choosing the optimal angular range, we would a posteriori choose the configuration that maximizes that number. In other words, we design the analysis to provide the best possible solution. In a different angular range, the data may agree completely with the prediction of $\Lambda$CDM.

A well-known strategy to account for the look-elsewhere effect is to study the PD of the realizations irrespective of their angular range: one searches for the maximum PD for each realization without caring about the angular range. Then, the fraction of these probabilities that are found to be lower than the maximum PD yielded by the data is a global p-value \cite{planckcollaboration2020Planck2018ResultsVIIa}.

Despite this, in this work we choose to follow a novel procedure to account for the look-elsewhere effect. We define a new estimator for the lack of correlation as the sum of all configurations of the angular range of $S_{\theta_{\rm min}, \theta_{\rm max}}$. For a generic observable X, it reads
\begin{equation}
    S^{\rm XX} \equiv \sum_{\qty{\theta_{\rm min}, \theta_{\rm max}}} \frac{S_{\theta_{\rm min}, \theta_{\rm max}}^{\rm XX}}{\expval{S_{\theta_{\rm min}, \theta_{\rm max}}^{\rm XX}}}\ .
    \label{eq: signi_estimator}
\end{equation}

This may be regarded as a marginalization of $S^{\rm XX}_{\theta_{\rm min}, \theta_{\rm max}}$ over angular information. Now, $S^{\rm XX}$ tells us whether a simulation of X has an anomalously low covariance, regardless of the angular range considered. This reasoning can be applied to each field that we considered in the previous analysis; thus, TT, TGW, GWGW, and their combinations. Note that we also normalize the values entering the sum by the mean of the unconstrained realizations, as done in eq.~\eqref{eq: joint_S} and~\eqref{eq: super_S}. This is because we do not want our results to be driven by the eventual presence of a high $S_{\theta_{\rm min}, \theta_{\rm max}}$ region, which would dominate the sum over angular ranges. 

Using this new estimator, we can study the significance of the lack-of-correlation anomaly taking into account the look-elsewhere effect. In fact, comparing the resulting values of $S$ with those obtained from the CMB data or from the constrained realizations of the CGWB, we compute the significance in terms of sigma. In other words, with this approach we can still affirm that a future observation of the CGWB falling outside the constrained curve indicates that we must revise our assumptions. However, we can also associate a significance to data points that fall within the predicted curve. In this way, we can study the effects of the inclusion of GWs on the actual assessment of the physical origin of the anomaly.

Finally, to quantify the significance we first compute the p-value of each realization (so the probability of getting an unconstrained realization with a lower value than the considered one), and then we translate the p-value in terms of a $\sigma$-distance w.r.t. a normal Gaussian. Given that we have a finite number of realizations, there may be cases in which no unconstrained realizations are found below a certain constrained value of $S$. In such a case, we recall that each percentage we obtain has an error of $0.01\%$, thus we associate to null p-value realizations a probability of $0.01\%$. This may underestimate the significance of the furthest values of $S$ w.r.t. the unconstrained realizations.

\section{Results} \label{sec: results}
In the following, we show the results assuming the most pessimistic case analyzed of $\ell_{\rm max} = 4$. 
Finally, in appendix~\ref{app: sensitivity} we show the other two cases of $\ell_{\rm max} =$ 6 and 10. 

\subsection{Optimal angles} \label{sec: res_optimal}
\begin{figure}[t]
    \centering
    \includegraphics[width = 0.49\textwidth]{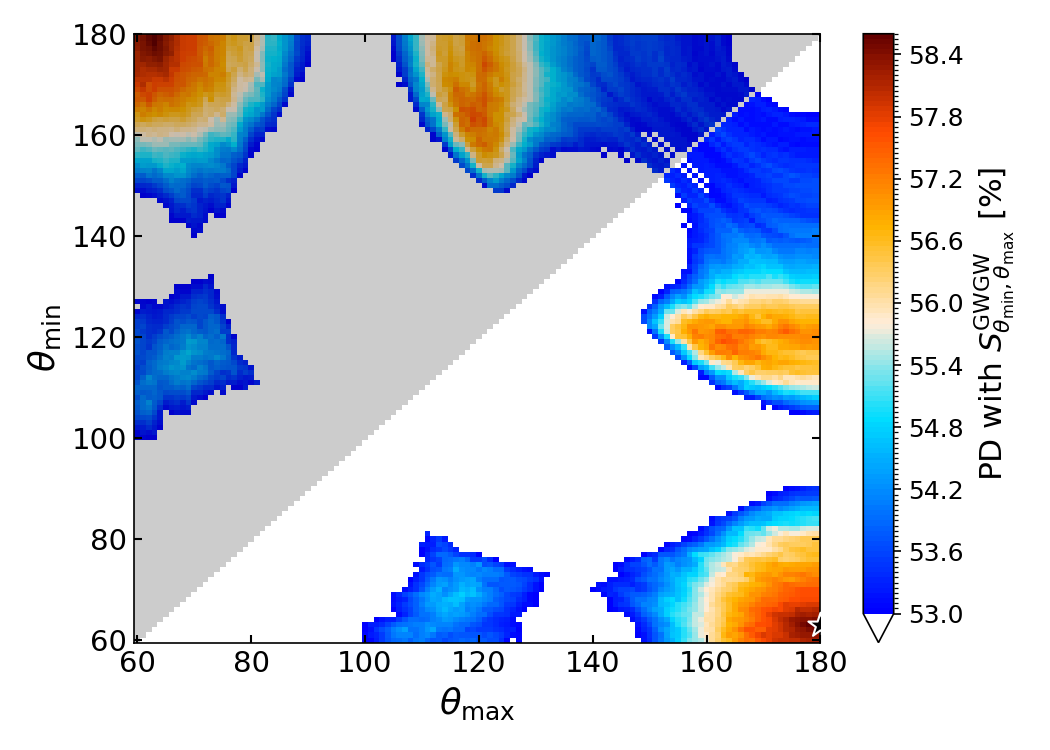}
    \includegraphics[width = 0.49\textwidth]{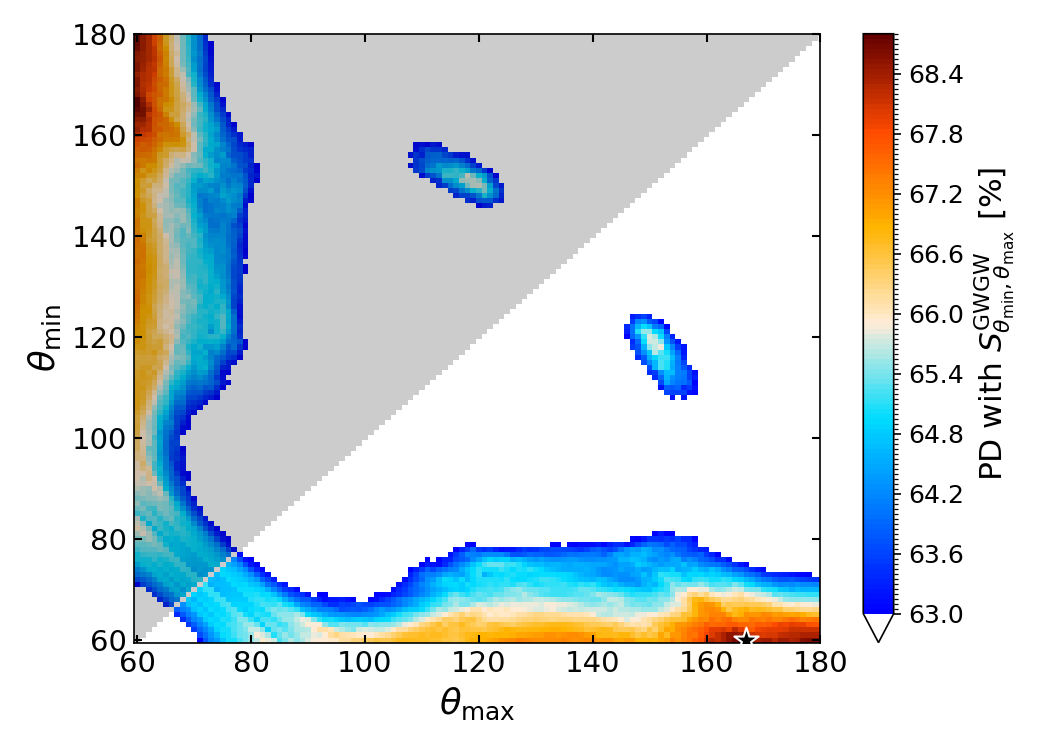}
    \caption{Optimal angles for GWGW. The left and right panels show respectively the results when we assume either full-sky or masked SMICA as our CMB observation. The star shows the best angular range. Here we assume $\ell_{\rm max}= 4$.}
    \label{fig: opt_GWGW}
\end{figure}

\begin{figure}[t]
    \centering
    \includegraphics[width = 0.49\textwidth]{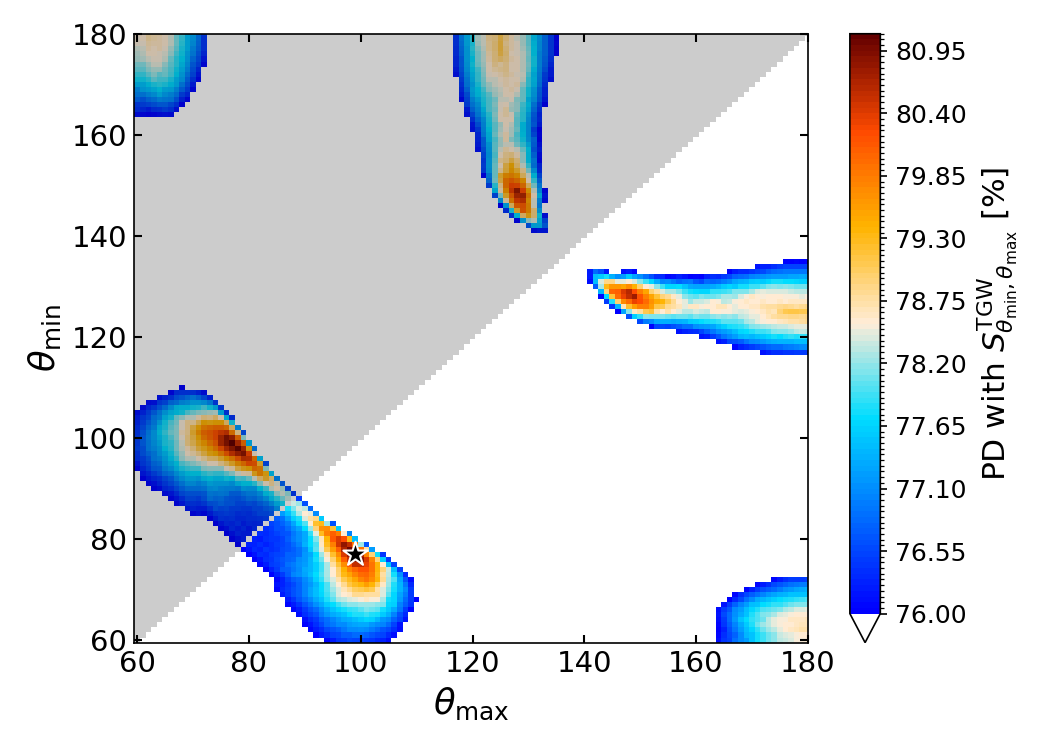}
    \includegraphics[width = 0.49\textwidth]{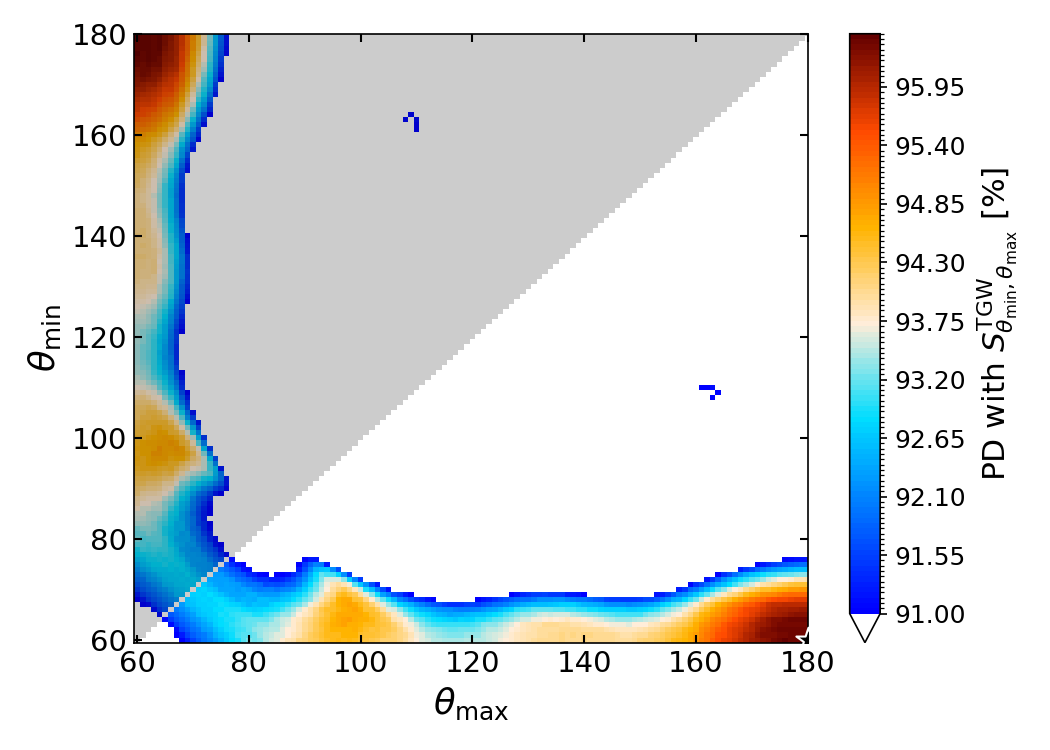}
    \caption{Optimal angles for TGW. The left and right panels show respectively the results when we assume either full-sky or masked SMICA as our CMB observation. The star shows the best angular range. Here we assume $\ell_{\rm max}= 4$.}
    \label{fig: opt_X}
\end{figure}

We start studying the optimal range of angles for each estimator as described in section~\ref{sec: metho_optimal}. 
We show the results for $S_{\theta_{\rm min}, \theta_{\rm max}}^{\rm GWGW}$ in figure~\ref{fig: opt_GWGW} and in figure~\ref{fig: opt_X} for $S_{\theta_{\rm min}, \theta_{\rm max}}^{\rm TGW}$. The left and right panels show the results for the full-sky and masked analysis, respectively. Note that every optimal angle plot we will show from now on is symmetric by construction; however, we add a gray-shaded region to emphasize that $\theta_{\rm min}$ cannot be greater than $\theta_{\rm max}$. 

The PDs of both GWGW and TGW change when passing from the full sky to masked analysis, showing that masking the galactic plane enhances the lack of correlation \cite{gruppuso2013LowVarianceLargescales, gruppuso2014TwopointCorrelationFunctionWMAP, nataleLackPowerAnomaly2019, planckcollaboration2020Planck2018ResultsVIIa}. Comparing this result with \cite{copi2013LargeAngleCMBSuppressionPolarizationa}, who performed the same analysis using the CMB E-mode polarization, we can appreciate how powerful GWs are in testing the fluke hypothesis. In fact, having a $\sim 90\%$ PD means that the distribution that TGW must follow if $\Lambda$CDM is correct is extremely peaked and localized. This maximizes the possibility of rejecting it in the event that a future observation happens to be outside that distribution, with the corresponding level of significance.

\begin{figure}[t]
    \centering
    \includegraphics[width = 0.49\textwidth]{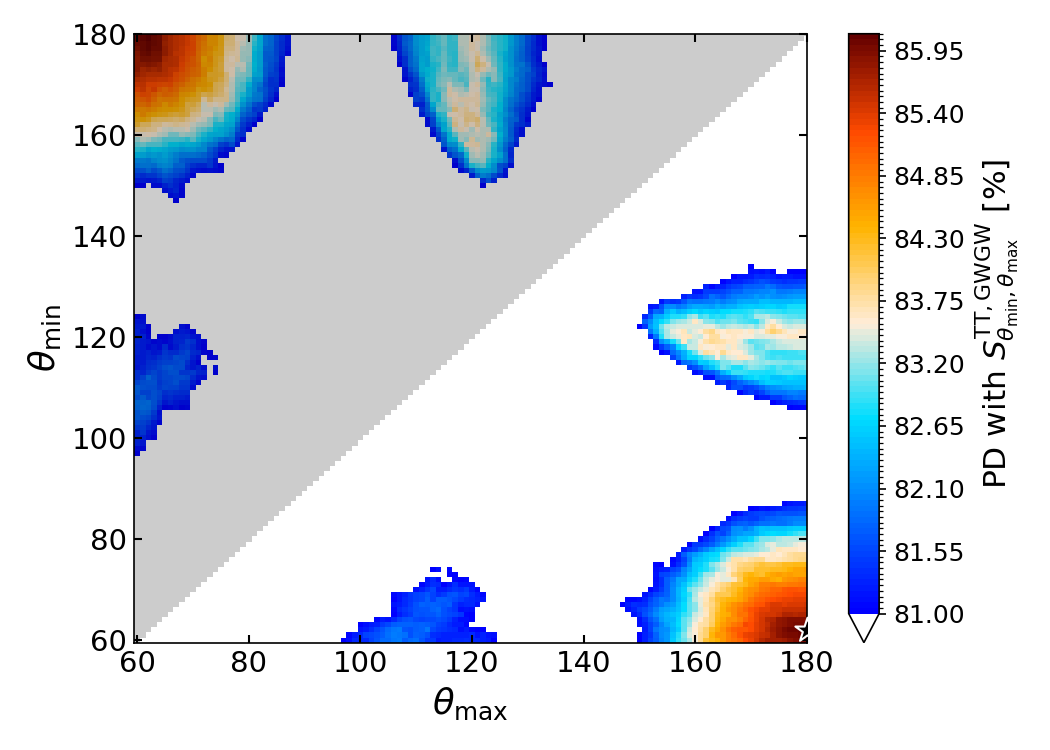}
    \includegraphics[width = 0.49\textwidth]{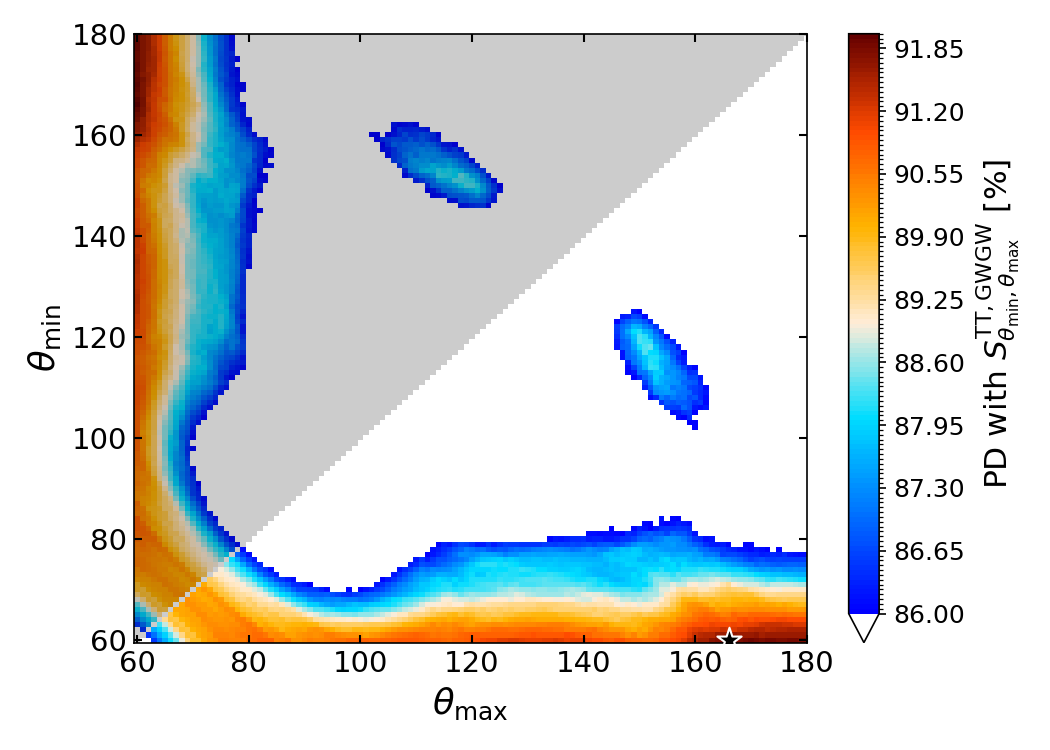}
    \caption{Optimal angles for the combination of TT and GWGW. The left and right panels show respectively the results when we assume either full-sky or masked SMICA as our CMB observation. The star shows the best angular range. Here we assume $\ell_{\rm max}= 4$.}
    \label{fig: opt_TGW}
\end{figure}

\begin{figure}[t]
    \centering
    \includegraphics[width = 0.49\textwidth]{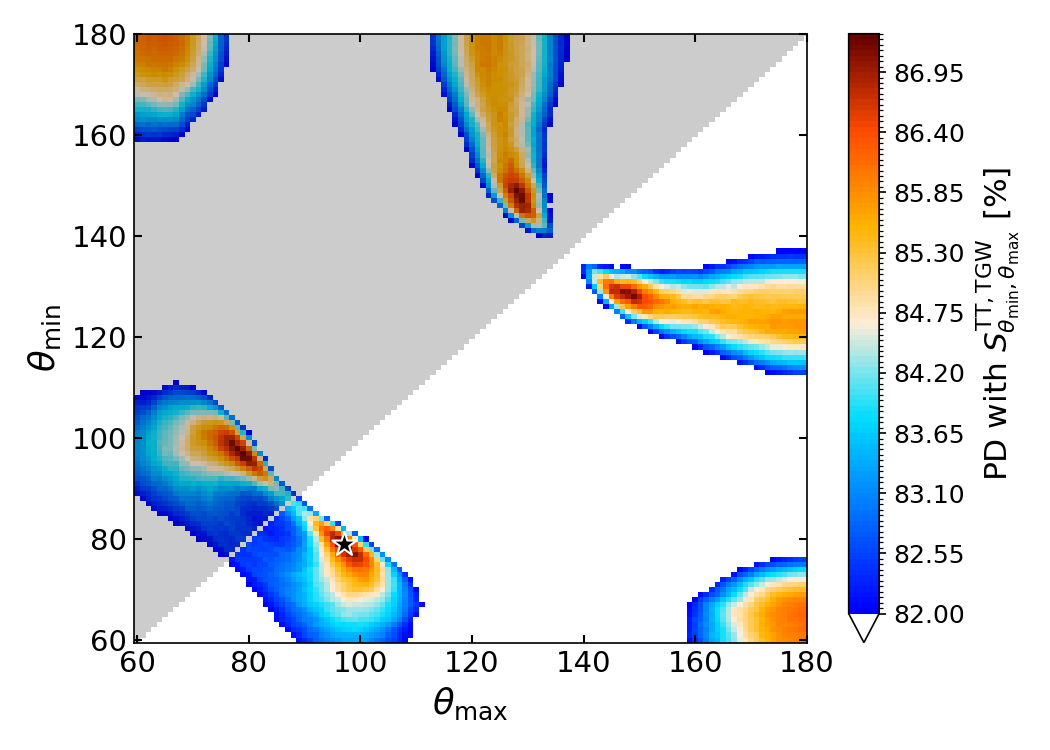}
    \includegraphics[width = 0.49\textwidth]{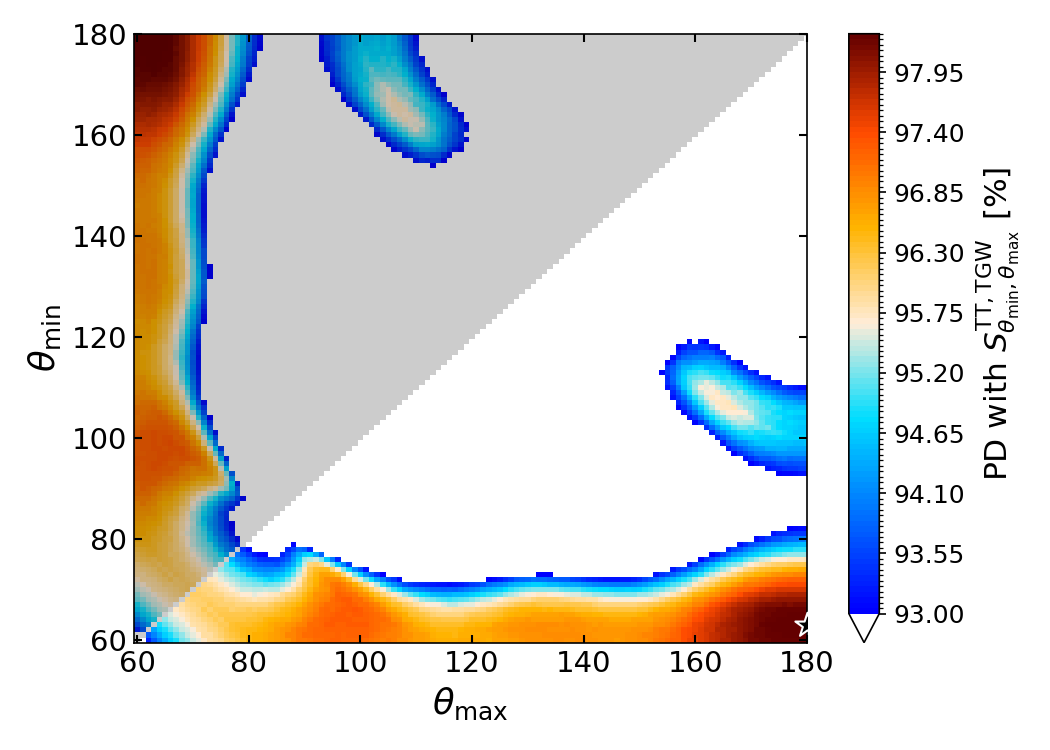}
    \caption{Optimal angles for the combination of TT and TGW. The left and right panels show respectively the results when we assume either full-sky or masked SMICA as our CMB observation. The star shows the best angular range. Here we assume $\ell_{\rm max}= 4$.}
    \label{fig: opt_TX}
\end{figure}

\begin{figure}[t]
    \centering
    \includegraphics[width = 0.49\textwidth]{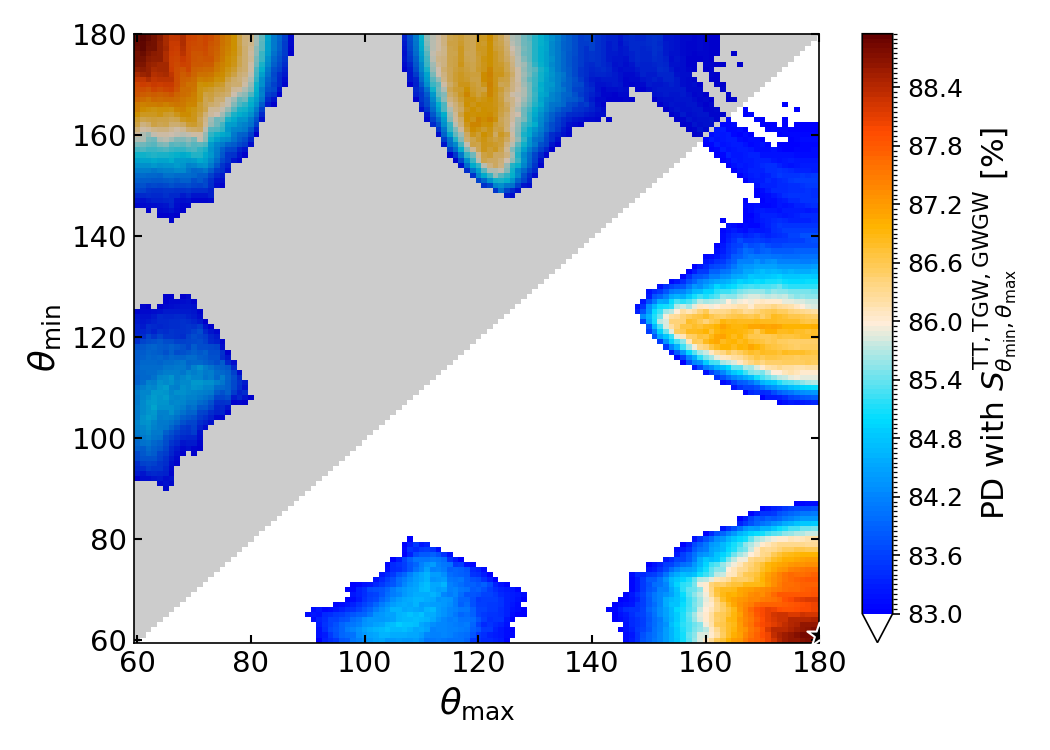}
    \includegraphics[width = 0.49\textwidth]{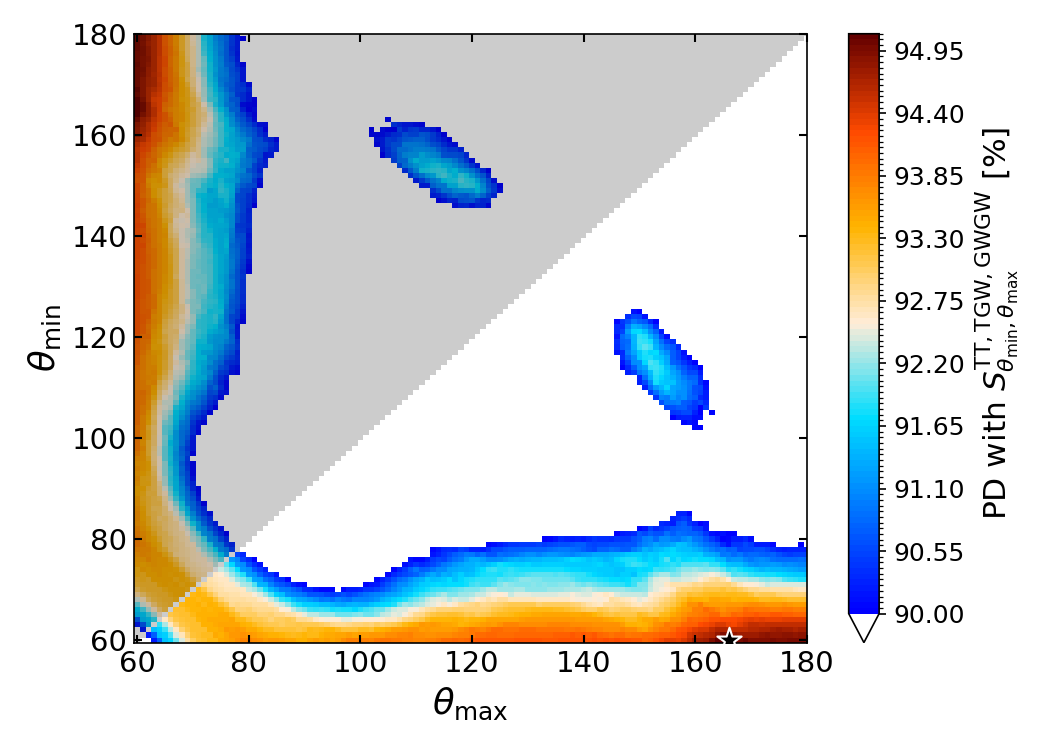}
    \caption{Optimal angles for the combination of TT, TGW and GWGW. The left and right panels show respectively the results when we assume either full-sky or masked SMICA as our CMB observation. The star shows the best angular range. Here we assume $\ell_{\rm max}= 4$.}
    \label{fig: opt_TXGW}
\end{figure}

We repeat the same analysis for $S^{\rm TT, GWGW}_{\theta_{\rm min}, \theta_{\rm max}}$, $S^{\rm TT, TGW}_{\theta_{\rm min}, \theta_{\rm max}}$ and $S^{\rm TT, TGW, GWGW}_{\theta_{\rm min}, \theta_{\rm max}}$; see figures~\ref{fig: opt_TGW}-\ref{fig: opt_TX}-\ref{fig: opt_TXGW}. Once again, the left and right panels show the full-sky and masked cases. Note that in every case there is a high-PD region near the range $\qty[60^\circ, 180^\circ]$, i.e. the range of the original estimator $S_{1/2}$ introduced by \citet{spergel2003FirstYearWilkinsonMicrowave}.

In this context, the PD reaches values well above $90\%$, especially for the masked case of $S^{\rm TT, TGW}_{\theta_{\rm min}, \theta_{\rm max}}$, where the optimal region reaches $\sim 98\%$. This proves that GWs may be crucial to test the fluke hypothesis and eventually reject the $\Lambda$CDM model. Interestingly, the best results are obtained by combining the CMB temperature with its cross-correlation with the CGWB. Instead, including the GWGW spectrum seems to reduce the ability to test the fluke hypothesis. All the results on the optimal angles are summarized in table~\ref{tab: optimal}.
\begin{table}[t]
 \centering
\begin{tabular}{l|cc|cc|cc|cc|cc}
    \toprule
    \multicolumn{11}{c}{Full-sky} \\
    \multicolumn{1}{c}{} & \multicolumn{2}{c}{$S^{\rm GWGW}_{\theta_{\rm min},\theta_{\rm max}}$} & \multicolumn{2}{c}{$S^{\rm TGW}_{\theta_{\rm min},\theta_{\rm max}}$} & \multicolumn{2}{c}{$S^{\rm TT, GWGW}_{\theta_{\rm min},\theta_{\rm max}}$} & \multicolumn{2}{c}{$S^{\rm TT, TGW}_{\theta_{\rm min},\theta_{\rm max}}$}  & \multicolumn{2}{c}{$S^{\rm TT, TGW, GWGW}_{\theta_{\rm min},\theta_{\rm max}}$} \\ \midrule
    & $\theta_{\rm min}$ & $\theta_{\rm max}$ & $\theta_{\rm min}$ & $\theta_{\rm max}$ & $\theta_{\rm min}$ & $\theta_{\rm max}$ & $\theta_{\rm min}$ & $\theta_{\rm max}$ & $\theta_{\rm min}$ & $\theta_{\rm max}$ \\
    $\ell_{\rm max} = 4$ &  $63^\circ$ &  $180^\circ$ &  $77^\circ$ &   $99^\circ$ &  $62^\circ$ &  $180^\circ$ &  $79^\circ$ &   $97^\circ$ &  $61^\circ$ &  $180^\circ$ \\
   $\ell_{\rm max} = 6$ &  $56^\circ$ &  $121^\circ$ &  $65^\circ$ &  $115^\circ$ &  $55^\circ$ &  $120^\circ$ &  $68^\circ$ &  $116^\circ$ &  $54^\circ$ &  $118^\circ$ \\
  $\ell_{\rm max} = 10$ &  $58^\circ$ &  $124^\circ$ &  $71^\circ$ &  $108^\circ$ &  $58^\circ$ &  $124^\circ$ &  $72^\circ$ &  $109^\circ$ &  $57^\circ$ &  $120^\circ$ \\
    \midrule
    \midrule
    \multicolumn{11}{c}{Mask-sky} \\
    \multicolumn{1}{c}{} & \multicolumn{2}{c}{$S^{\rm GWGW}_{\theta_{\rm min},\theta_{\rm max}}$} & \multicolumn{2}{c}{$S^{\rm TGW}_{\theta_{\rm min},\theta_{\rm max}}$} & \multicolumn{2}{c}{$S^{\rm TT, GWGW}_{\theta_{\rm min},\theta_{\rm max}}$} & \multicolumn{2}{c}{$S^{\rm TT, TGW}_{\theta_{\rm min},\theta_{\rm max}}$}  & \multicolumn{2}{c}{$S^{\rm TT, TGW, GWGW}_{\theta_{\rm min},\theta_{\rm max}}$} \\ \midrule
    & $\theta_{\rm min}$ & $\theta_{\rm max}$ & $\theta_{\rm min}$ & $\theta_{\rm max}$ & $\theta_{\rm min}$ & $\theta_{\rm max}$ & $\theta_{\rm min}$ & $\theta_{\rm max}$ & $\theta_{\rm min}$ & $\theta_{\rm max}$ \\
   $\ell_{\rm max} = 4$ &  $60^\circ$ &  $167^\circ$ &  $60^\circ$ &  $180^\circ$ &  $60^\circ$ &  $166^\circ$ &  $63^\circ$ &  $180^\circ$ &  $60^\circ$ &  $166^\circ$ \\
   $\ell_{\rm max} = 6$ &  $50^\circ$ &   $95^\circ$ &  $54^\circ$ &  $129^\circ$ &  $42^\circ$ &  $122^\circ$ &  $55^\circ$ &  $180^\circ$ &  $40^\circ$ &  $122^\circ$ \\
  $\ell_{\rm max} = 10$ &  $74^\circ$ &   $77^\circ$ &  $61^\circ$ &  $132^\circ$ &  $60^\circ$ &   $88^\circ$ &  $62^\circ$ &  $134^\circ$ &  $59^\circ$ &   $88^\circ$ \\
    \bottomrule
\end{tabular}
\caption{Optimal angles for every observable and combination of them.}
\label{tab: optimal}
\end{table}

\begin{figure}[t]
    \centering
    \includegraphics[width = 0.49\textwidth]{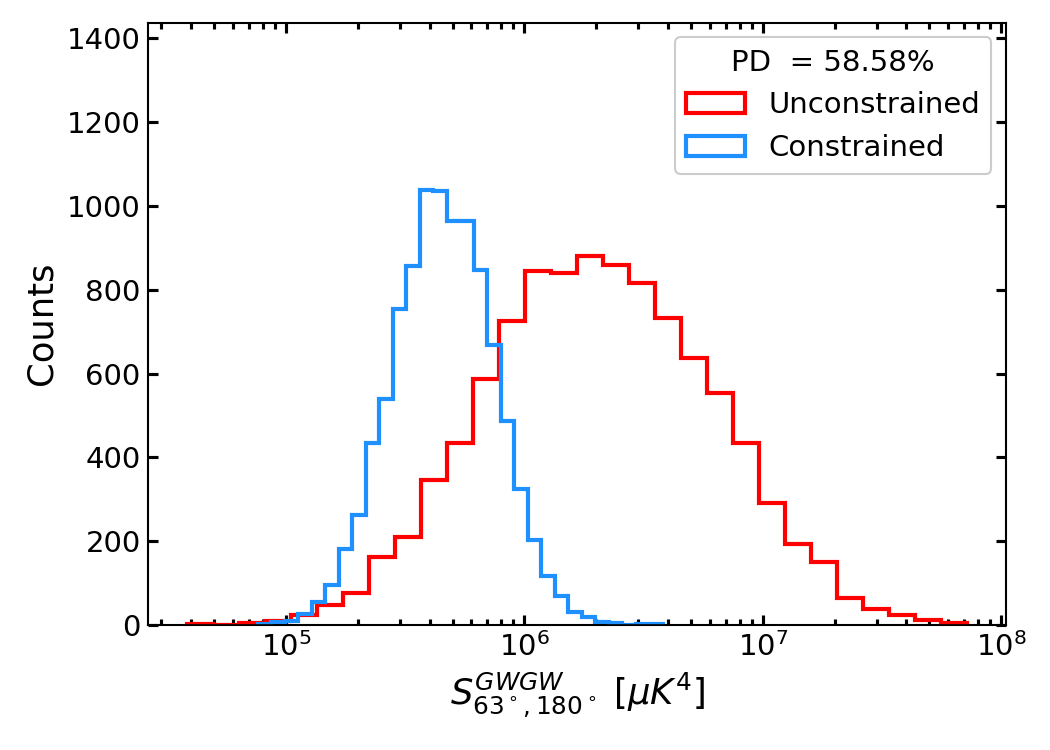}
    \includegraphics[width = 0.49\textwidth]{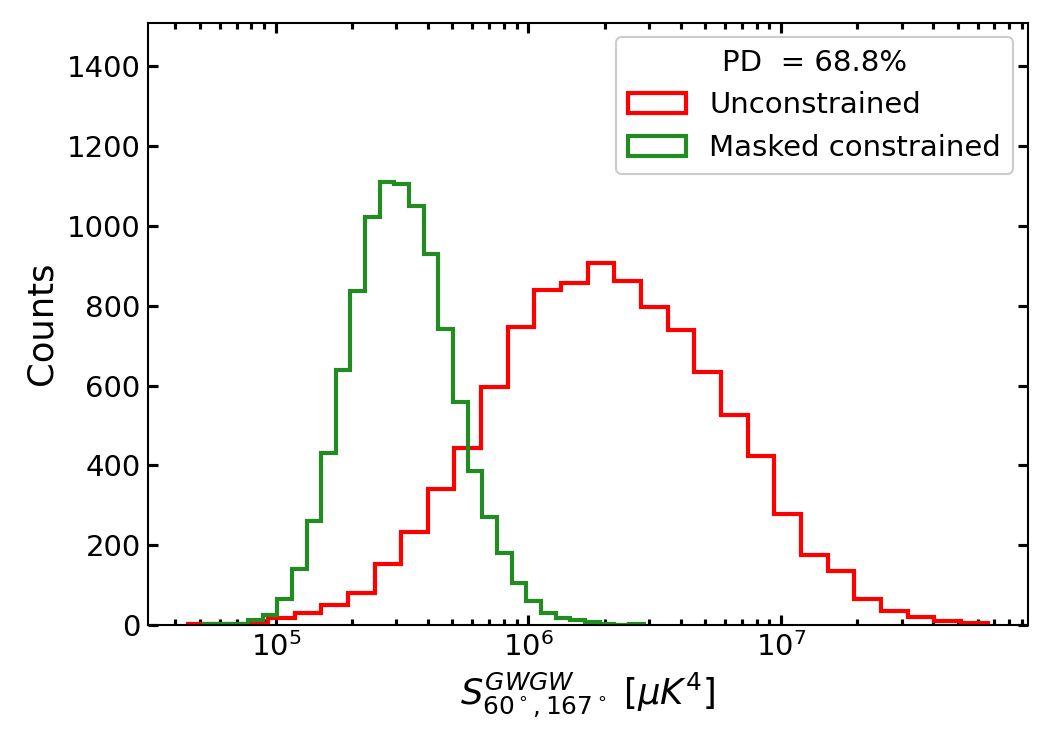}
    \caption{$S^{\rm GWGW}_{\theta_{\rm min}, \theta_{\rm max}}$ distributions for the full-sky and masked cases. The angles $\theta_{\rm min}, \theta_{\rm max}$ are chosen to be the optimal ones of both cases. Here we assume $\ell_{\rm max}= 4$.}
    \label{fig: hist_GWGW}
\end{figure}

\begin{figure}[t]
    \centering
    \includegraphics[width = 0.49\textwidth]{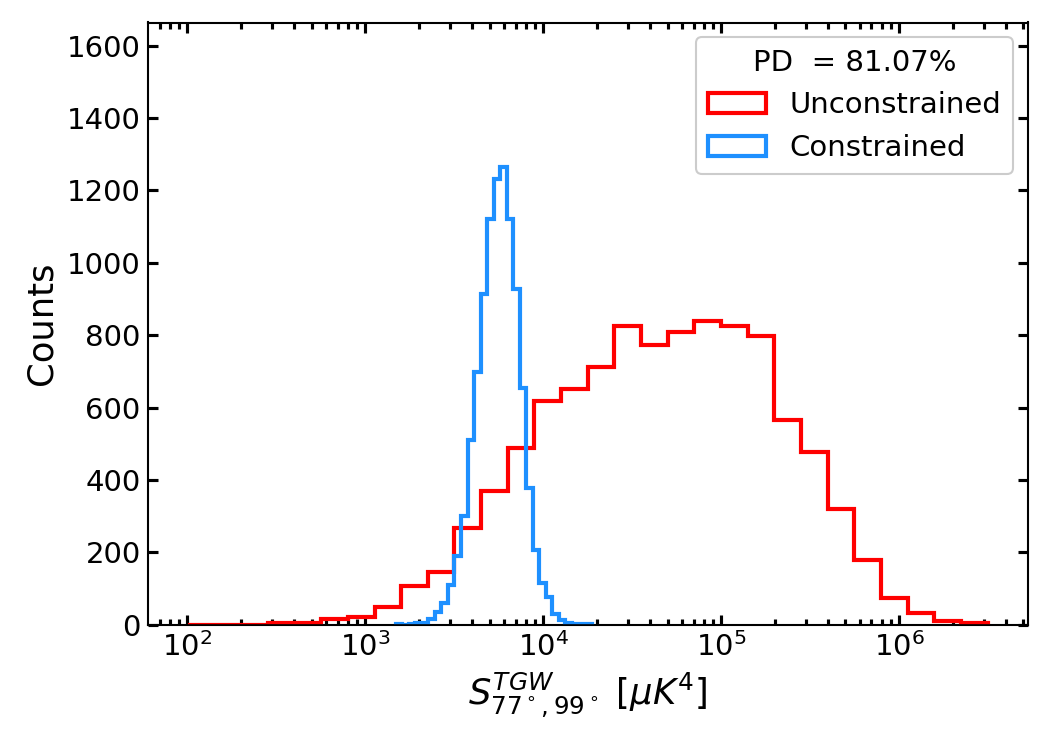}
    \includegraphics[width = 0.49\textwidth]{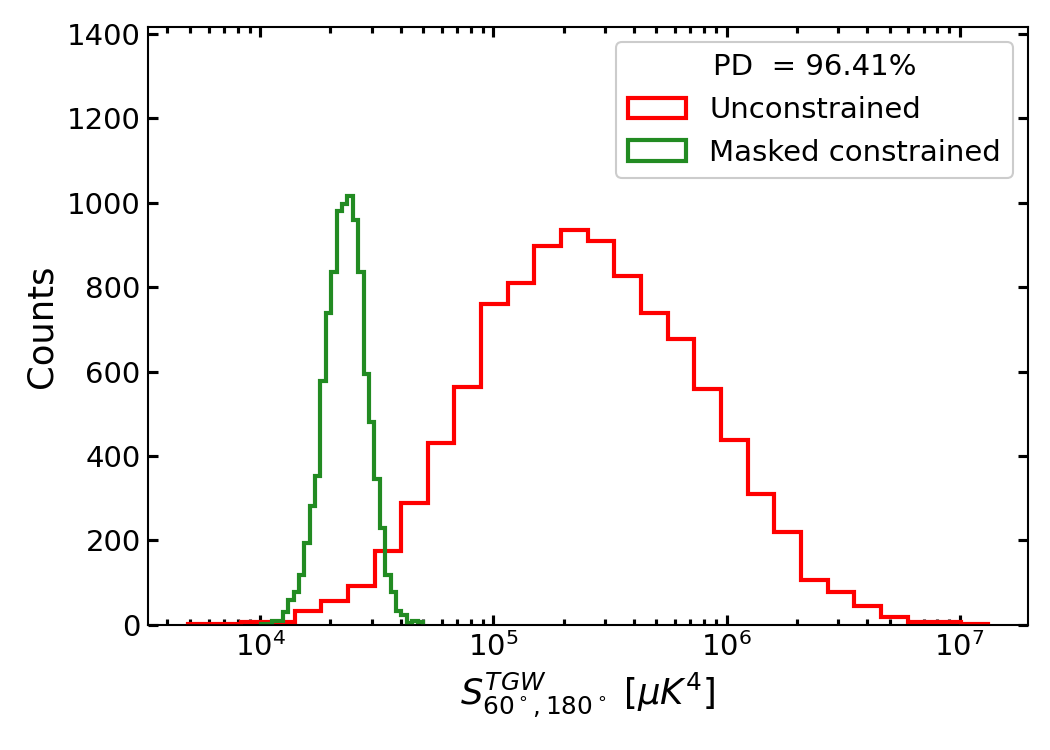}
    \caption{$S^{\rm TGW}_{\theta_{\rm min}, \theta_{\rm max}}$ distributions for the full-sky and masked cases. The angles $\theta_{\rm min}, \theta_{\rm max}$ are chosen to be the optimal ones of both cases. Here we assume $\ell_{\rm max}= 4$.}
    \label{fig: hist_X}
\end{figure}

Assuming that we choose the optimal angular range for the full-sky and masked case, figure~\ref{fig: hist_GWGW} shows the correspondent distributions of $S^{\rm GWGW}_{\theta_{\rm min}, \theta_{\rm max}}$. 
Doing the same for TGW, we instead obtain the figure~\ref{fig: hist_X}.
Note that TGW gives PDs consistently higher than the ones from GWGW. This suggests that TGW is a better probe to test the fluke hypothesis. In fact, even without combining more than one probe (see the following paragraphs), TGW achieves a PD of 96.41\% under the most pessimistic assumption of $\ell_{max=4}$. 

\begin{figure}[t]
    \centering
    \includegraphics[width = 0.49\textwidth]{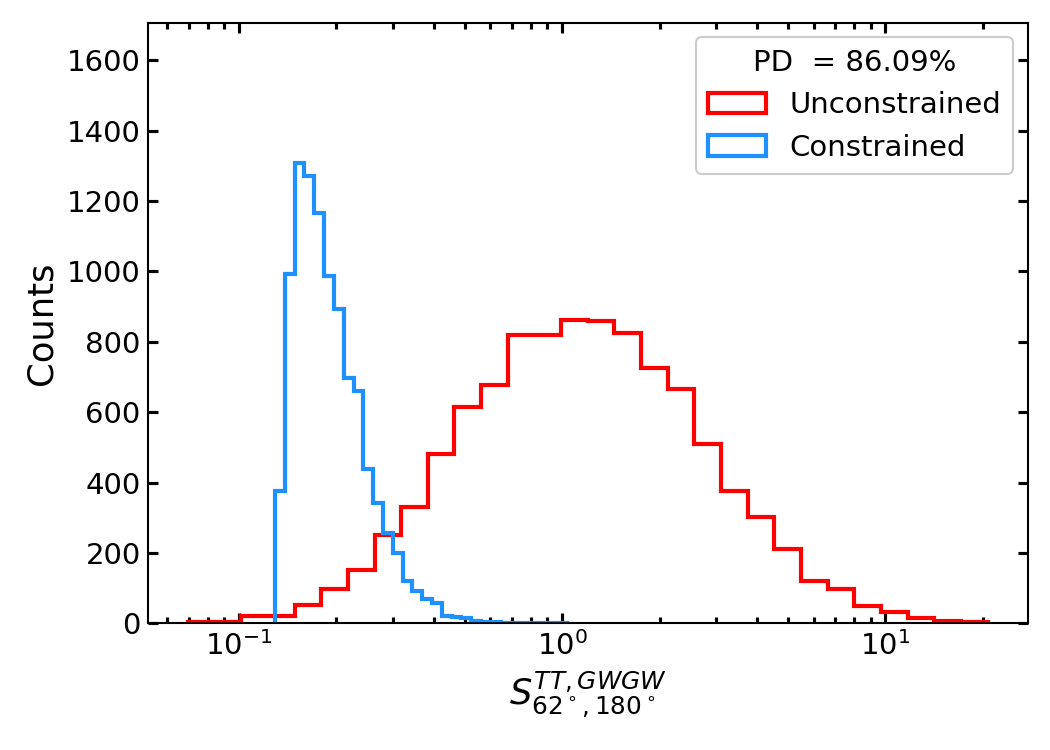}
    \includegraphics[width = 0.49\textwidth]{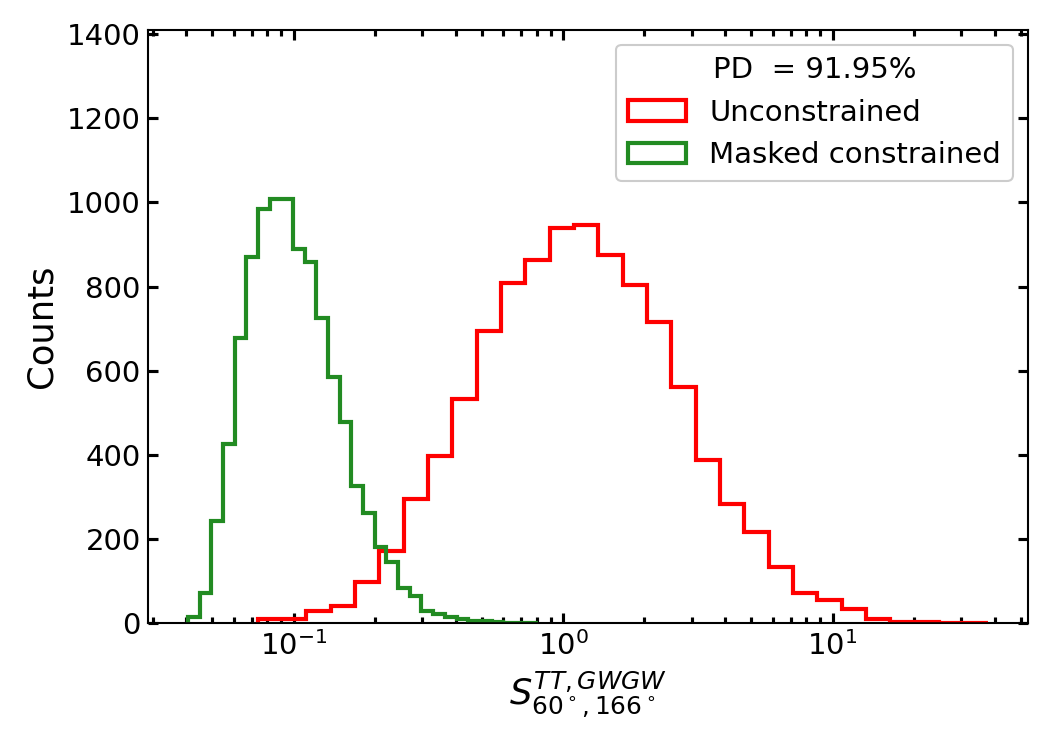}
    \caption{$S^{\rm TT, GWGW}_{\theta_{\rm min}, \theta_{\rm max}}$ distributions for the full-sky and masked cases. The angles $\theta_{\rm min}, \theta_{\rm max}$ are chosen to be the optimal ones of both cases. Here we assume $\ell_{\rm max}= 4$.}
    \label{fig: hist_TGW}
\end{figure}

\begin{figure}[t]
    \centering
    \includegraphics[width = 0.49\textwidth]{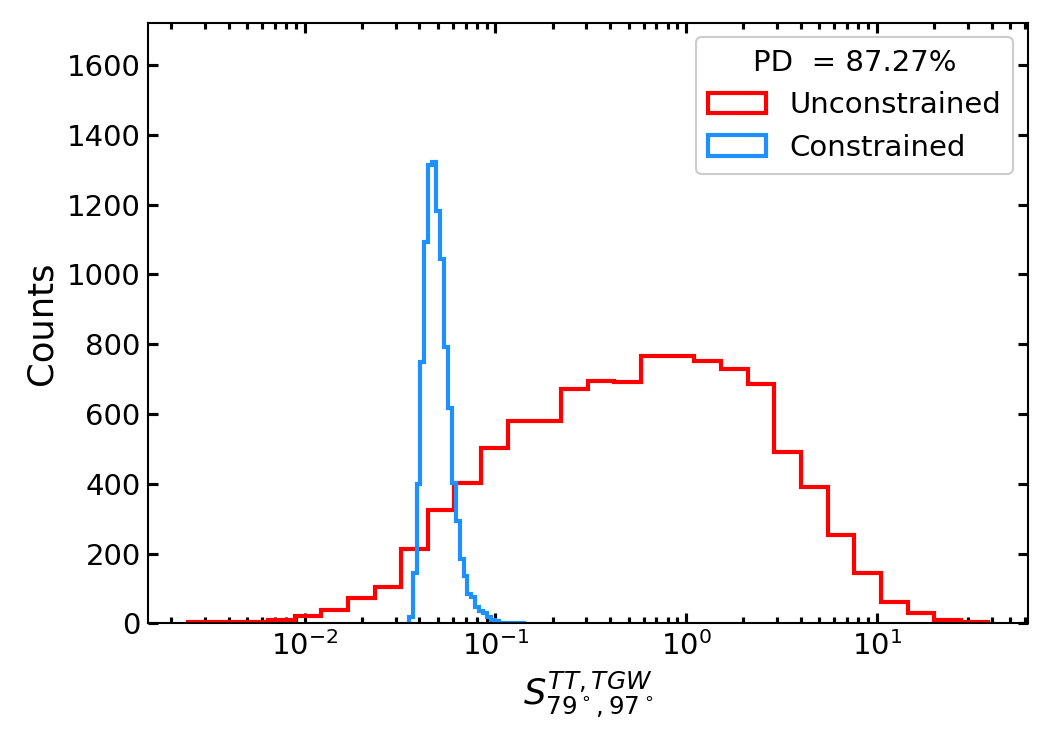}
    \includegraphics[width = 0.49\textwidth]{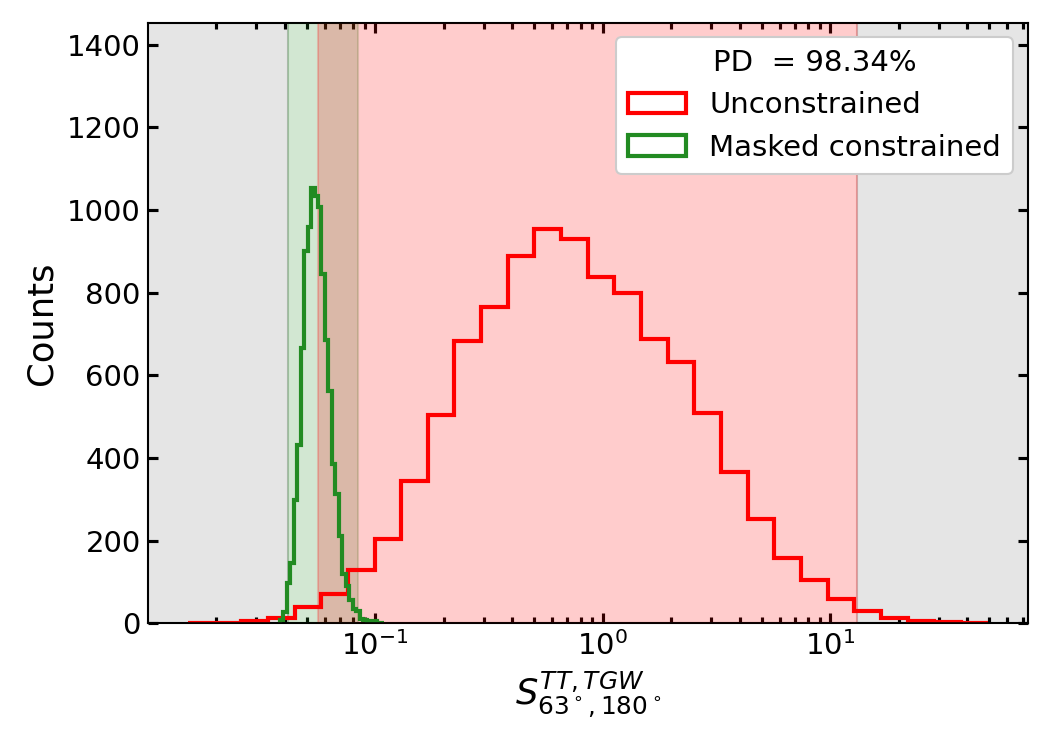}
    \caption{$S^{\rm TT, TGW}_{\theta_{\rm min}, \theta_{\rm max}}$ distributions for the full-sky and masked cases. The angles $\theta_{\rm min}, \theta_{\rm max}$ are chosen to be the optimal ones of both cases. Here we assume $\ell_{\rm max}= 4$. The right plot also shows shaded areas corresponding to different conclusions that one may draw if a future observation happens to be in those (see the end of section~\ref{sec: res_optimal}).}
    \label{fig: hist_TX}
\end{figure}

\begin{figure}[t]
    \centering
    \includegraphics[width = 0.49\textwidth]{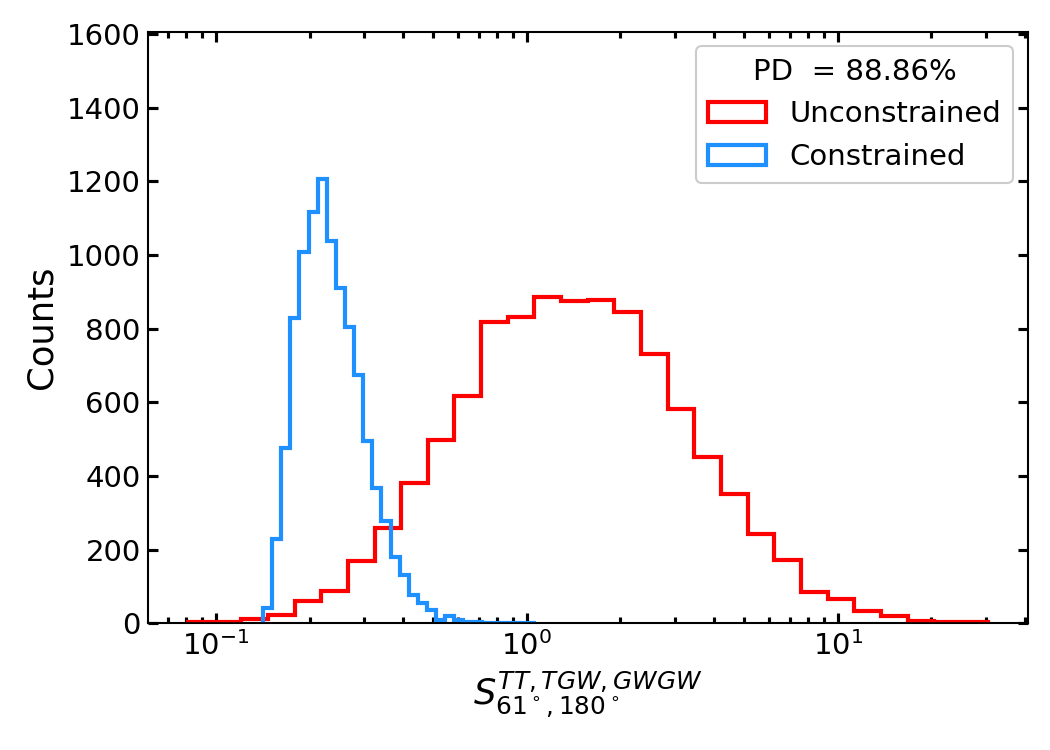}
    \includegraphics[width = 0.49\textwidth]{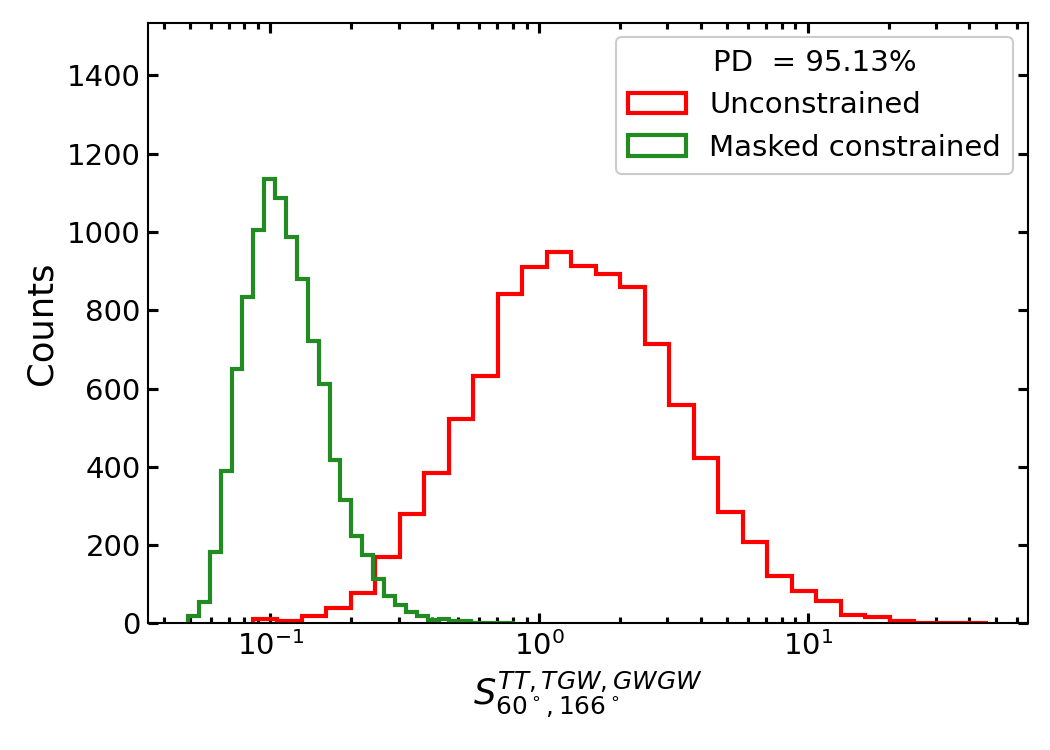}
    \caption{$S^{\rm TT, TGW, GWGW}_{\theta_{\rm min}, \theta_{\rm max}}$ distributions for the full-sky and masked cases. The angles $\theta_{\rm min}, \theta_{\rm max}$ are chosen to be the optimal ones of both cases. Here we assume $\ell_{\rm max}= 4$.}
    \label{fig: hist_TXGW}
\end{figure}

We repeat the same for the joint analyses of the CMB temperature, CGWB, and their cross-correlation. We obtain figure~\ref{fig: hist_TGW} using $S^{\rm TT, GWGW}_{\theta_{\rm min}, \theta_{\rm max}}$; doing the same for $S^{\rm TT, TGW}_{\theta_{\rm min}, \theta_{\rm max}}$ and $S^{\rm TT, TGW, GWGW}_{\theta_{\rm min}, \theta_{\rm max}}$ results in figure~\ref{fig: hist_TX}-\ref{fig: hist_TXGW}, respectively.
All the results on PDs are summarized in table~\ref{tab: PDs}. This shows some interesting features: looking at the masked case, not only we notice an overall consistency on the PDs obtained changing $\ell_{\rm max}$ on each probe, but also there seems to be a certain pattern on how different probes perform at each $\ell_{\rm max}$. Indeed, ranking probes with an ascending PD-ordering, the pattern remains the same for all the masked case. This does not hold for the full-sky case.

\begin{table}[t]
 \centering
\begin{tabular}{l|c|c|c|c|c}
    \toprule
    \multicolumn{6}{c}{Full-sky} \\
    \multicolumn{1}{c}{} & \multicolumn{1}{c}{$S^{\rm GWGW}_{\theta_{\rm min},\theta_{\rm max}}$} & \multicolumn{1}{c}{$S^{\rm TGW}_{\theta_{\rm min},\theta_{\rm max}}$} & \multicolumn{1}{c}{$S^{\rm TT, GWGW}_{\theta_{\rm min},\theta_{\rm max}}$} & \multicolumn{1}{c}{$S^{\rm TT, TGW}_{\theta_{\rm min},\theta_{\rm max}}$}  & \multicolumn{1}{c}{$S^{\rm TT, TGW, GWGW}_{\theta_{\rm min},\theta_{\rm max}}$} \\ \midrule
    & PD & PD & PD & PD & PD \\
    $\ell_{\rm max} = 4$ &  $58.58\%$ &  $81.07\%$ &  $86.09\%$ &  $87.27\%$ &  $88.86\%$ \\
   $\ell_{\rm max} = 6$ &  $60.75\%$ &  $90.44\%$ &  $87.61\%$ &  $94.29\%$ &  $90.35\%$ \\
  $\ell_{\rm max} = 10$ &  $62.27\%$ &  $88.99\%$ &  $88.51\%$ &   $93.4\%$ &  $91.22\%$ \\
    \midrule
    \midrule
    \multicolumn{6}{c}{Mask-sky} \\
    \multicolumn{1}{c}{} & \multicolumn{1}{c}{$S^{\rm GWGW}_{\theta_{\rm min},\theta_{\rm max}}$} & \multicolumn{1}{c}{$S^{\rm TGW}_{\theta_{\rm min},\theta_{\rm max}}$} & \multicolumn{1}{c}{$S^{\rm TT, GWGW}_{\theta_{\rm min},\theta_{\rm max}}$} & \multicolumn{1}{c}{$S^{\rm TT, TGW}_{\theta_{\rm min},\theta_{\rm max}}$}  & \multicolumn{1}{c}{$S^{\rm TT, TGW, GWGW}_{\theta_{\rm min},\theta_{\rm max}}$} \\ \midrule
    & PD & PD & PD & PD & PD \\
    $\ell_{\rm max} = 4$ &   $68.8\%$ &  $96.41\%$ &  $91.95\%$ &  $98.34\%$ &  $95.13\%$ \\
   $\ell_{\rm max} = 6$ &  $73.12\%$ &  $96.98\%$ &  $94.23\%$ &   $98.9\%$ &  $96.36\%$ \\
  $\ell_{\rm max} = 10$ &  $65.24\%$ &   $96.9\%$ &  $88.57\%$ &  $98.89\%$ &  $91.94\%$ \\
    \bottomrule
\end{tabular}
\caption{PDs for every observable and combination of them.}
\label{tab: PDs}
\end{table}

How can these results be used in the presence of a measurement of the CGWB? Consider our highest PD assuming $\ell_{\rm max}=4$, thus the one given by the masked $S^{\rm TT, TGW}_{63^\circ,180^\circ}$ analysis. Exploiting the right panel of figure~\ref{fig: hist_TX} we show different shaded areas corresponding to the different conclusions that can be drawn. Firstly, these regions are obtained by computing the range of each histogram that encapsulates the 99\% percent of the simulations. This identifies five different parts of this plot:
\begin{itemize}
    \item two gray regions corresponding to the values of $S^{\rm TT, TGW}_{63^\circ,180^\circ}$ that are not consistent with neither the constrained nor unconstrained realizations;
    \item the green and red regions, where the observation falls within the constrained or unconstrained histograms;
    \item the intersection of the green and red regions (resulting in a darker region), in which we are not sure if the eventual observation follows the constrained or unconstrained distribution.
\end{itemize}
Depending on the region a future measurement of the CGWB will fall in, we can conclude the following:
\begin{itemize}
    \item If it falls into the gray region, we may conclude that $\Lambda$CDM cannot explain the observed value of $S^{\rm TT, TGW}_{63^\circ,180^\circ}$. Therefore, we need to find a more comprehensive model that can explain this.
    \item If it falls inside the green one, we can say that our observation is well-explained by our model; however, we cannot say anything more that this (being consistent, the fluke hypothesis remains valid).
    \item If the measurement is in the red region, we can draw two different conclusions: either the $\Lambda$CDM model is unable to describe the observations and the fluke hypothesis can be rejected, or the CGWB signal is not correlated (constrained) to the current measurements of the CMB temperature anisotropies (this indeed is an assumption).
    \item If the observation falls at the intersection, it could be following the $\Lambda$CDM prediction (green curve), preventing us from rejecting the fluke hypothesis.
\end{itemize}

This reasoning can also be applied to the other fields, or combinations of them, knowing that the distributions at the optimal angle maximize our capability to get useful information from the CGWB.

In conclusion, in appendix~\ref{app: sensitivity} we show the optimal angles and the distributions in the optimal range assuming the more optimistic cases of $\ell_{\rm max} = 6$ and $\ell_{\rm max} = 10$. 

\subsection{Significance of the anomaly}\label{sec: signi_res}

Until now, we have explored the fluke hypothesis, finding that GWs can be crucial in rejecting at least some of the assumptions of the $\Lambda$CDM model. Despite this, as mentioned above, we cannot conclude anything regarding the actual significance of the anomaly with such an analysis. 

At the end of the previous section, we mentioned that in the green region and at the intersection of the green and red ones of figure~\ref{fig: hist_TX} we cannot draw any meaningful conclusion since the fluke hypothesis still holds.

Here we look at the same problem, but from the perspective of the significance of the anomaly, thus making use of the newly defined estimator in eq.~\eqref{eq: signi_estimator}.

\begin{figure}[t]
    \centering
    \includegraphics[width = 0.49\textwidth]{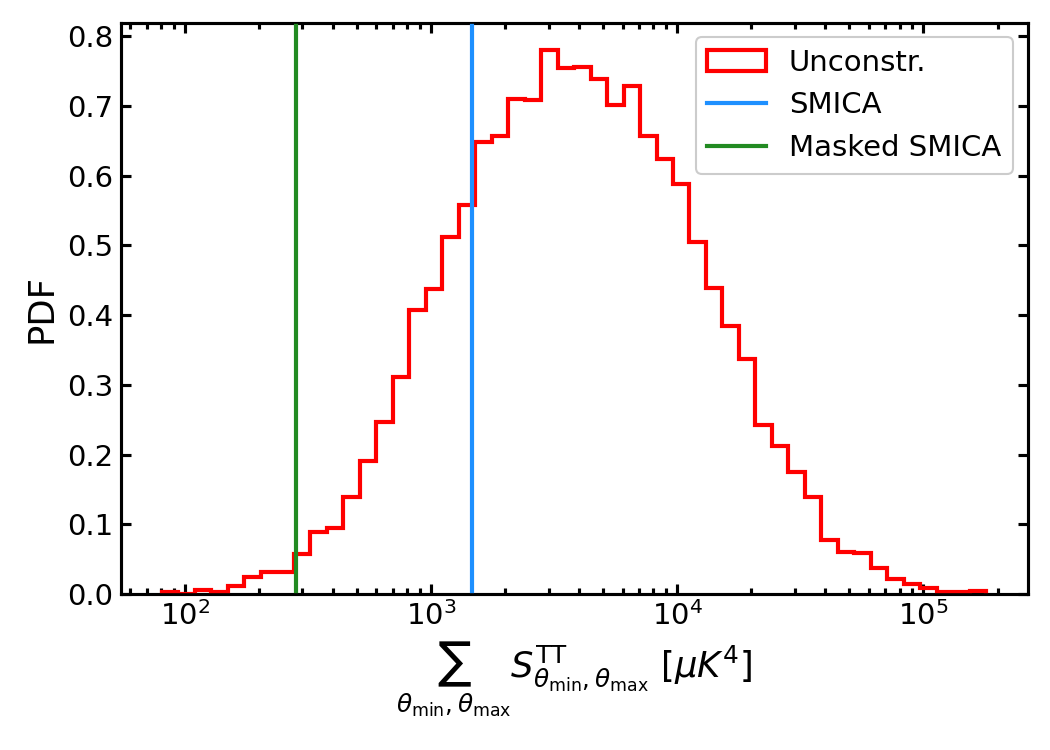}
    \caption{Value of $S^{\rm TT}$ for full-sky and masked SMICA and 10000 $\Lambda$CDM realizations of CMB temperature ($\ell_{\rm max}=4$).}
    \label{fig:signi_TT}
\end{figure}

Applying this to the SMICA maps, we obtain figure~\ref{fig:signi_TT}.
We can see that the SMICA is characterized by a low angular covariance irrespective of the angular range considered, since each score consistently lowers than most $\Lambda$CDM realizations. In terms of significance, full-sky and masked SMICA correspond respectively to $0.82\sigma$ and $2.41\sigma$. This confirms what has already been found in the literature, i.e. masking increases the significance of the anomaly, meaning that high-latitude points drive it.

\begin{figure}[t]
    \centering
    \includegraphics[width = 0.49\textwidth]{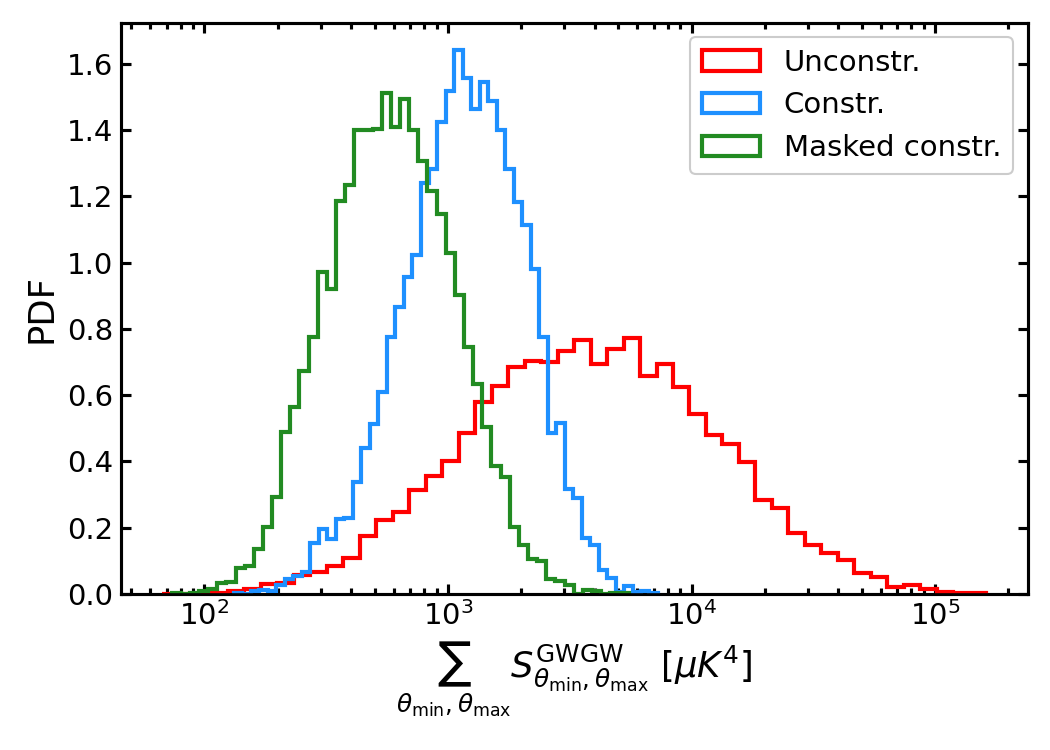}
    \includegraphics[width = 0.49\textwidth]{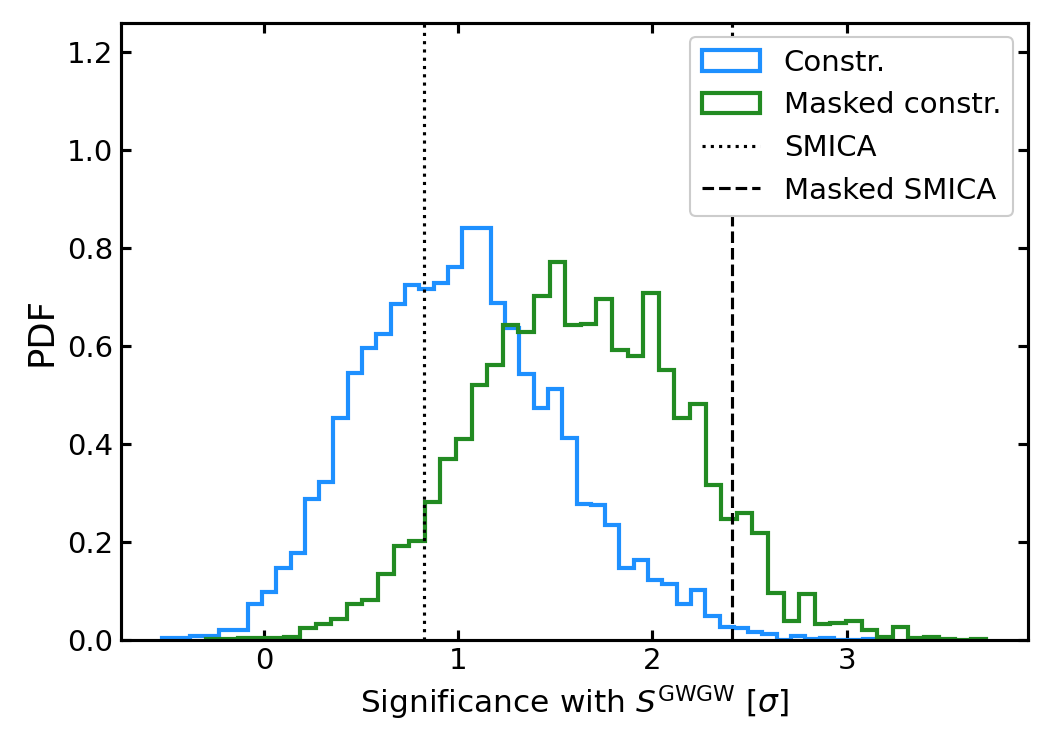}
    \caption{On the left panel, the value of $S^{\rm GWGW}$ for full-sky and masked constrained realizations of the CGWB and the $\Lambda$CDM realizations ($\ell_{\rm max}=4$). On the right panel, corresponding significance in terms of $\sigma$ w.r.t. the unconstrained realizations. The dotted and dashed vertical lines indicate the full-sky and masked SMICA-alone significance, respectively.}
    \label{fig:signi_GWGW}
\end{figure}

We now apply this reasoning to the CGWB. In this case, the role of data is played by our constrained realizations. Figure~\ref{fig:signi_GWGW} shows the results; the left panel depicts the values of $S^{\rm GWGW}$ and the right one their significance corresponding to the $\Lambda$CDM realizations.

\begin{figure}[t]
    \centering
    \includegraphics[width = 0.49\textwidth]{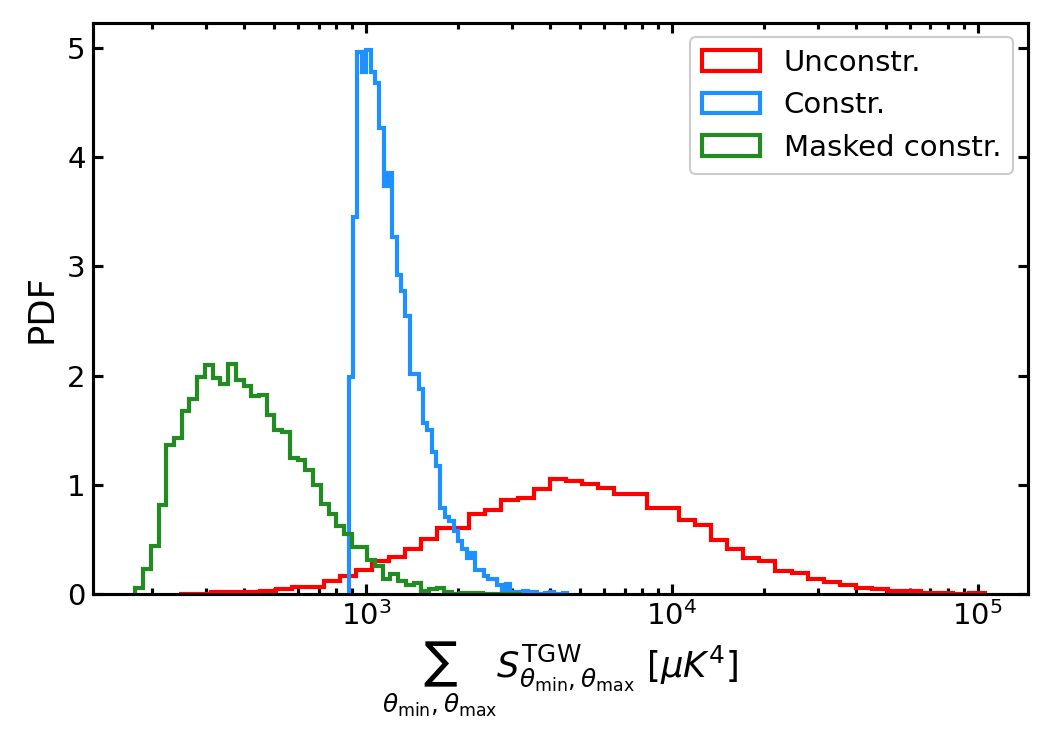}
    \includegraphics[width = 0.49\textwidth]{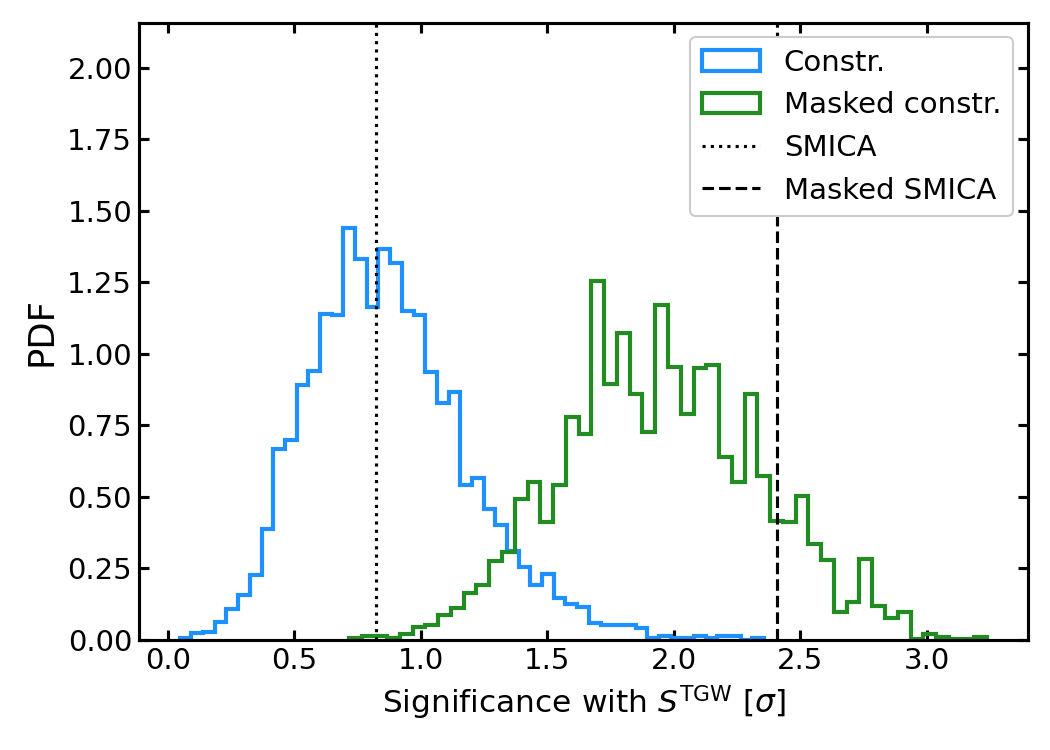}
    \caption{On the left panel, the value of $S^{\rm TGW}$ for full-sky and masked constrained realizations of the CGWB and the $\Lambda$CDM realizations ($\ell_{\rm max}=4$). On the right panel, corresponding significance in terms of $\sigma$ w.r.t. the unconstrained realizations. The dotted and dashed vertical lines indicate the full-sky and masked SMICA-alone significance, respectively.}
    \label{fig:signi_X}
\end{figure}

Repeating the procedure for TGW, we obtain the figure~\ref{fig:signi_X}.
In addition, in this case, TGW seems to perform similarly w.r.t. GWGW, providing consistent values for the significance. However, both of them essentially fail in increasing the significance that one can get from TT. 

\begin{figure}[t]
    \centering
    \includegraphics[width = 0.49\textwidth]{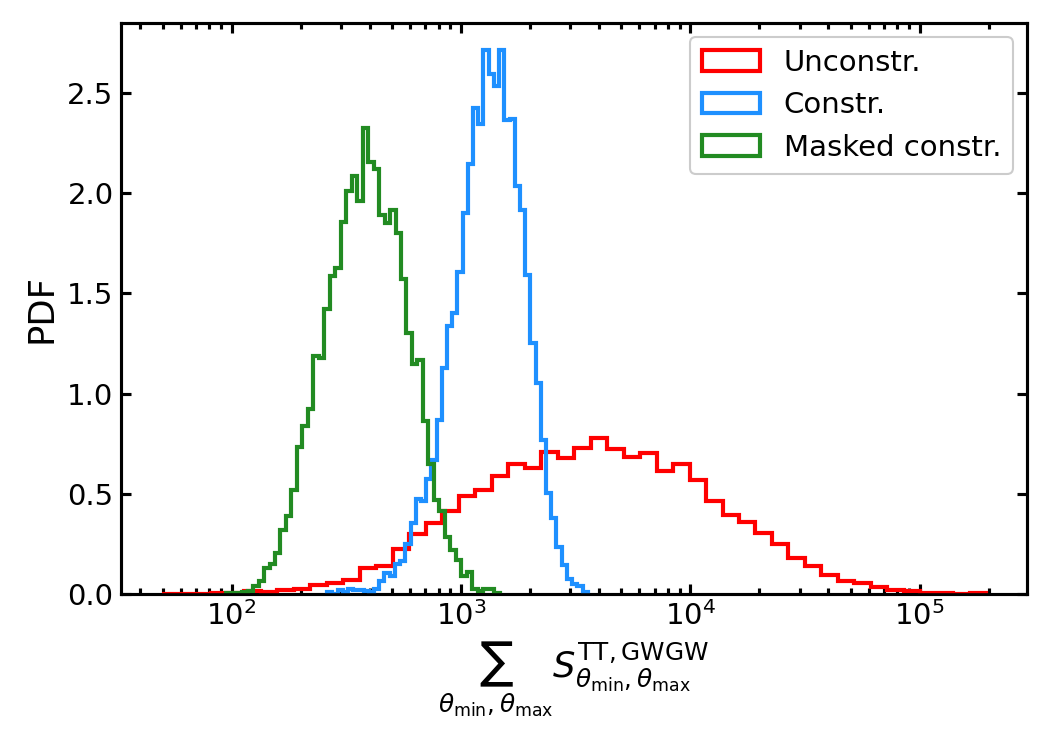}
    \includegraphics[width = 0.49\textwidth]{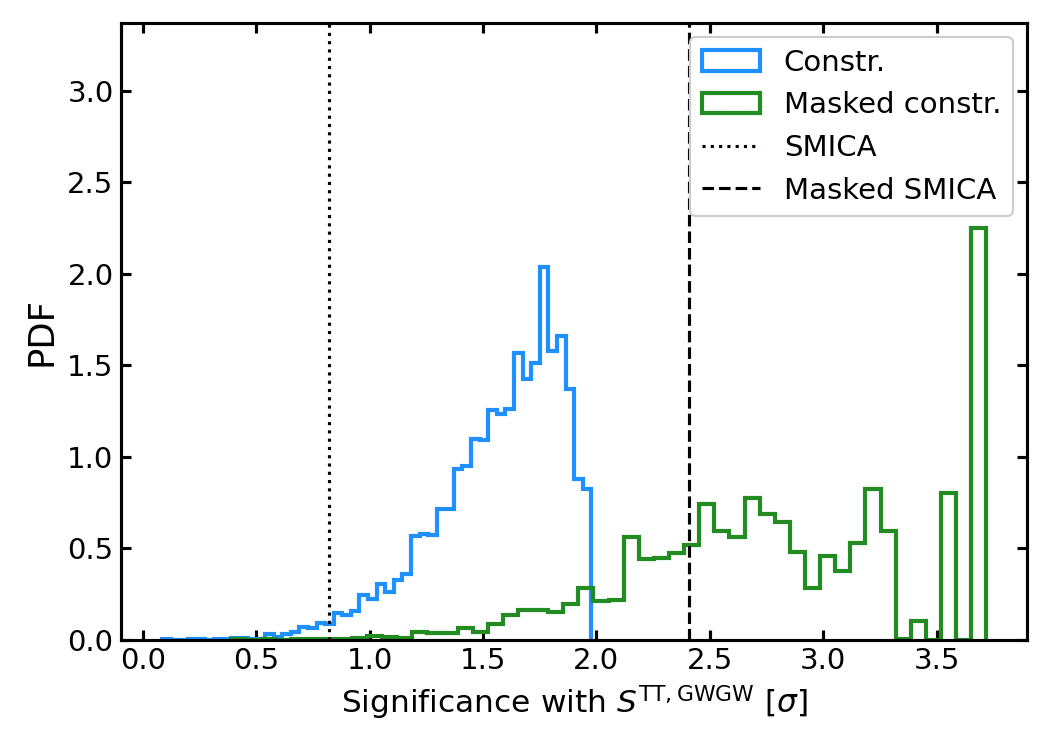}
    \caption{On the left panel, the value of $S^{\rm TT, GWGW}$ for full-sky and masked constrained realizations of the CGWB and the $\Lambda$CDM realizations ($\ell_{\rm max}=4$). On the right panel, corresponding significance in terms of $\sigma$ w.r.t. the unconstrained realizations. The dotted and dashed vertical lines indicate the full-sky and masked SMICA-alone significance, respectively.}
    \label{fig:signi_TTGW}
\end{figure}

\begin{figure}[t]
    \centering
    \includegraphics[width = 0.49\textwidth]{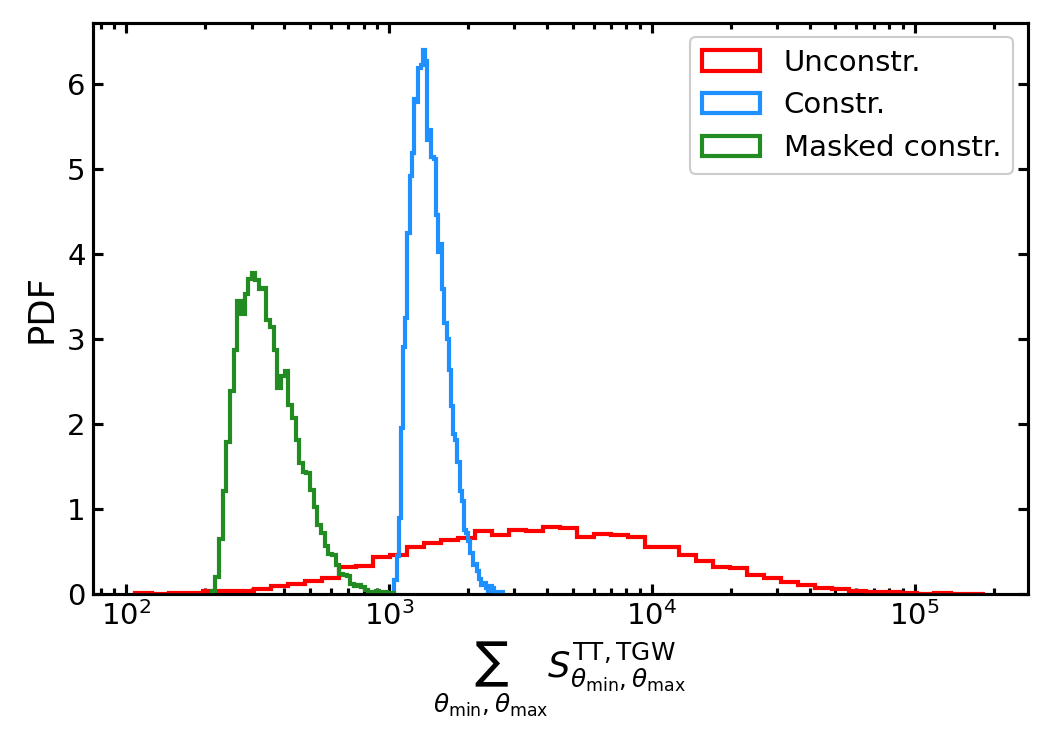}
    \includegraphics[width = 0.49\textwidth]{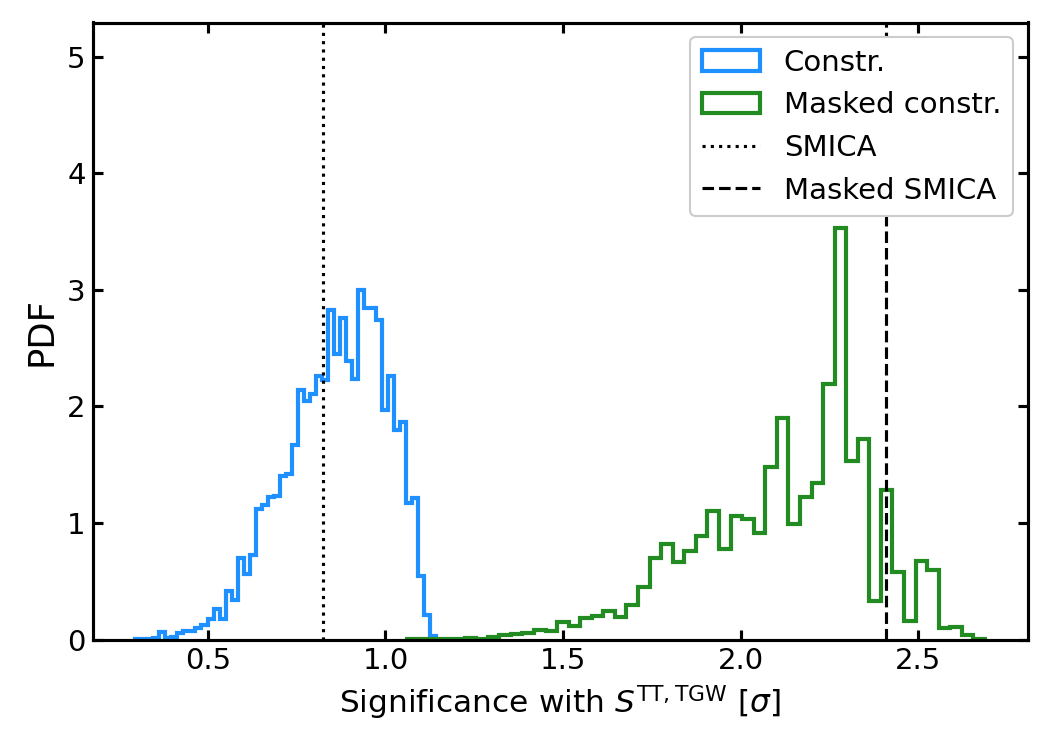}
    \caption{On the left panel, the value of $S^{\rm TT, TGW}$ for full-sky and masked constrained realizations of the CGWB and the $\Lambda$CDM realizations ($\ell_{\rm max}=4$). On the right panel, corresponding significance in terms of $\sigma$ w.r.t. the unconstrained realizations. The dotted and dashed vertical lines indicate the full-sky and masked SMICA-alone significance, respectively.}
    \label{fig:signi_TX}
\end{figure}

\begin{figure}[t]
    \centering
    \includegraphics[width = 0.49\textwidth]{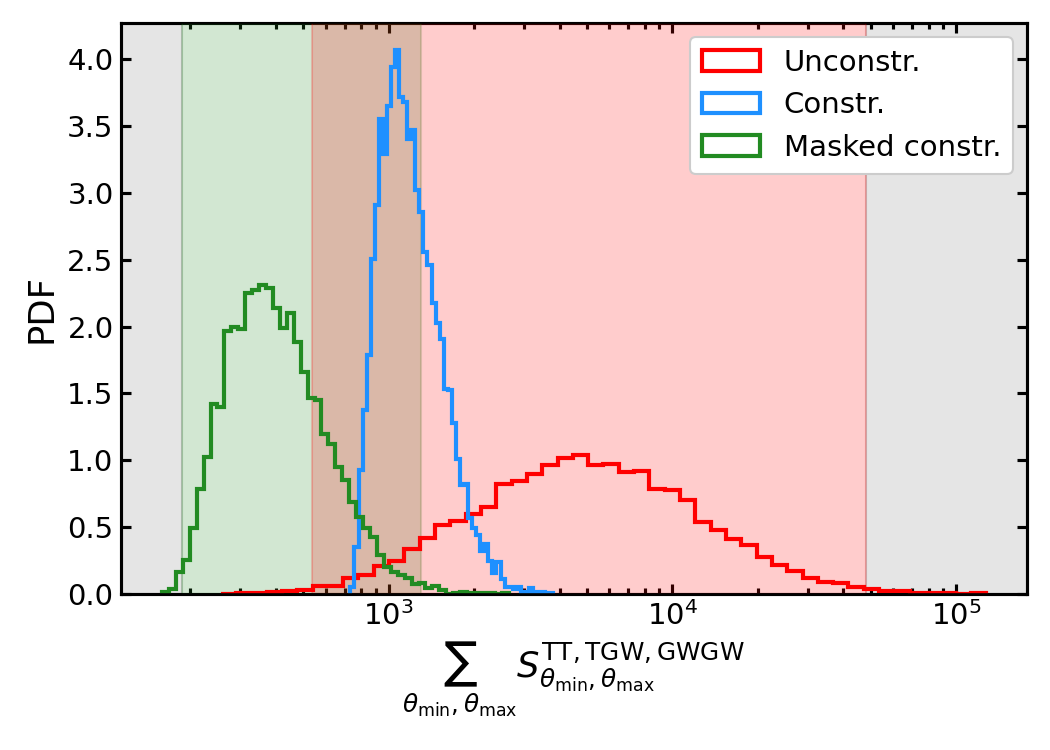}
    \includegraphics[width = 0.49\textwidth]{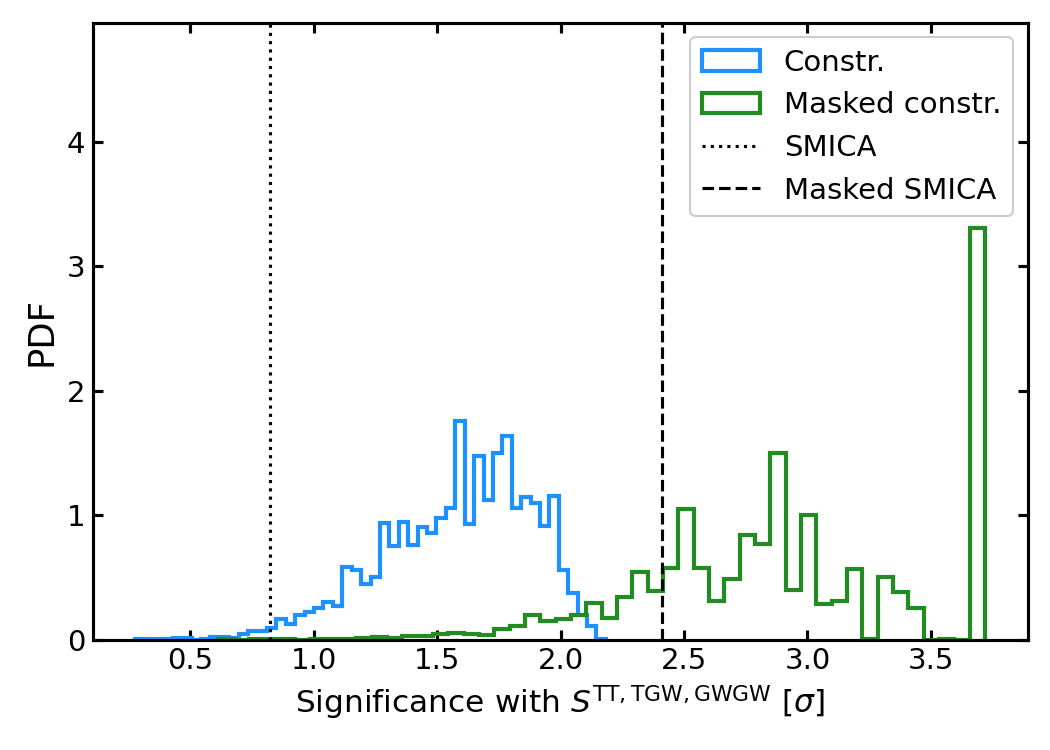}
    \caption{On the left panel, the value of $S^{\rm TT, TGW, GWGW}$ for full-sky and masked constrained realizations of the CGWB and the $\Lambda$CDM realizations ($\ell_{\rm max}=4$). On the right panel, corresponding significance in terms of $\sigma$ w.r.t. the unconstrained realizations. The dotted and dashed vertical lines indicate the full-sky and masked SMICA-alone significance, respectively.}
    \label{fig:signi_TTXGW}
\end{figure}

Going to the joint analyzes, we obtain figure~\ref{fig:signi_TTGW},~\ref{fig:signi_TX} and~\ref{fig:signi_TTXGW} for $S^{\rm TT, GWGW}$, $S^{\rm TT, TGW}$ and
$S^{\rm TT, TGW,GWGW}$, respectively.

Once again, we ask ourselves: how can these results be used in the presence of an eventual measurement of the CGWB? The answer is essentially the same as the one at the end of section~\ref{sec: res_optimal}, with an important exception. In fact, we already know how to interpret a future observation falling in the red or gray region. However, this time we can associate each constrained realization with a significance in the form of sigma distance. Therefore, the green region corresponds to high-significance realizations (where one can claim that the anomaly is the result of a physical phenomenon), while the intersection corresponds to low-significance ones (where further investigation is needed to assess the origin of the anomaly). In other words, having accounted for the look-elsewhere effect allows us to get useful information on all the possible values of the estimator but the intersection.

Talking about the actual results, we summarize them in table~\ref{tab: signi}. Here, we compare the performances of the various combinations of observables by counting how many constrained realizations reach a better significance w.r.t. SMICA alone.

\begin{table}[t]
 \centering
\begin{tabular}{l|c|c|c|c|c|c}
    \toprule
    \multicolumn{7}{c}{Full-sky} \\
    \multicolumn{1}{c}{} & \multicolumn{1}{c}{Signi. $S^{\rm SMICA}$} & \multicolumn{1}{c}{$S^{\rm GWGW}$} & \multicolumn{1}{c}{$S^{\rm TGW}$} & \multicolumn{1}{c}{$S^{\rm TT, GWGW}$} & \multicolumn{1}{c}{$S^{\rm TT, TGW}$}  & \multicolumn{1}{c}{$S^{\rm TT, TGW, GWGW}$} \\ \midrule
   $\ell_{\rm max} = 4$ &  $0.82\sigma$ &  $63.13\%$ &  $52.95\%$ &  $98.34\%$ &  $62.93\%$ &  $98.81\%$ \\
   $\ell_{\rm max} = 6$ &   $1.3\sigma$ &  $38.49\%$ &  $39.33\%$ &  $97.59\%$ &  $54.59\%$ &  $98.59\%$ \\
  $\ell_{\rm max} = 10$ &  $1.18\sigma$ &  $47.15\%$ &  $45.02\%$ &   $98.3\%$ &  $57.67\%$ &  $99.01\%$ \\
    \midrule
    \midrule
    \multicolumn{7}{c}{Mask-sky} \\
    \multicolumn{1}{c}{} & \multicolumn{1}{c}{Signi. $S^{\rm SMICA}$} & \multicolumn{1}{c}{$S^{\rm GWGW}$} & \multicolumn{1}{c}{$S^{\rm TGW}$} & \multicolumn{1}{c}{$S^{\rm TT, GWGW}$} & \multicolumn{1}{c}{$S^{\rm TT, TGW}$}  & \multicolumn{1}{c}{$S^{\rm TT, TGW, GWGW}$} \\ \midrule
    $\ell_{\rm max} = 4$ &  $2.41\sigma$ &  $7.57\%$ &  $12.34\%$ &  $72.53\%$ &  $7.78\%$ &  $81.67\%$ \\
   $\ell_{\rm max} = 6$ &  $2.37\sigma$ &  $1.96\%$ &   $1.23\%$ &  $81.34\%$ &  $0.31\%$ &  $90.11\%$ \\
  $\ell_{\rm max} = 10$ &  $2.17\sigma$ &  $0.13\%$ &   $0.46\%$ &  $58.33\%$ &  $0.15\%$ &  $70.08\%$ \\
    \bottomrule
\end{tabular}
\caption{Percentage of constrained realizations that improve the significance of SMICA, which is shown in the first column.}
\label{tab: signi}
\end{table}

For $\ell_{\rm max} =4$, despite what we find in section~\ref{sec: res_optimal},  the best combination seems to be TT + TGW + GWGW, which achieves the 98.81\% and the 81.67\% of the constrained realizations, improving the significance of SMICA (full-sky and masked, respectively). Thus, to determine the actual significance of the anomaly with GWs and to assess the physical origin of the anomaly, it is crucial to observe the autospectrum. This in fact brings the majority of information when combined with TT, as shown by the last and second to last columns of the table~\ref{tab: signi}.

\section{Conclusions}\label{sec:conclusions}

Since COBE \cite{boggessCOBEMissionIts1992, bennett19964YearCOBEDMRCosmic, hinshaw19962PointCorrelationsCOBEDMR}, we have measured a low two-point angular correlation function of the CMB temperature on large scales. This feature has been reassessed both by WMAP \cite{cruz2011AnomalousVarianceWMAPdata, gruppuso2014TwopointCorrelationFunctionWMAP, monteserin2008LowCMBVarianceWMAP} and Planck \cite{planckcollaboration2014Planck2013ResultsXXIII, planckcollaboration2016Planck2015ResultsXVI, planckcollaboration2020Planck2018ResultsVIIa} suggesting that it is not the product of some systematic, given that the three experiments are independent in this regard. Still, it is not clear whether the so-called lack-of-correlation anomaly is the product of some non-standard physics or whether it is the manifestation of the fact that we live in a rare realization of the $\Lambda$CDM model. 
Since we already have a cosmic-variation-limited measurement of CMB temperature on low and intermediate scales, this latter possibility has to be explored with some observable other than temperature. An example is the E-mode polarization of CMB photons, which is correlated with the temperature and can provide new information on the anomaly \cite{copi2013LargeAngleCMBSuppressionPolarizationa, billi2019PolarisationTracerCMBanomalies, yoho2015MicrowaveBackgroundPolarizationProbe, chiocchetta2021LackofcorrelationAnomalyCMBlargea}. 
In this work, we study the ability of the CGWB to shed light on this matter \cite{albaPrimordialGravityWave2016, contaldiAnisotropiesGravitationalWave2017, geller2018PrimordialAnisotropiesGravitationalWave, bartoloCharacterizingCosmologicalGravitational2020, bartoloAnisotropiesNonGaussianityCosmological2019, valbusadallarmiImprintRelativisticParticles2021}. This is done by exploiting both the autospectrum of the CGWB and its cross-spectrum with CMB temperature (hereby named TGW). In fact, we know that this primordial signal has a great degree of correlation with the temperature of the CMB \cite{ricciardone2021CrosscorrelatingAstrophysicalCosmologicalGravitational}. Thus, we can produce both constrained and unconstrained realizations of the CGWB using the SMICA temperature map as our CMB observation. Since we know that the lack-of-correlation anomaly is enhanced when the galactic plane is removed \cite{gruppuso2013LowVarianceLargescales, gruppuso2014TwopointCorrelationFunctionWMAP, nataleLackPowerAnomaly2019}, this is done while considering both full-sky and masked \textit{Planck}'s SMICA. These maps are smoothed with a Gaussian beam with FWHM $=2^\circ$ and degraded to $N_{\rm side}=64$. We have not shown this in the main body of this work, but we also repeated the analysis assuming a smoothing of FWHM $=0.92^\circ$ (thus the grid scale corresponding to $N_{\rm side}=64$) and FWHM $=4^\circ$. As expected, the former case brings some differences caused by pixelization effects, which is why it is usually advisable to smooth maps with a beam two or three times bigger than the grid scale. Instead, the latter produces identical results w.r.t. our choice of $2^\circ$, proving that it is sufficiently large. When performing the analysis, we must also take into account that GWs are notably difficult to observe. Thus, we consider three different choices of what the maximum multipole is that we can observe in a signal-dominated way, i.e. $\ell_{\rm max} = 4, 6$ and 10. Depending on the assumption one makes for the monopole radiation of GWs, this can be obtained with one of the future GW interferometers or using a combination of them (for example, LISA \cite{amaro-seoaneLaserInterferometerSpace2017, barausseProspectsFundamentalPhysics2020, bartolo2016ScienceSpacebasedInterferometerLISA, capriniReconstructingSpectralShape2019, pieroniForegroundCleaningTemplatefree2020}, DECIGO \cite{kawamuraJapaneseSpaceGravitational2006}, ET \cite{sathyaprakashScientificPotentialEinstein2012, maggioreScienceCaseEinstein2020} and CE \cite{abbottExploringSensitivityNext2017}).

Summarizing the methodology followed in this work, in section~\ref{sec: dataset} we define the dataset we exploit to perform our analysis, while in section~\ref{sec: constrained} we generalize the full-sky expression of the constrained realizations to account for the multipole couplings brought by the presence of a mask (see also appendix~\ref{app: window}). Then, after having defined the key quantity on which the lack-of-correlation manifests, i.e., the two-point angular correlation function (see section~\ref{sec: ang_corr}), in section~\ref{sec: quantify} we follow \citet{copi2013LargeAngleCMBSuppressionPolarizationa} to define an estimator to quantify the anomaly. In particular, by integrating the two-point angular correlation function (squared) over a certain angular range $\qty[\theta_{\rm min}, \theta_{\rm max}]$, one can define the quantity named $S_{\theta_{\rm min}, \theta_{\rm max}}$ for each field considered (in our case TT, TGW, and GWGW).
Furthermore, \cite{chiocchetta2021LackofcorrelationAnomalyCMBlargea} defines an estimator capable of combining those of two different observables. Therefore, in our case, we define three combined estimators $S^{\rm TT, GWGW}_{\theta_{\rm min}, \theta_{\rm max}}$, $S^{\rm TT, TGW}_{\theta_{\rm min}, \theta_{\rm max}}$, and $S^{\rm TT, TGW, GWGW}_{\theta_{\rm min}, \theta_{\rm max}}$, which encode the information of all the observables considered. In section~\ref{sec: metho_optimal}, we follow again \citet{copi2013LargeAngleCMBSuppressionPolarizationa}, to define a way to maximize the amount of information that we can obtain by adding CGWB to the estimate. In particular, computing the constrained and unconstrained realizations of the CGWB (considering both the full-sky and the masked version of SMICA), we count how many unconstrained realizations can recover higher values of $S^{\rm GWGW}_{\theta_{\rm min}, \theta_{\rm max}}$ w.r.t. the 99th percentile of the constrained ones. This defines what we call the Percentage Displacement (PD) of the two distributions. When this PD is maximal, we can say that the CGWB is as sensitive as possible to what is observed in the CMB (in terms of lack of correlation). The specific angular range found is named the ``optimal angular range''. The same procedure is also performed for $S^{\rm TGW}_{\theta_{\rm min}, \theta_{\rm max}}$, $S^{\rm TT, GWGW}_{\theta_{\rm min}, \theta_{\rm max}}$, $S^{\rm TT, TGW}_{\theta_{\rm min}, \theta_{\rm max}}$ and $S^{\rm TT, TGW,GWGW}_{\theta_{\rm min}, \theta_{\rm max}}$. Finally, in section~\ref{sec: signi} we define a new estimator for the lack-of-correlation which takes into account the so-called look-elsewhere effect. Indeed, to study the significance of the anomaly, we must find a way to marginalize the angular-range information so that we recover the anomaly irrespective of the particular range. In this way, we are able to provide a forecast of the improvement brought about by the CGWB in terms of the significance of the lack-of-correlation anomaly.

Summarizing now the results, in section~\ref{sec: res_optimal} we report the results for the optimal angular ranges. We find that, in general, passing from the full-sky treatment to the masked treatment increases the PDs obtained. This confirms that the anomaly seems to increase in significance together with the angle from the galactic plane. In addition, when considering one field at a time, TGW seems to be a consistently better probe to test the fluke hypothesis w.r.t. GWGW. Furthermore, using different combinations of fields, we show that TT + TGW is the best combination to test the fluke hypothesis. In fact, even in the most pessimistic case of $\ell_{\rm max} = 4$ we obtain a PD of 96.41\% in the optimal angular range $\qty[63^\circ, 180^\circ]$ (see table~\ref{tab: optimal}-\ref{tab: PDs}). This PD can be compared with the ones in \cite{copi2013LargeAngleCMBSuppressionPolarizationa} regarding E-mode polarization, showing that GWs are actually much more restrictive in testing the fluke hypothesis. As discussed in section~\ref{sec: res_optimal}, this means that this combination of observables is extremely good for testing this hypothesis, maximizing the probability of rejecting the $\Lambda$CDM model in the event that a future observation happens to be outside the distribution shown in figure~\ref{fig: hist_TX}, with the corresponding level of significance. Regarding the comparison with different assumptions on $\ell_{\rm max}$, we note that the best results in terms of PDs are obtained with $\ell_{\rm max} = 6$, suggesting that the lack of correlation signal lives in the first six multipoles (see appendix~\ref{app: sensitivity} for the plots). Overall, we also show that there always seems to be a high-PD region near the range where the original $S_{1/2}$ estimator is defined \cite{spergel2003FirstYearWilkinsonMicrowave}. As mentioned in section~\ref{sec: res_optimal}, another interesting feature of this analysis is depicted in table~\ref{tab: PDs}: the masked case seems to give more consistent results in terms of PDs against a change of $\ell_{\rm max}$ or used probes. Indeed, there seems to be a fixed order of probes at each $\ell_{\rm max}$ when we rank them for ascending PD. This does not hold for the full-sky case. Then, one might argue that the masked sky provides a more faithful representation of reality due to its stronger consistency. Another feature emerges when changing the $\ell_{rm max}$ of the analysis (see appendix~\ref{app: sensitivity}) Indeed, increasing the number of multipoles acts in different ways on the optimal regions: the GWGW realizations start concentrating towards the $\approx 75^\circ$ scale and the TGW abandon $\theta_{\rm max} \simeq 180^\circ$ in favor of $\approx 130^\circ$. Hence, the various combined estimators report a combination of the preferred regions of the two, or three, probes involved.

In section~\ref{sec: signi_res}, we show the results for the actual significance of the anomaly. Firstly, when applying our newly defined estimator to the CMB temperature alone, we obtain a significance of $0.82\sigma$ and $2.41\sigma$ for full-sky and masked SMICA respectively. This confirms that masking the sky greatly enhances the significance of this anomaly (in our case of a factor three). Also, table~\ref{tab: signi} shows the significance of TT alone when increasing the number of multipoles. It peaks at $\ell_{\rm max} = 6$, suggesting in accordance with the optimal angular range analysis that the anomaly lives in that multipole range. Focusing then on the CGWB contributions, despite what we find analyzing the optimal angular ranges, we point out that the autospectrum GWGW is crucial to obtain a good level of significance when including a CGWB observation. The best results in this sense are given by the full combination TT + TGW + GWGW, which provides 98.81\% (81.67\%) of the realizations improving the current significance w.r.t. full-sky (masked) SMICA (see table~\ref{tab: signi}). Unlike what we could conclude with the analysis of optimal angles alone, using the new estimator for the significance, not only can we reject the $\Lambda$CDM model if a measurement of the CGWB falls outside the predicted distribution, but also we can associate an actual significance to a measurement following those curves. Comparing again different assumptions on $\ell_{\rm max}$ and focusing on the full combination of observables, the significance remains fairly stable in all cases in the full-sky case. Also, we observe that the constrained realizations of GWGW and TGW are always centered on the significance of SMICA alone (see appendix~\ref{app: sensitivity}). For the masked case, the results are behaving more complexly. Starting from $\ell_{\rm max} = 6$, having included two extra multipoles seems to act as we expect, assuming that the anomaly lives in the first six multipoles as our analysis suggests. Indeed, comparing the $S$-estimator distributions with the results shown in the main body for $\ell_{\rm max} =4$, they appear to be in the same relative position w.r.t. the unconstrained realizations, but with a shrunken dispersion (resulting in more peaked histograms). Thus, the same ``anomalous signal'' is present and gets better constrained by the higher number of multipoles. This is also testified by the increase in the significance obtained (see table~\ref{tab: signi}). When considering the case of $\ell_{\rm max} = 10$, the situation gets flipped over. The $S$-estimator distributions approach tends to move toward the mean of the unconstrained distribution (even getting superimposed to the full-sky case for GWGW alone), resulting in an overall loss of significance. Together with the optimal range analysis and the significance of TT alone, this suggests the following interpretation: if the anomaly actually lives in the first six multipoles, we may expect that including more and more non-anomalous multipoles should decrease the significance since these would distribute as the $\Lambda$CDM ones. In other words, since we are summing over the considered multipole range (see eq.~\eqref{eq: S12}), these new multipoles are pushing the overall distribution toward the standard behavior.

Even though this is the first time the CGWB is exploited to explore the fluke hypothesis for one of the known anomalies, here we show that it can be used effectively to shed light on other ones. Indeed, even considering the very first multipoles of the CGWB power spectrum it is possible to get a great amount of information on these anomalies. For example, \cite{galloni2022TestStatisticalIsotropyUniversea} already showed that it can help assess the presence of a dipolar modulation that accounts for the so-called hemispherical power asymmetry. Furthermore, considering instead the Astrophysical Gravitational Wave Background (AGWB), \cite{dallarmi2022DipoleAstrophysicalGravitationalWaveBackground} explores the possibility of extracting the dipole anisotropy. This is related to another branch of possible departures from the standard cosmological model inspired by the recent emergence of several anomalously large dipoles in many large-scale structure tracers \cite{gibelyou2012DipolesSky, rubart2013CosmicRadioDipoleNVSS, tiwari2016RevisitingNVSSNumbercount, chen2016AngularTwopointCorrelationNVSS, bengaly2018ProbingCosmologicalPrinciplecounts, secrest2021TestCosmologicalPrincipleQuasarsa}.

In this paper we use \texttt{NaMaster} to compute the angular power spectrum of masked skies (see section~\ref{sec: dataset}). In particular, this approach is applied whenever we need the spectrum of masked SMICA to compute the estimators or when we need to compute the spectrum of TGW while masking the CMB sky. Although this provides an unbiased estimate of the spectra, it does not minimize their variance. For this reason, it would be interesting to explore this anomaly with an analogous analysis employing some maximum likelihood estimator, which instead allows one to get a minimal variance estimate of the spectra \cite{shi2023TestingCMBAnomaliesEmode, tegmark1997HowMeasureCMBpower, vanneste2018QuadraticEstimatorCMBcrosscorrelation}. We leave this for future work. 

While writing this paper, another interesting aspect to explore in the time ahead has been emphasized by \citet{hansen2023CosmicMicrowaveBackgroundanomalies}. They show the evidence of the presence of an extra-galactic foreground on top of the CMB temperature data. Indeed, if their claim is found to be correct, it would mean that the variance measured on large scales is actually enhanced by this signal. Therefore, mitigating this foreground from our maps would bring the variance of the first CMB multipoles even lower than its current value, suggesting that the significance of the lack-of-correlation anomaly might increase (even assuming TT alone). By extension, if this hypothesis is correct, the CGWB might become crucial to boost the significance to the level of an actual tension. In fact, in section~\ref{sec: signi_res} we show that in some cases the CGWB is expected to provide a significance near the $4\sigma$ level, even with current data.

Concluding, in this work we just consider the CMB temperature and the CGWB. Instead, we know that CGWB also shares a correlation with E-mode polarization \cite{ricciardone2021CrosscorrelatingAstrophysicalCosmologicalGravitational, braglia2021ProbingPreRecombinationPhysicsCrossCorrelation}. Thus, this framework can be extended to all three of them. In this context, future experiments such as LiteBIRD \cite{hazumiLiteBIRDSatelliteStudies2019} could be crucial to finally assess the physical origin of this anomaly, given that LiteBIRD is expected to be fully cosmic-variance-limited on the large-scale polarization \cite{litebirdcollaboration2022ProbingCosmicInflationLiteBIRDa}.

\acknowledgments

The authors thank Adrià Gómez-Valent, Frode Kristian Hansen and Javier Duque Carron for their valuable comments and discussions. This work is based on observations obtained with Planck (\url{http://www.esa.int/Planck}), an ESA science mission with instruments and contributions directly funded by ESA Member States, NASA, and Canada. Some of the results in this paper have been derived using the following packages: \texttt{classy} \cite{lesgourguesCosmicLinearAnisotropy2011, blasCosmicLinearAnisotropy2011}, \texttt{healpy} \cite{gorskiHEALPixFrameworkHighResolution2005}, \texttt{NaMaster} \cite{alonso2019UnifiedPseudoEllframework}, \texttt{Matplotlib} \cite{hunterMatplotlib2DGraphics2007}, \texttt{SciPy} \cite{virtanen2020SciPyFundamentalAlgorithmsscientific}, \texttt{Numba} \cite{lam2015NumbaLLVMbasedPythonJIT}, \texttt{NumPy} \cite{harrisArrayProgrammingNumPy2020} and \texttt{Ray} \cite{moritz2018RayDistributedFrameworkEmerging}. G.G., M.B., and N.B. acknowledge support from the COSMOS network (\url{www.cosmosnet.it}) through the ASI (Italian Space Agency) Grants 2016-24-H.0, 2016-24-H.1-2018 and 2020-9-HH.0. 
G.G. and M.B. acknowledge financial support from the INFN InDark initiative.
A.R. acknowledges financial support from the Supporting Talent in ReSearch@University of Padova (STARS@UNIPD) for the project ``Constraining Cosmology and Astrophysics with Gravitational Waves, Cosmic Microwave Background and Large-Scale Structure cross-correlations''.

\bibliographystyle{apsrev}
\bibliography{My_Library}

\begin{appendix} 
\section{Window function in a partial sky}\label{app: window}
In section~\ref{sec: methodology} we generalize the formula for constrained realizations to the cut-sky case. In doing so, we use the window function of a mask, so here we give more details on this quantity.

Assume to have an observable field $X$, which gets decomposed using spherical harmonics. This provides an efficient description of an observation in terms of the $a_{\ell m}$ coefficients. However, these will have different values and different statistics depending on the fraction of the sky one is able to observe. In particular, the relation between full-sky (FS) and cut-sky (CS) coefficients is \cite{gorski1994DeterminingSpectrumPrimordialInhomogeneity, mortlock2002AnalysisCosmicMicrowavebackground}
\begin{equation}
\begin{aligned}
    a_{\ell m}^{\rm CS} & = \int_{cut-sky} X(\theta, \phi) Y_{\ell m}^*(\theta, \phi)  \dd(\cos\theta)\dd\phi \\
    & = \int_{cut-sky} \sum_{\ell' m'} a_{\ell' m'}^{\rm FS} Y_{\ell' m'}(\theta, \phi) Y_{\ell m}^*(\theta, \phi)  \dd(\cos\theta)\dd\phi \\
    & = \sum_{\ell' m'} a_{\ell' m'}^{\rm FS} \int_{cut-sky}  Y_{\ell' m'}(\theta, \phi) Y_{\ell m}^*(\theta, \phi)  \dd(\cos\theta)\dd\phi \\ \ .
\end{aligned}
\label{eq: cs_vs_fs}
\end{equation}
In the full-sky case, the integral gives $\delta_{\ell \ell'}\delta_{m m'}$ thank to the orthogonality of the spherical harmonics on the complete sky. However, on a portion of the sphere, they are not orthogonal, so $\ell-\ell'$ couplings will emerge. 

We can write eq.~\eqref{eq: cs_vs_fs} as a function of a window function $W$ defined by the particular mask we are using as
\begin{equation}
    a_{\ell m}^{\rm CS} = a_{\ell' m'}^{\rm FS} W_{\ell m}^{\ell' m'}\ ,
\end{equation}
where repeated indexes are summed over. Thus, in order to estimate the full-sky coefficients from a partial sky observation we must compute $W_{\ell m}^{\ell' m'}$. Since one typically works with maps divided in pixels, we can write the discretized window function as a sum over the unmasked pixels $p$ \cite{mortlock2002AnalysisCosmicMicrowavebackground}
\begin{equation}
    W_{\ell m}^{\ell' m'} = \sum_{p\ \in\ CS} Y_{\ell' m'}(p) Y_{\ell m}^*(p) \Omega_p\ ,
\label{eq: window_discrete}
\end{equation}
where $\Omega_p$ is the angular area of the pixel and the spherical harmonics are evaluated in the center of each pixel. This sum will depend on the considered pixelization ($N_{\rm side}$ parameter of \texttt{Healpy}), but it will tend to the continuous integration as the pixel size goes to zero. Note that in the case of very aggressive masks, the window function may become singular.

An alternative route to obtain the window function is to use partial sky realizations of the considered field. Indeed, looking at the covariance of the $a_{\ell m}^{\rm CS}$ we can write (here we drop the indexes for the sake of notation)
\begin{equation}
\begin{aligned}
    \expval{\qty(a^{\rm CS})^* a^{\rm CS}} & = W^*\expval{\qty(a^{\rm FS})^* a^{\rm FS}}W \\
    &=W^*DW \\
    & = \qty(D^{1/2}W)^* \qty(D^{1/2}W)\ ,
\end{aligned}
\label{eq: CS_cov}
\end{equation}
where $D$ is a matrix with the angular power spectrum in the diagonal elements (see eq.~\eqref{eq: FS_cov}). In the last line, we obtain the classic definition of the Cholesky decomposition. This shows that having an empirical covariance obtained from N cut-sky realizations of $X$ is equivalent to having computed the window function of the mask while knowing the angular power spectrum of the observable. In particular, we can write
\begin{equation}
\begin{aligned}
   D^{1/2}W &= Chol.\qty(Cov^{\rm CS}) \\ 
   W &= D^{-1/2} \times Chol.\qty(Cov^{\rm CS})\ ,
\end{aligned}
\end{equation}
obtaining the window function from the cut-sky realizations. Obviously, this alternative way will depend on the realizations used and their number.
\begin{figure}[t]
    \centering
    \includegraphics[width = 0.49\textwidth]{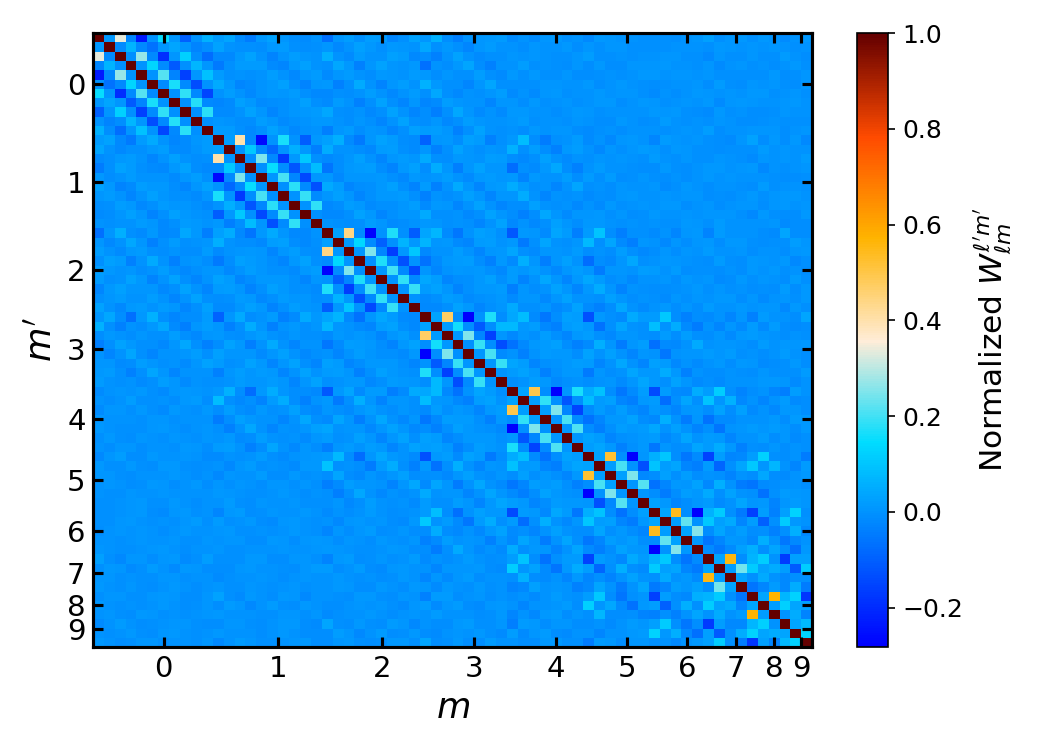}
    \includegraphics[width = 0.49\textwidth]{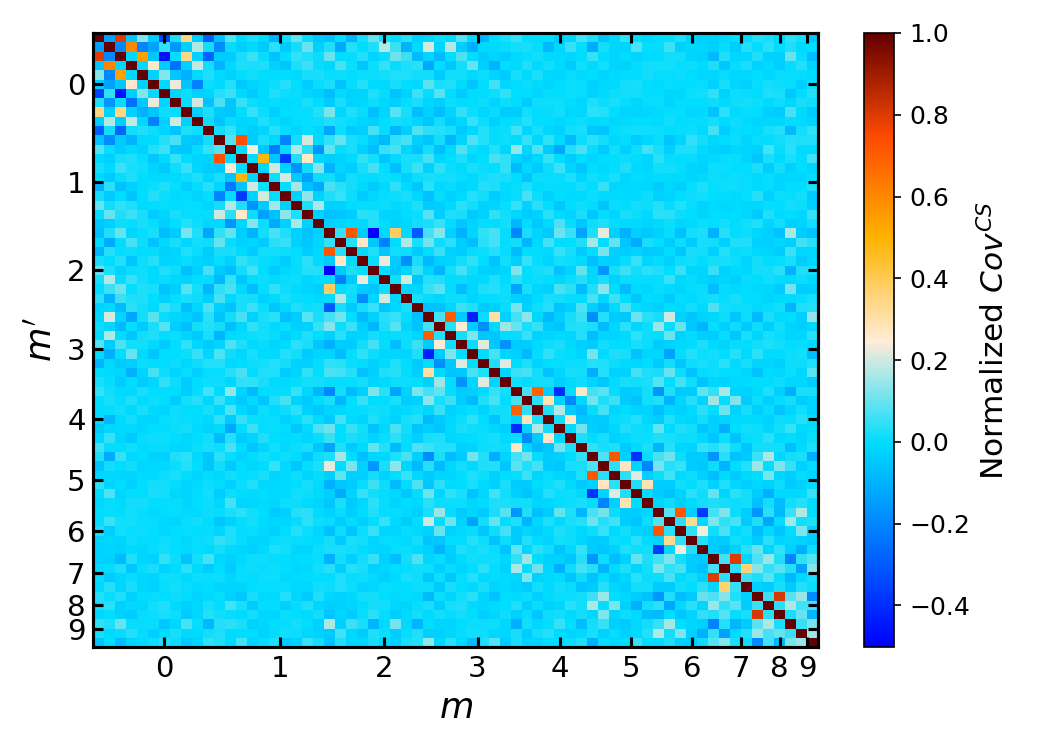}
    \caption{The left panel shows the window function obtained from the common intensity mask at $N_{\rm side} = 64$ and computed from eq.~\eqref{eq: window_discrete}. The right panel shows the covariance of the cut-sky spherical harmonic coefficients, as defined in eq.~\eqref{eq: CS_cov}. For both panels, we mark the $m$-blocks, following the usual indexing of the $a_{\ell m}$s (note that the last block for $m=10$ is not labeled for visualization purposes). Here we assume $\ell_{\rm max}= 10$.}
    \label{fig: window}
\end{figure} 

At this point, we can specify the window function to our case, hence to the common intensity mask of \textit{Planck} (see section~\ref{sec: dataset}). We compute the window function from eq.~\eqref{eq: window_discrete} and using $N_{\rm side} = 64$. Figure~\ref{fig: window} shows the real part of the obtained matrix assuming $\ell_{\rm max} = 10$ and of the full covariance matrix $W^*DW$ (the imaginary part of both is negligible w.r.t. the real one) normalized to the diagonal elements to emphasize the correlation structures. The figure shows that the window function is not block diagonal. Instead, coefficients with different $m$ will get slightly correlated. Indeed, \citet{mortlock2002AnalysisCosmicMicrowavebackground} show that in the case of symmetric, constant latitude cuts, the window function goes exactly to a block diagonal matrix. Still, the elements correlating different orientations are negligible with respect to elements belonging to the same $m$. To highlight this feature, we labeled in the figure the different $m$-blocks, which contain the $\ell-\ell'$ correlations. 

\subsection{Sky variance of the reconstructed coefficients}

\begin{figure}[t]
    \centering
    \includegraphics[width = 0.49\textwidth]{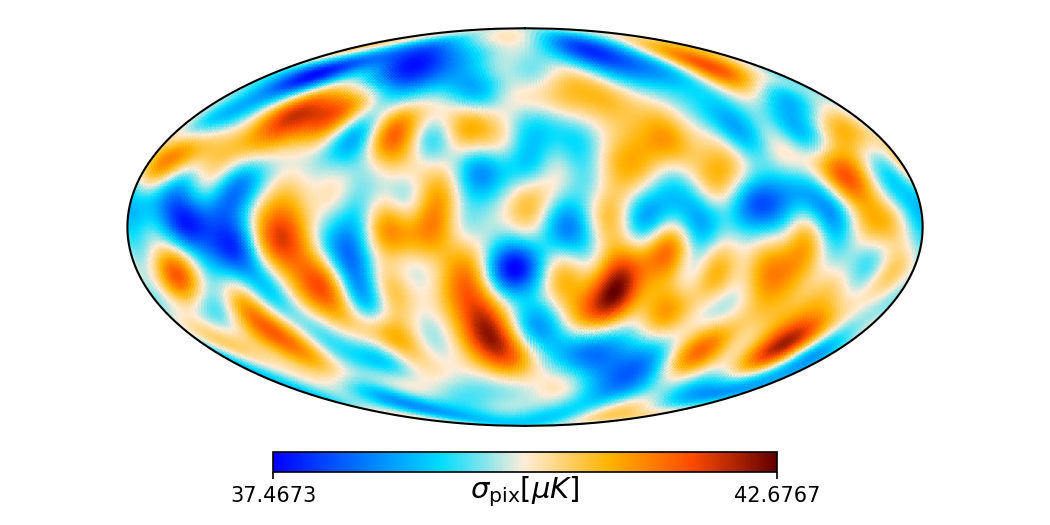}
    \includegraphics[width = 0.49\textwidth]{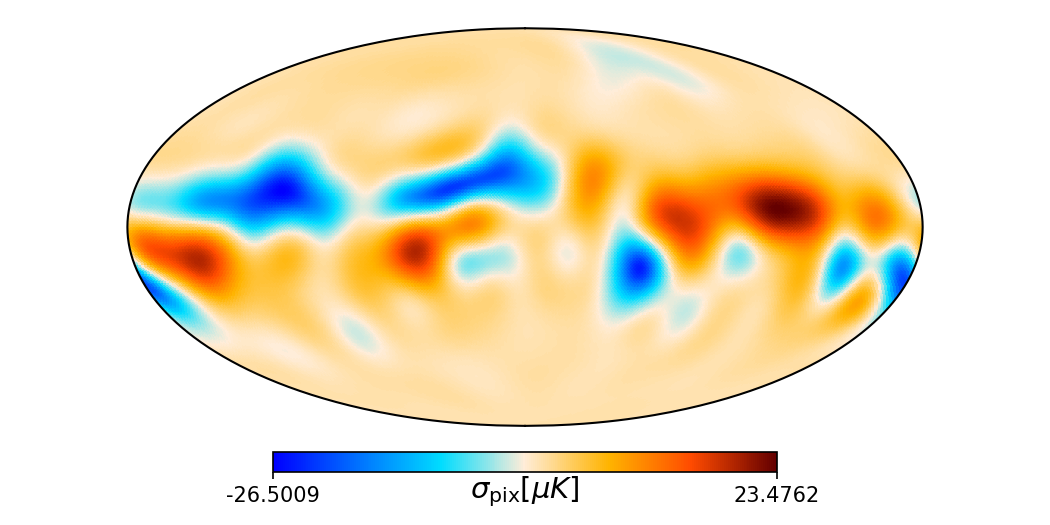}
    \caption{Pixel standard deviation across the sky for the true full-sky coefficients and the reconstructed ones.}
    \label{fig: variance_true_vs_recon}
\end{figure}

As shown in eq.~\eqref{eq: new_constrained}, we use this window function to reconstruct the full-sky spherical harmonics coefficients, however this must come at a cost. Indeed, assuming to have a very aggressive mask, it is intuitive to expect that not all the harmonic modes can be recovered, especially if the typical angular scale of those is comparable with the width of the mask. In more mathematical terms, the window function would be singular, thus one cannot invert it to compute the reconstructed full-sky coefficients from the cut-sky ones \cite{mortlock2002AnalysisCosmicMicrowavebackground}. In our case, the $W$ is invertible, however we expect the reconstructed coefficients to be more noisy than the true ones. To verify this, we generate 1000 full-sky realizations, we mask each and we reconstruct the $\hat{a}_{\ell m}^{\rm FS}$. Then, we compute the standard deviation of each pixel to investigate whether an extra scatter is present w.r.t. the full-sky dispersion. Focusing on the $\ell_{\rm max} = 10$ case, figure~\ref{fig: variance_true_vs_recon} shows the results: on the left we plot the standard deviation pixel by pixel of the true $a_{\ell m}^{\rm FS}$, while on the right we plot the extra standard deviation induced by the reconstruction procedure.

Note that this contribution concentrates around masked areas of the sky (see figure~\ref{fig: maps}), suggesting that we pay the price of a full-sky description by increasing the noisiness of those regions, as expected.

\section{Angular correlation functions for the CGWB}\label{app: ang_corr_cgwb}

\begin{figure}[t]
    \centering
    \includegraphics[width = 0.49\textwidth]{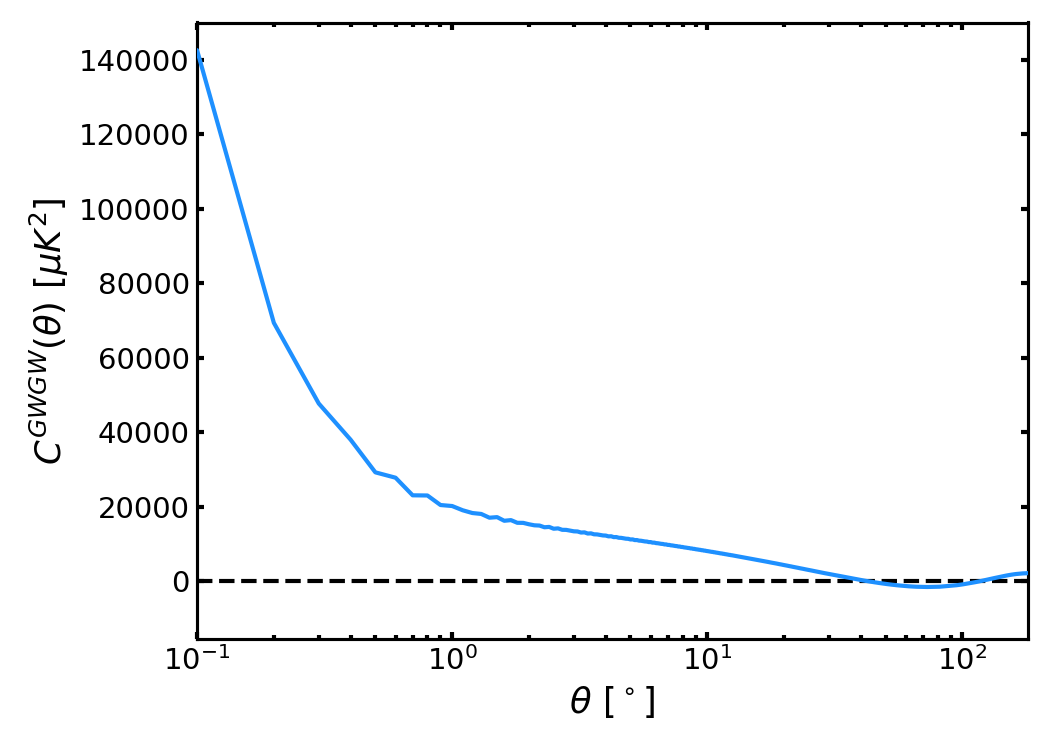}
    \includegraphics[width = 0.49\textwidth]{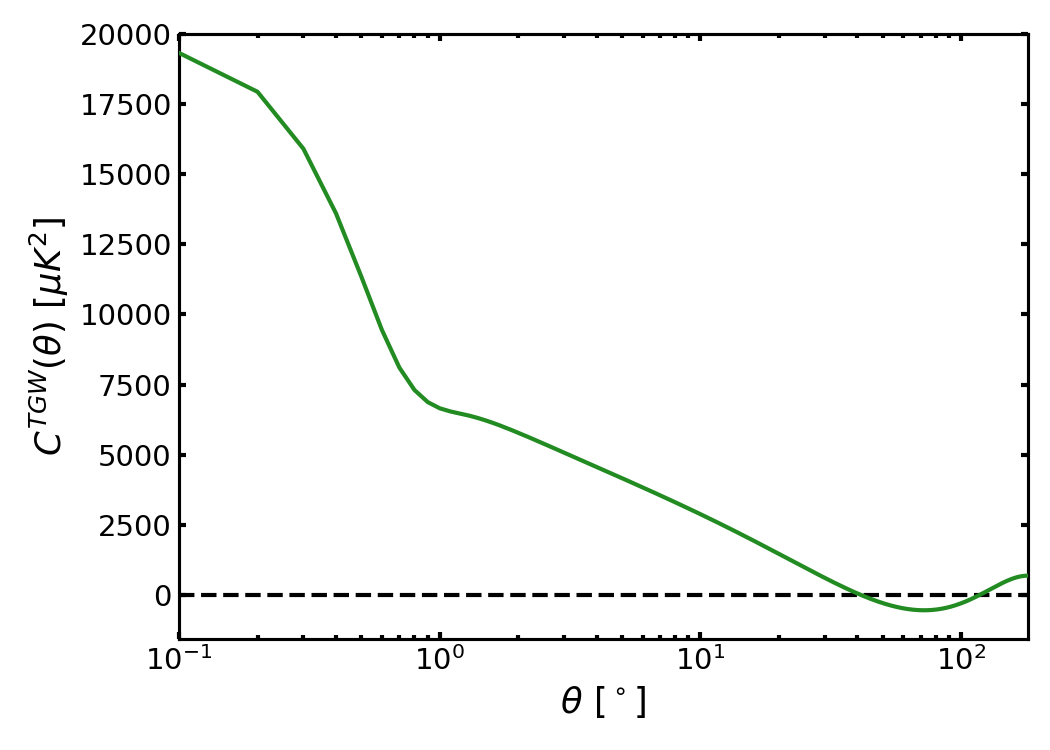}
    \caption{Angular correlation function $C(\theta)$ of GWGW (left) and TGW (right) assuming $\ell_{\rm max} = 2000$.}
    \label{fig: high_ell_ang_corr}
\end{figure}

In section~\ref{sec: ang_corr} we show the angular correlation functions for GWGW and TGW, however, we limit the angular range to the one accessible imposing $\ell_{\rm max} = 4$. This is done to account for the performance of future GW interferometers. Despite this, it is interesting to have a look at the 2-point angular correlation function on all scales, as predicted by $\Lambda$CDM. Indeed, the angular power spectrum of the CGWB is very similar to the one of CMB temperature due to their high correlation, but they diverge on small scales ($\ell \gtrapprox 100$). Hence, figure~\ref{fig: high_ell_ang_corr} shows $C(\theta)$ for GWGW and TGW when assuming $\ell_{\rm max} = 2000$.
Indeed, note that both the GWGW and the TGW present the typical ``sinusoidal'' behavior on large scales. Instead, on small ones GWGW tend to very high values since the CGWB spectrum is not dumped as the CMB temperature one.

\section{Optimistic GWs detection}\label{app: sensitivity}

Here, we show some of the results of the same analysis performed in the main body if we change the assumption on $\ell_{\rm max}$. Indeed, we just showed the most pessimistic case of $\ell_{\rm max} = 4$. For the sake of brevity, we cannot show everything as we did in the main body. Instead, we will show exclusively the results when we assume masked SMICA to be our CMB observation.

\subsection{Optimal angles with \texorpdfstring{$\ell_{\rm max} = 6$}{TEXT}}

\begin{figure}[t]
    \centering
    \includegraphics[width = 0.49\textwidth]{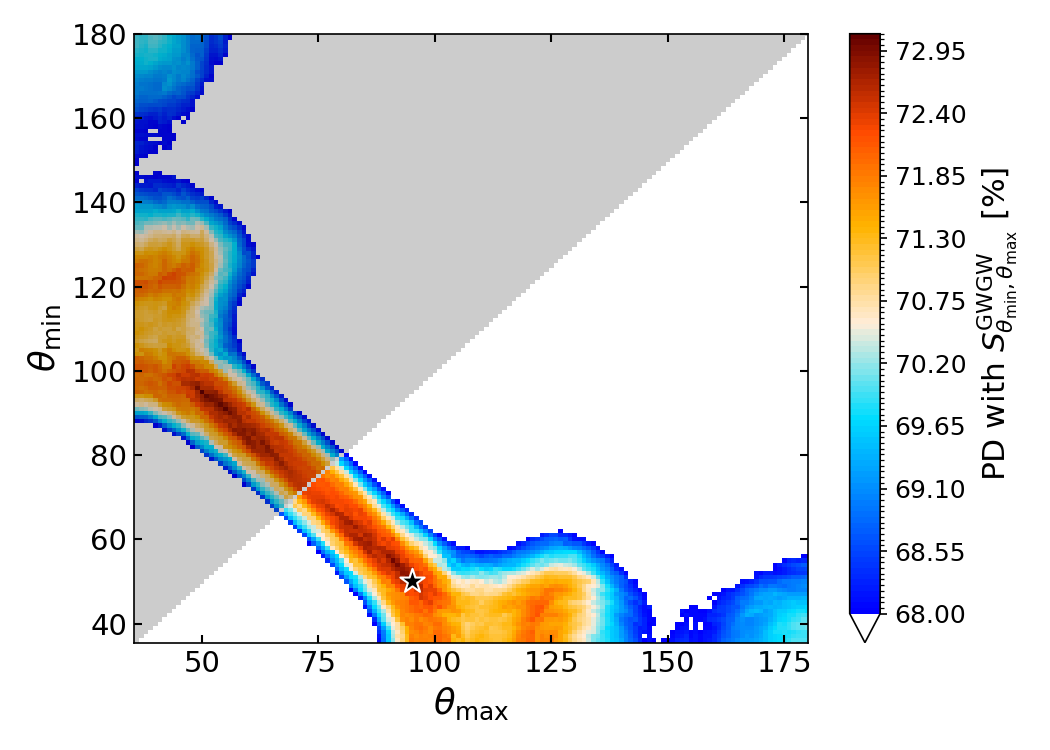}
    \includegraphics[width = 0.49\textwidth]{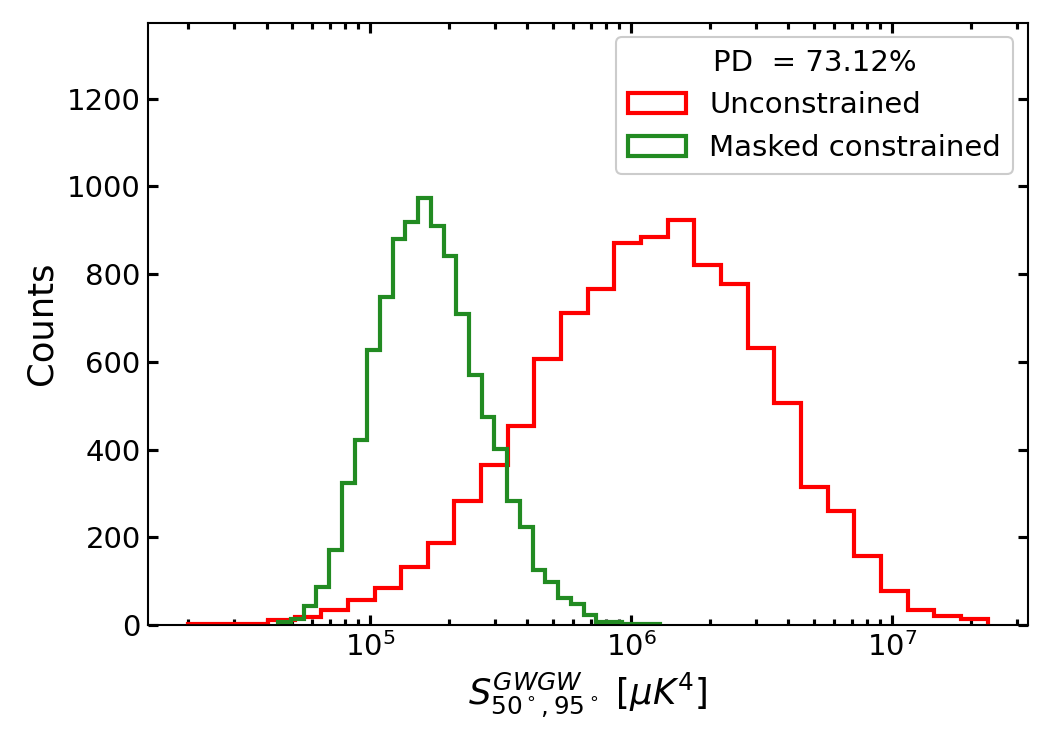}
    \caption{Optimal angles for GWGW. The left and right panels show respectively optimal angles analysis and the results at the optimal range. Here we assume $\ell_{\rm max}= 6$.}
    \label{fig: masked_GWGW_lmax6}
\end{figure}

\begin{figure}[t]
    \centering
    \includegraphics[width = 0.49\textwidth]{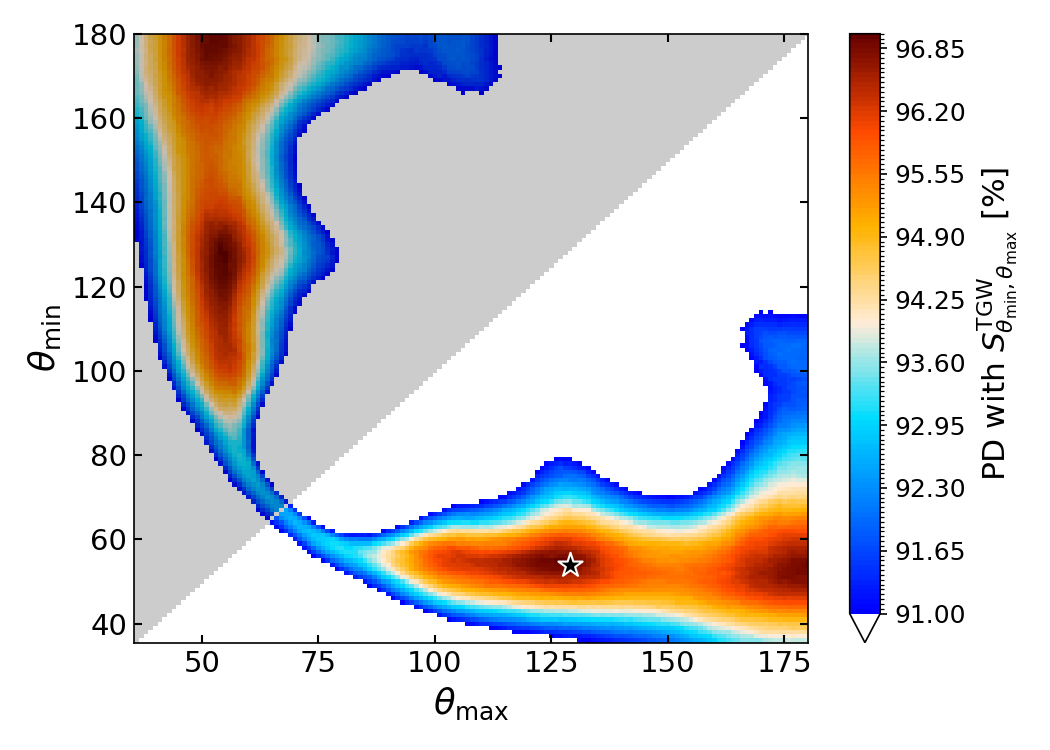}
    \includegraphics[width = 0.49\textwidth]{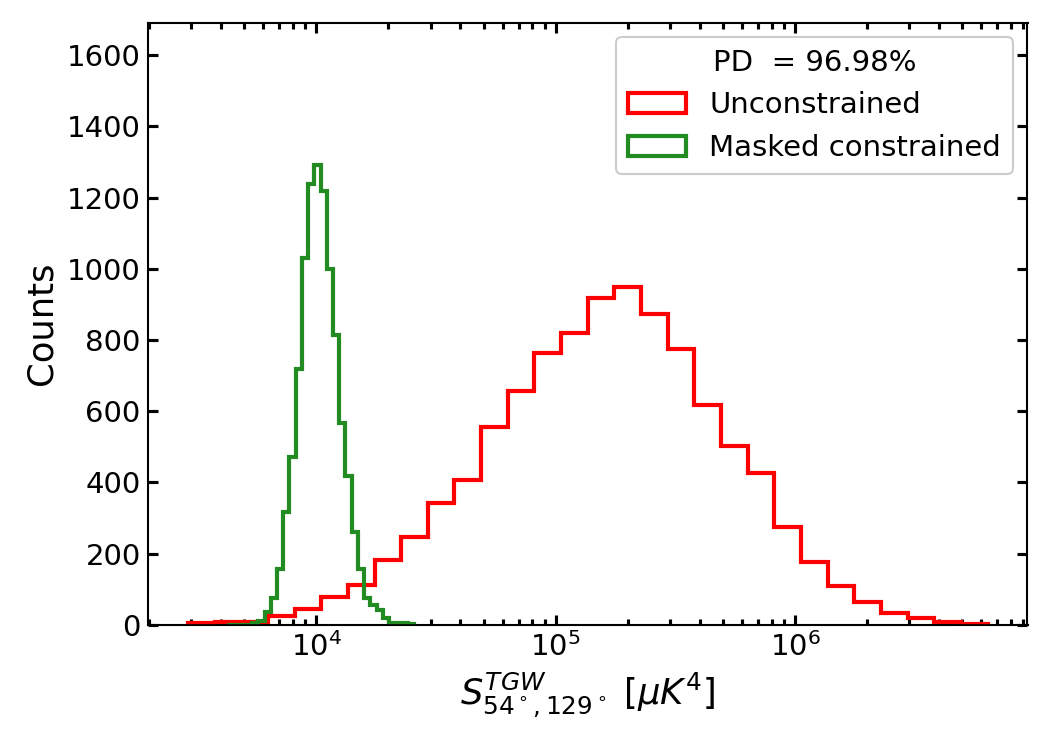}
    \caption{Optimal angles for TGW. The left and right panels show respectively optimal angles analysis and the results at the optimal range. Here we assume $\ell_{\rm max}= 6$.}
    \label{fig: masked_X_lmax6}
\end{figure}

We show here the results of the optimal angles analysis for $S_{\theta_{\rm min}, \theta_{\rm max}}^{\rm GWGW}$ on figure~\ref{fig: masked_GWGW_lmax6} and on figure~\ref{fig: masked_X_lmax6} for $S_{\theta_{\rm min}, \theta_{\rm max}}^{\rm TGW}$.

\begin{figure}[t]
    \centering
    \includegraphics[width = 0.49\textwidth]{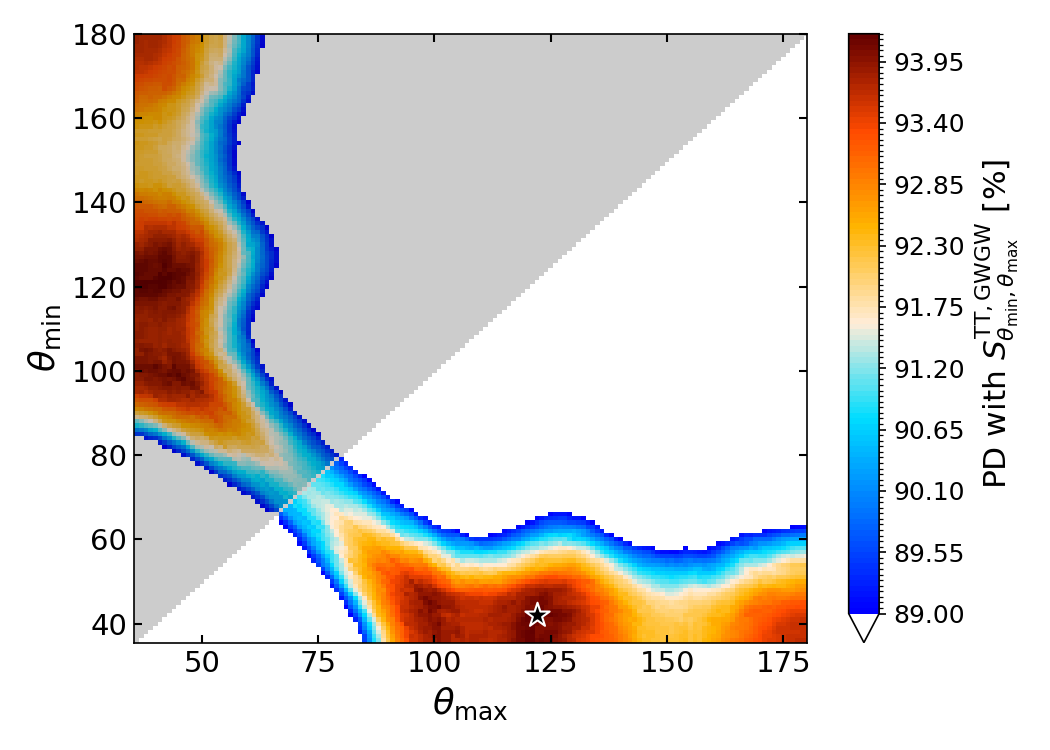}
    \includegraphics[width = 0.49\textwidth]{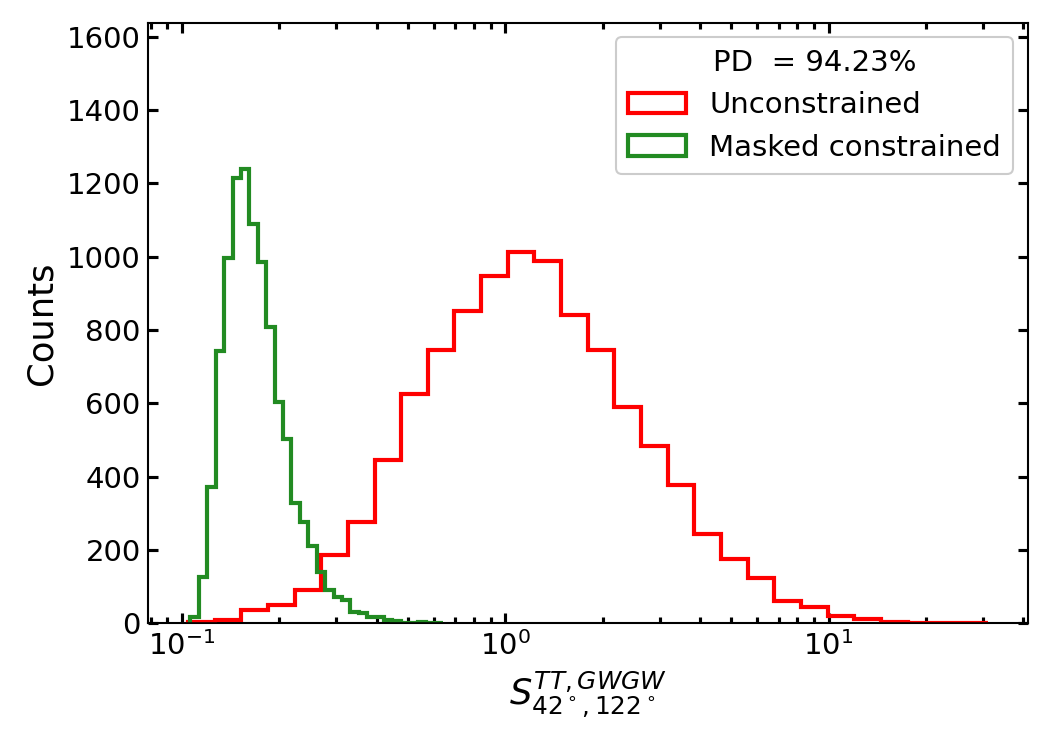}
    \caption{Optimal angles for the combination of TT and GWGW. The left and right panels show respectively optimal angles analysis and the results at the optimal range. Here we assume $\ell_{\rm max}= 6$.}
    \label{fig: masked_TTGW_lmax6}
\end{figure}

\begin{figure}[t]
    \centering
    \includegraphics[width = 0.49\textwidth]{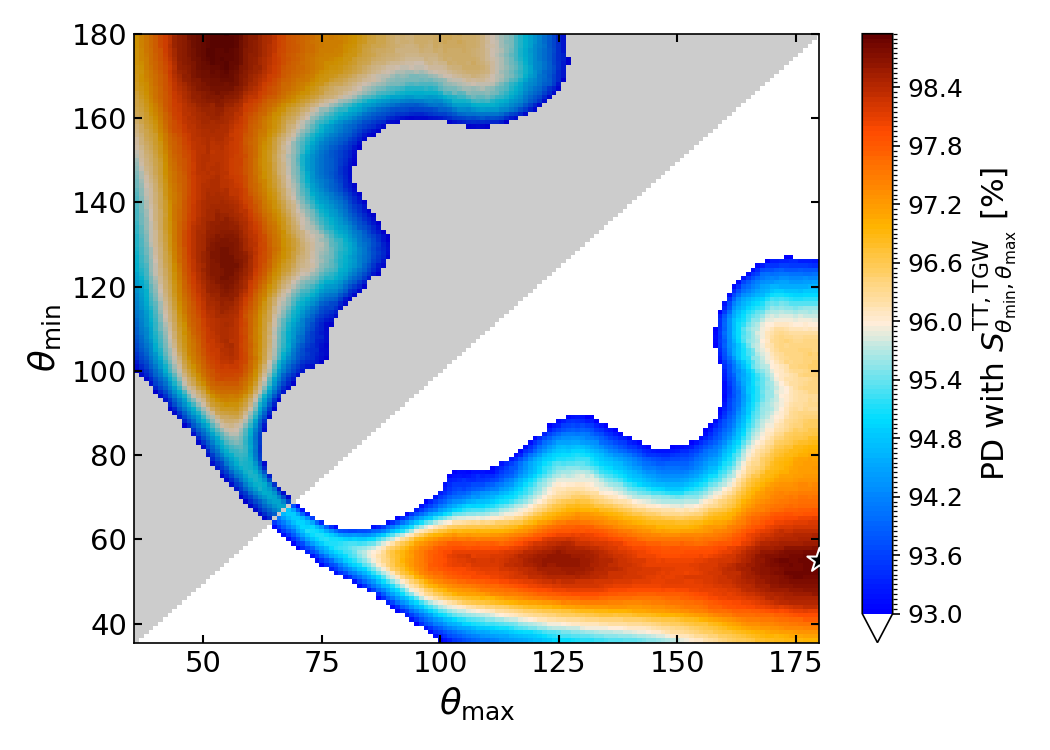}
    \includegraphics[width = 0.49\textwidth]{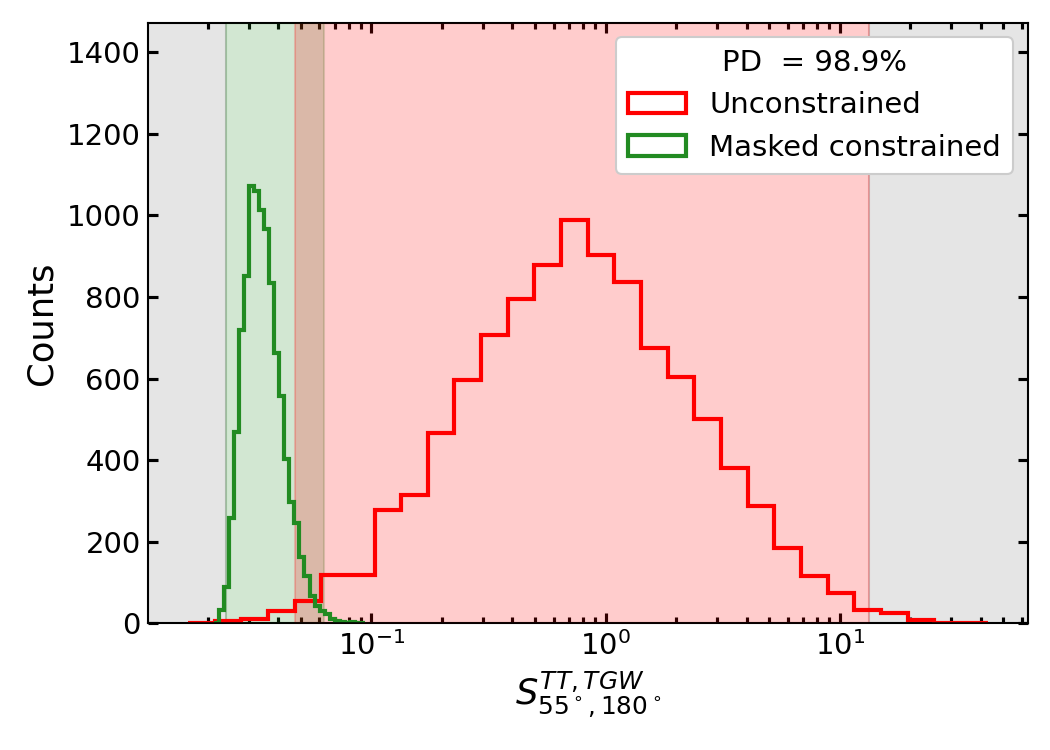}
    \caption{Optimal angles for the combination of TT and TGW. The left and right panels show respectively optimal angles analysis and the results at the optimal range. Here we assume $\ell_{\rm max}= 6$.}
    \label{fig: masked_TTX_lmax6}
\end{figure}

\begin{figure}[t]
    \centering
    \includegraphics[width = 0.49\textwidth]{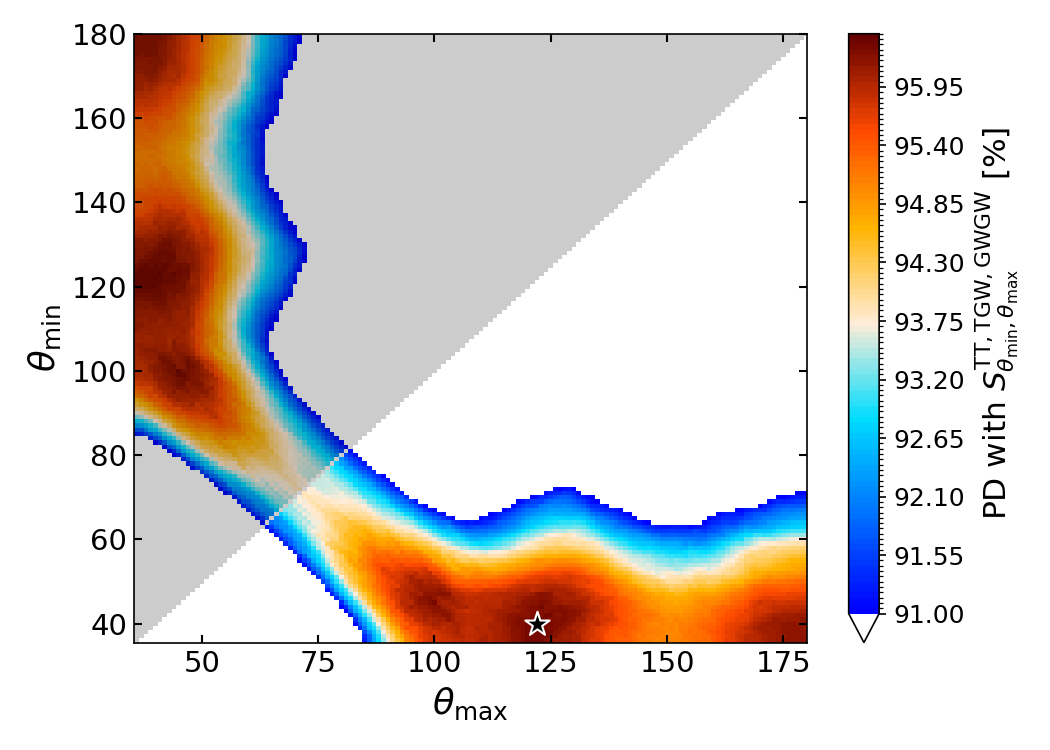}
    \includegraphics[width = 0.49\textwidth]{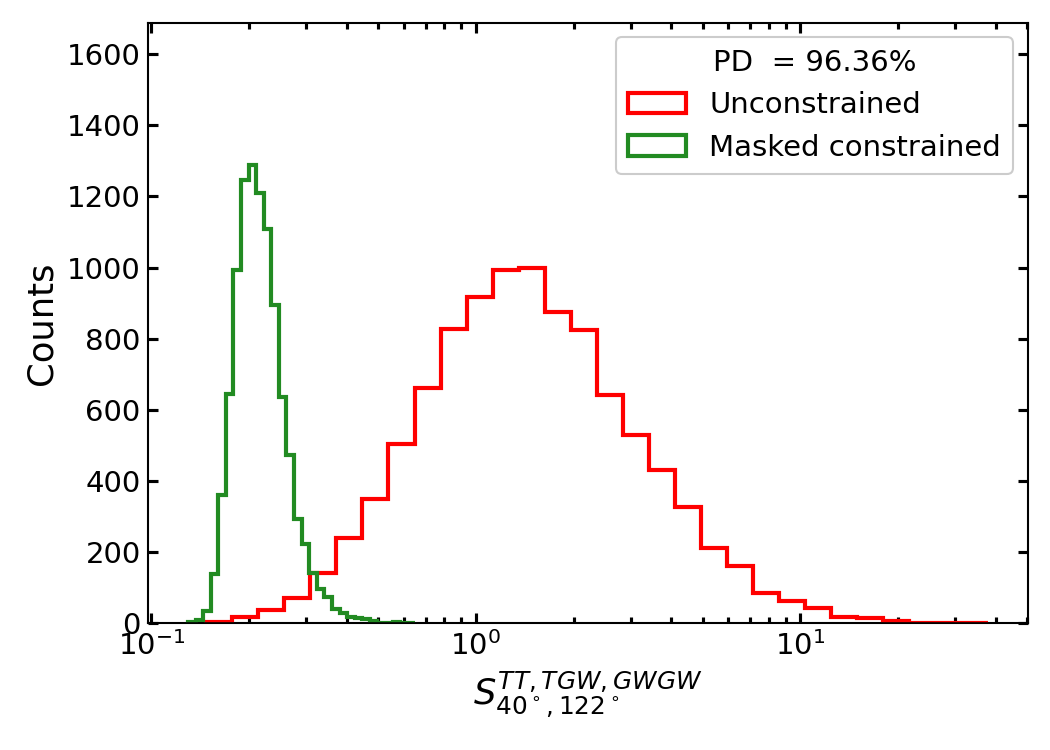}
    \caption{Optimal angles for the combination of TT, TGW, and GWGW. The left and right panels show respectively optimal angles analysis and the results at the optimal range. Here we assume $\ell_{\rm max}= 6$.}
    \label{fig: masked_TTXGW_lmax6}
\end{figure}

Instead, figures~\ref{fig: masked_TTGW_lmax6}-\ref{fig: masked_TTX_lmax6}-\ref{fig: masked_TTXGW_lmax6} show the same respectively for $S^{\rm TT, GWGW}_{\theta_{\rm min}, \theta_{\rm max}}$, $S^{\rm TT, TGW}_{\theta_{\rm min}, \theta_{\rm max}}$ and $S^{\rm TT, TGW, GWGW}_{\theta_{\rm min}, \theta_{\rm max}}$. As mentioned in section~\ref{sec: res_optimal}, there always seems to be relatively high-PD region where the original $S_{1/2}$ estimator is defined, i.e. between $60^\circ$ and $180^\circ$.


\subsection{Significance with \texorpdfstring{$\ell_{\rm max} = 6$}{TEXT}}

\begin{figure}[t]
    \centering
    \includegraphics[width = 0.49\textwidth]{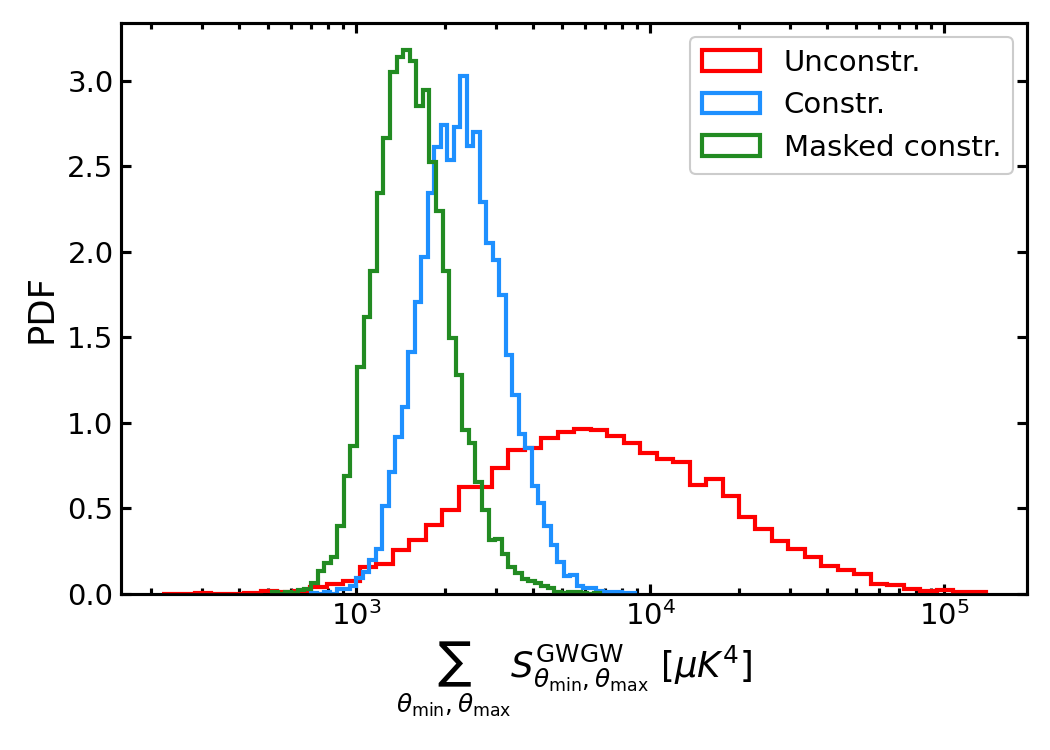}
    \includegraphics[width = 0.49\textwidth]{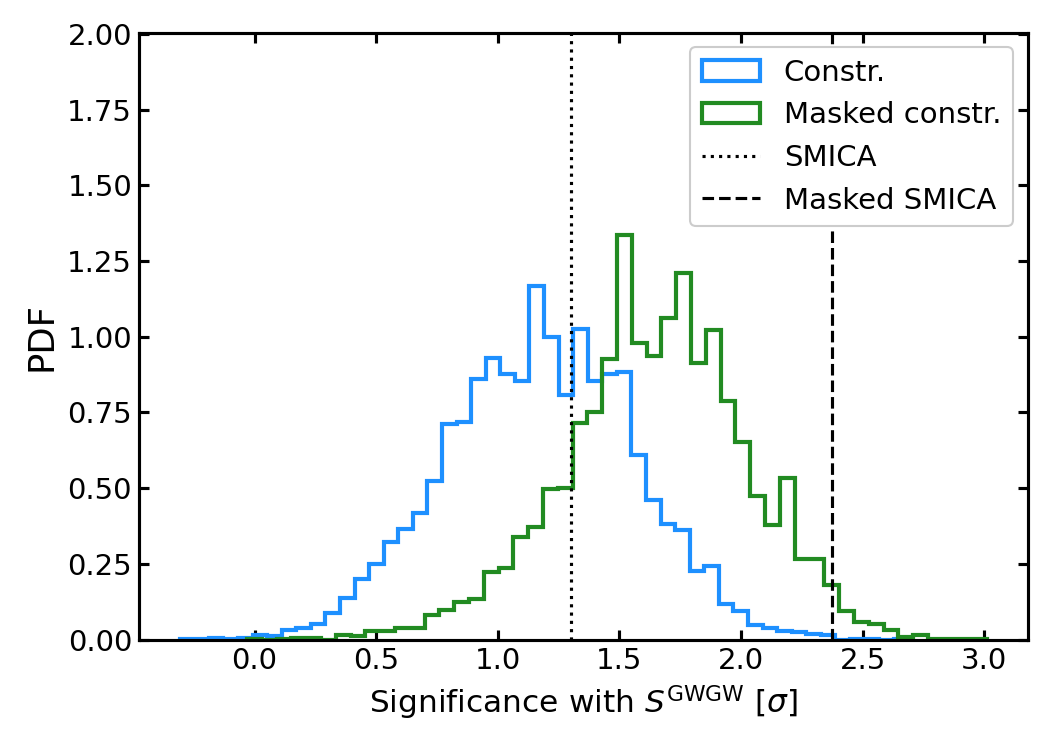}
    \caption{On the left panel, the value of $S^{\rm GWGW}$ for full-sky and masked constrained realizations of the CGWB and the $\Lambda$CDM realizations ($\ell_{\rm max}=6$). On the right panel, corresponding significance in terms of $\sigma$ w.r.t. the unconstrained realizations. The dotted and dashed vertical lines indicate the full-sky and masked SMICA-alone significance, respectively.}
    \label{fig:signi_GWGW_lmax6}
\end{figure}

\begin{figure}[t]
    \centering
    \includegraphics[width = 0.49\textwidth]{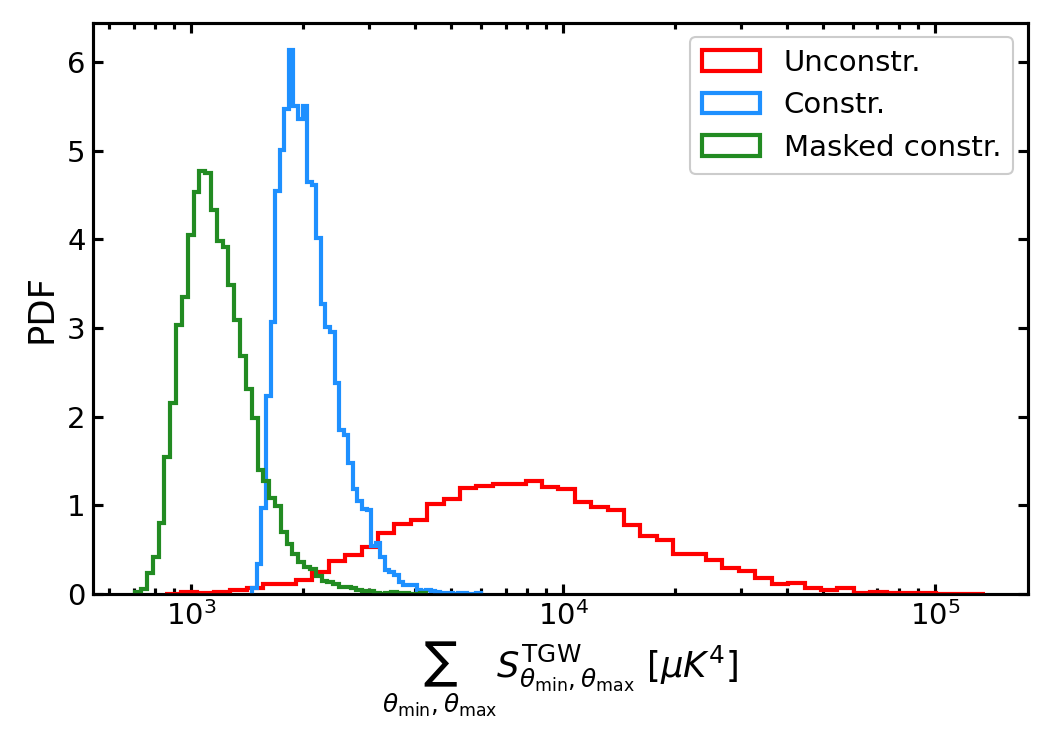}
    \includegraphics[width = 0.49\textwidth]{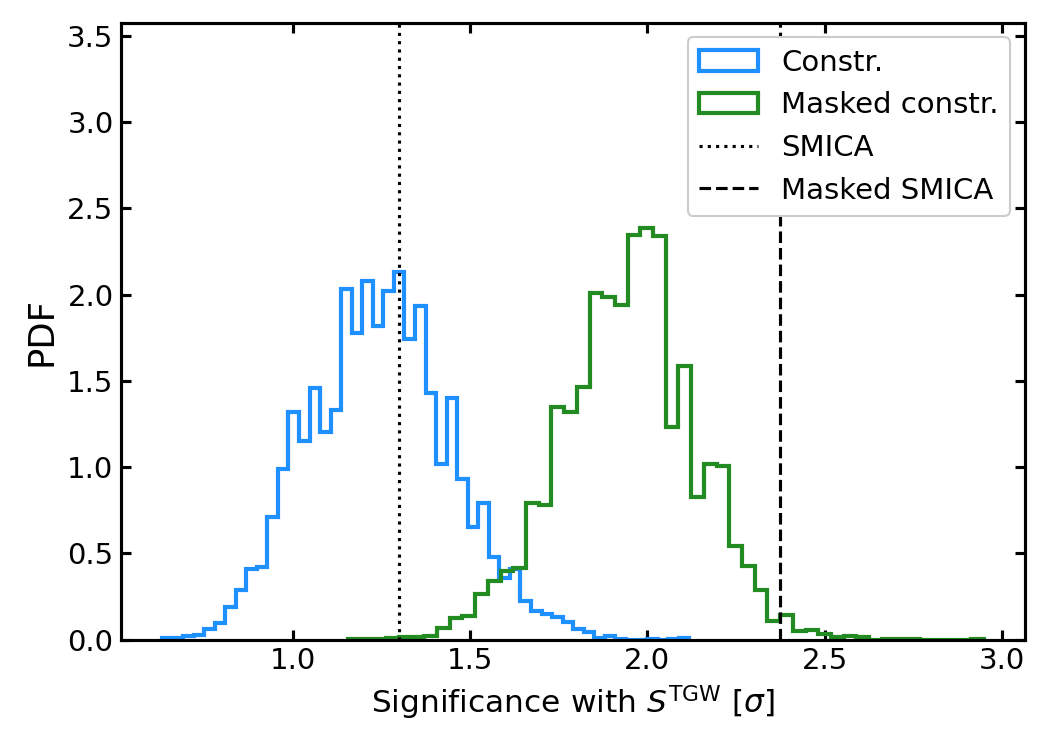}
    \caption{On the left panel, the value of $S^{\rm TGW}$ for full-sky and masked constrained realizations of the CGWB and the $\Lambda$CDM realizations ($\ell_{\rm max}=6$). On the right panel, corresponding significance in terms of $\sigma$ w.r.t. the unconstrained realizations. The dotted and dashed vertical lines indicate the full-sky and masked SMICA-alone significance, respectively.}
    \label{fig:signi_X_lmax6}
\end{figure}

As done in section~\ref{sec: signi}, we report here the significance analysis performed using the newly defined estimator of eq.~\eqref{eq: signi_estimator}. Here, we can also show the full-sky results given that we can plot both in the same figure. Starting from $S^{\rm GWGW}$, we obtain figure~\ref{fig:signi_GWGW_lmax6}, while for $S^{\rm TGW}$ we get figure~\ref{fig:signi_X_lmax6}.


\begin{figure}[t]
    \centering
    \includegraphics[width = 0.49\textwidth]{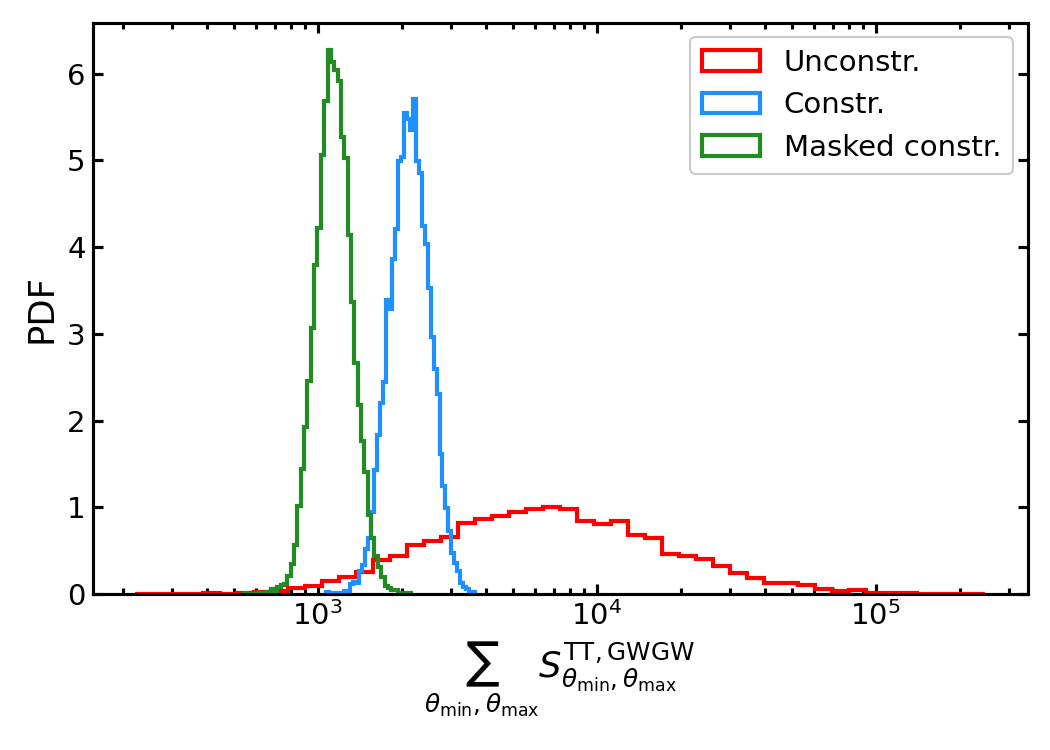}
    \includegraphics[width = 0.49\textwidth]{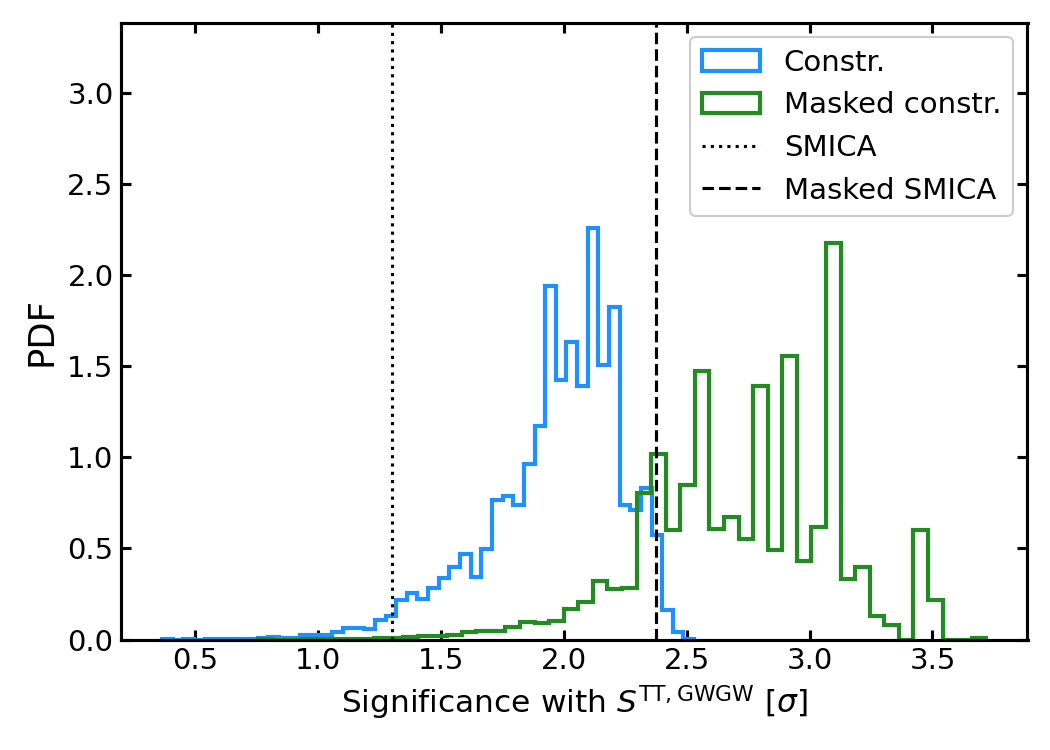}
    \caption{On the left panel, the value of $S^{\rm TT, GWGW}$ for full-sky and masked constrained realizations of the CGWB and the $\Lambda$CDM realizations ($\ell_{\rm max}=6$). On the right panel, corresponding significance in terms of $\sigma$ w.r.t. the unconstrained realizations. The dotted and dashed vertical lines indicate the full-sky and masked SMICA-alone significance, respectively.}
    \label{fig:signi_TTGW_lmax6}
\end{figure}

\begin{figure}[t]
    \centering
    \includegraphics[width = 0.49\textwidth]{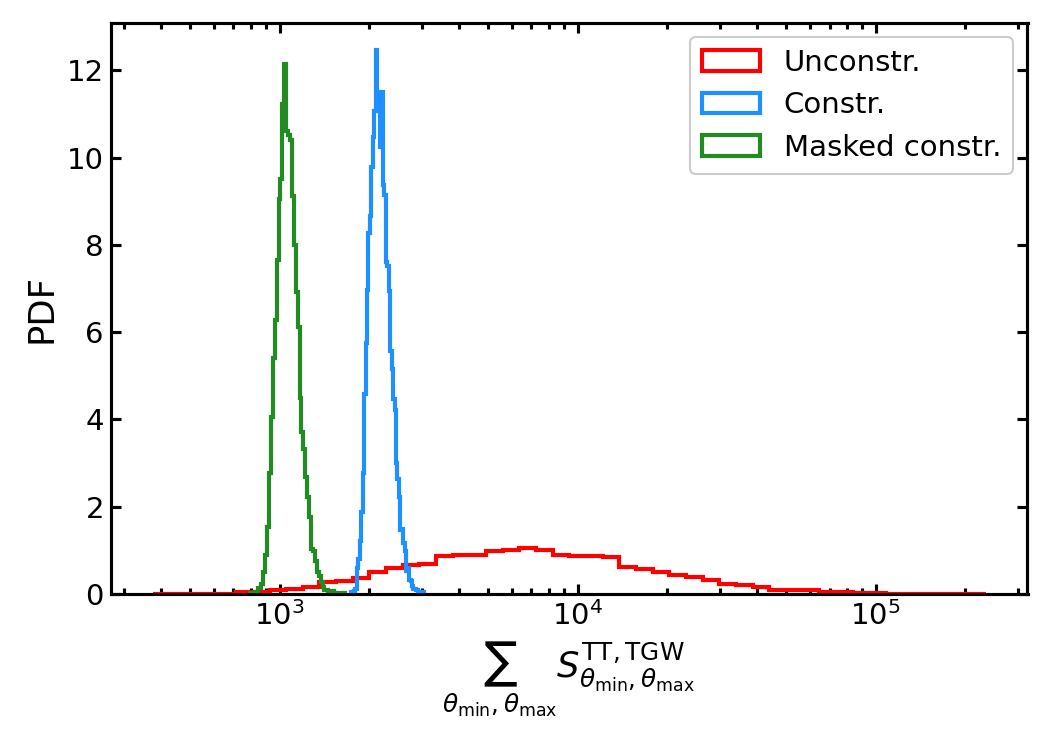}
    \includegraphics[width = 0.49\textwidth]{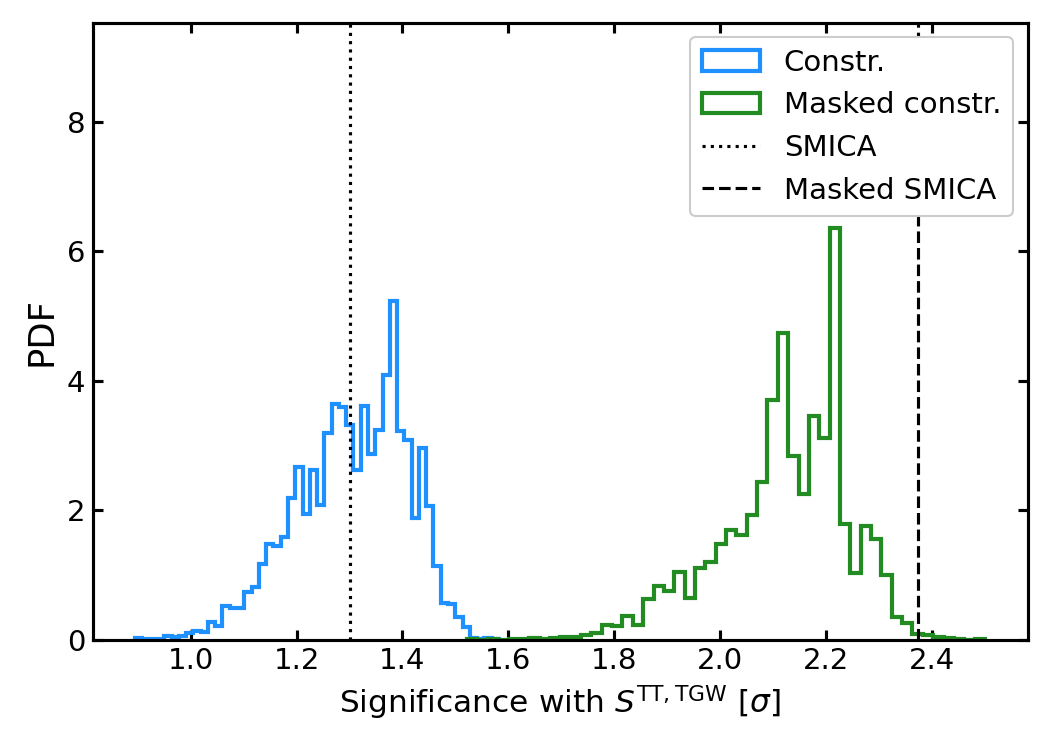}
    \caption{On the left panel, the value of $S^{\rm TT, TGW}$ for full-sky and masked constrained realizations of the CGWB and the $\Lambda$CDM realizations ($\ell_{\rm max}=6$). On the right panel, corresponding significance in terms of $\sigma$ w.r.t. the unconstrained realizations. The dotted and dashed vertical lines indicate the full-sky and masked SMICA-alone significance, respectively.}
    \label{fig:signi_TTX_lmax6}
\end{figure}

\begin{figure}[t]
    \centering
    \includegraphics[width = 0.49\textwidth]{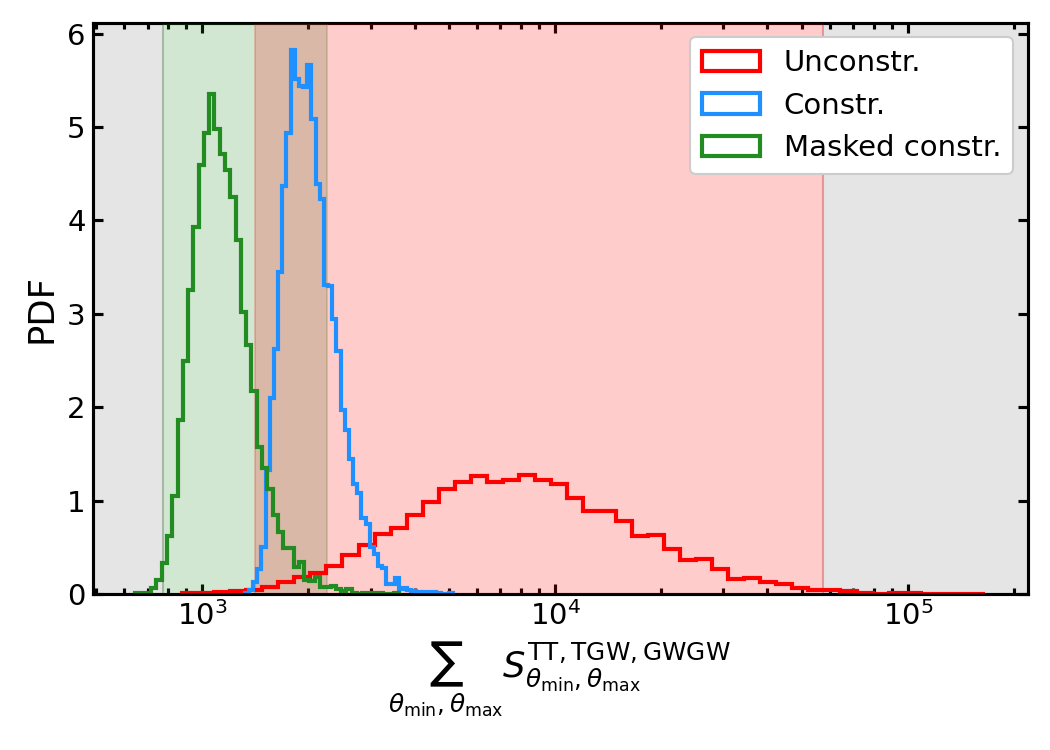}
    \includegraphics[width = 0.49\textwidth]{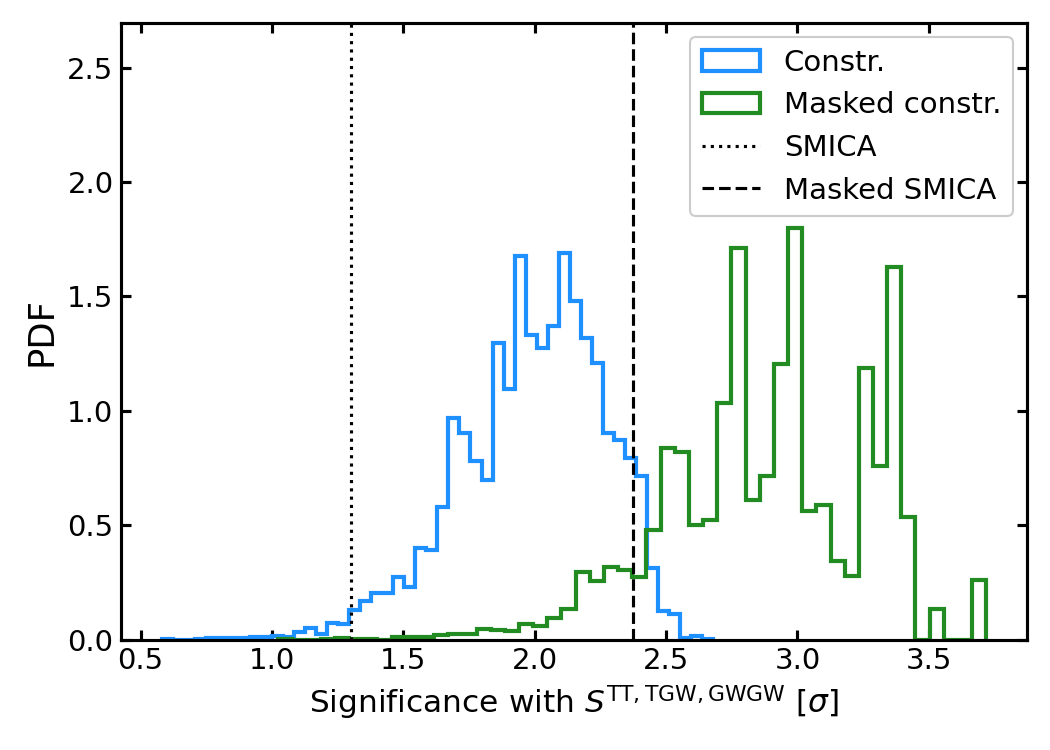}
    \caption{On the left panel, the value of $S^{\rm TT, TGW, GWGW}$ for full-sky and masked constrained realizations of the CGWB and the $\Lambda$CDM realizations ($\ell_{\rm max}=6$). On the right panel, corresponding significance in terms of $\sigma$ w.r.t. the unconstrained realizations. The dotted and dashed vertical lines indicate the full-sky and masked SMICA-alone significance, respectively.}
    \label{fig:signi_TTXGW_lmax6}
\end{figure}

Switching to the multi-field analysis, we obtain figures~\ref{fig:signi_TTGW_lmax6}-\ref{fig:signi_TTX_lmax6}-\ref{fig:signi_TTXGW_lmax6} for $S^{\rm TT, GWGW}$, $S^{\rm TT, TGW}$, and $S^{\rm TT, TGW, GWGW}$.


\subsection{Optimal angles with \texorpdfstring{$\ell_{\rm max} = 10$}{TEXT}}

\begin{figure}[t]
    \centering
    \includegraphics[width = 0.49\textwidth]{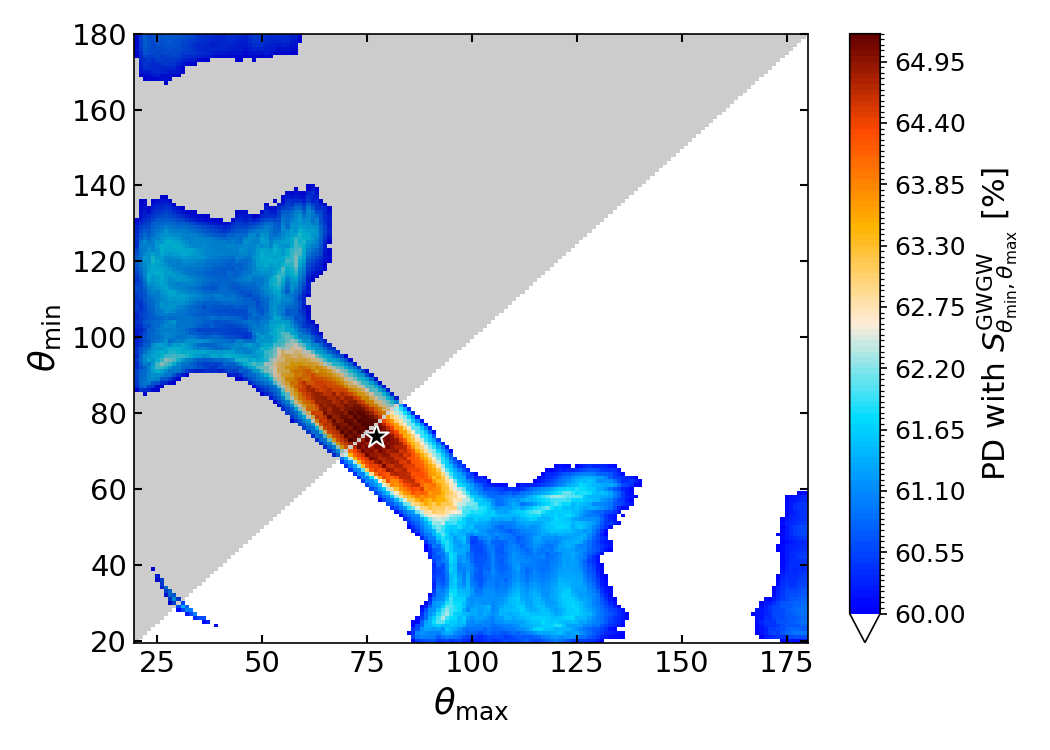}
    \includegraphics[width = 0.49\textwidth]{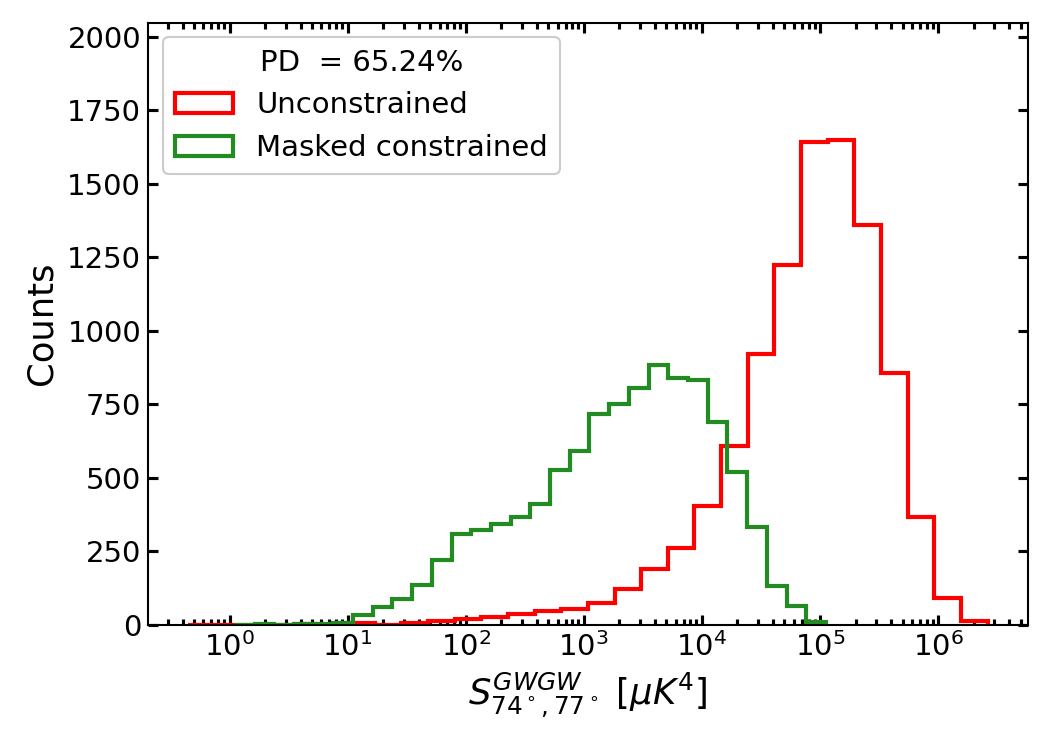}
    \caption{Optimal angles for GWGW. The left and right panels show respectively optimal angles analysis and the results at the optimal range. Here we assume $\ell_{\rm max}= 10$.}
    \label{fig: masked_GWGW_lmax10}
\end{figure}

\begin{figure}[t]
    \centering
    \includegraphics[width = 0.49\textwidth]{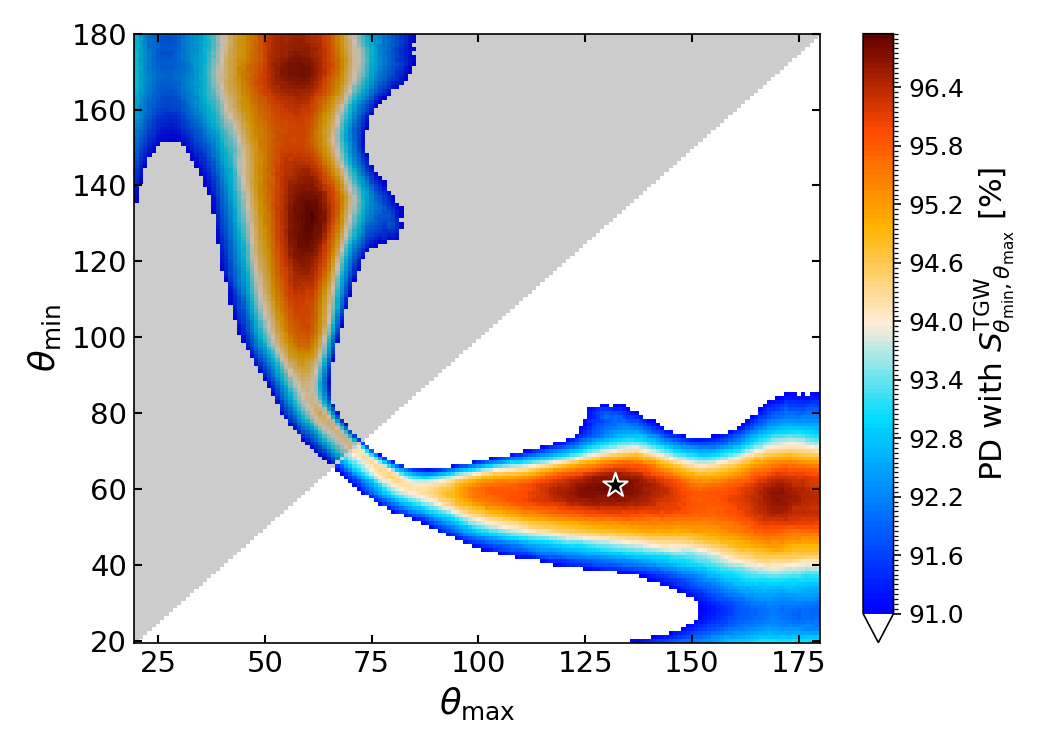}
    \includegraphics[width = 0.49\textwidth]{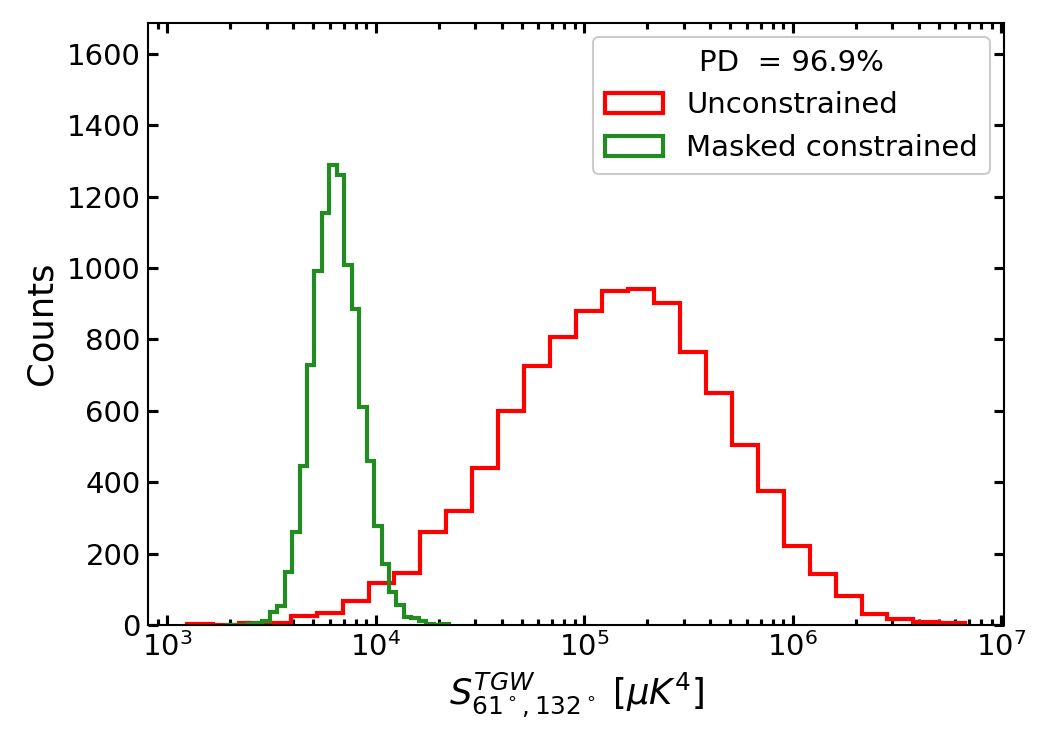}
    \caption{Optimal angles for TGW. The left and right panels show respectively optimal angles analysis and the results at the optimal range. Here we assume $\ell_{\rm max}= 10$.}
    \label{fig: masked_X_lmax10}
\end{figure}

We repeated the same thing for $\ell_{\rm max} = 10$. Firstly, for $S_{\theta_{\rm min}, \theta_{\rm max}}^{\rm GWGW}$ we obtain figure~\ref{fig: masked_GWGW_lmax10} and figure~\ref{fig: masked_X_lmax10} for $S_{\theta_{\rm min}, \theta_{\rm max}}^{\rm TGW}$.

\begin{figure}[t]
    \centering
    \includegraphics[width = 0.49\textwidth]{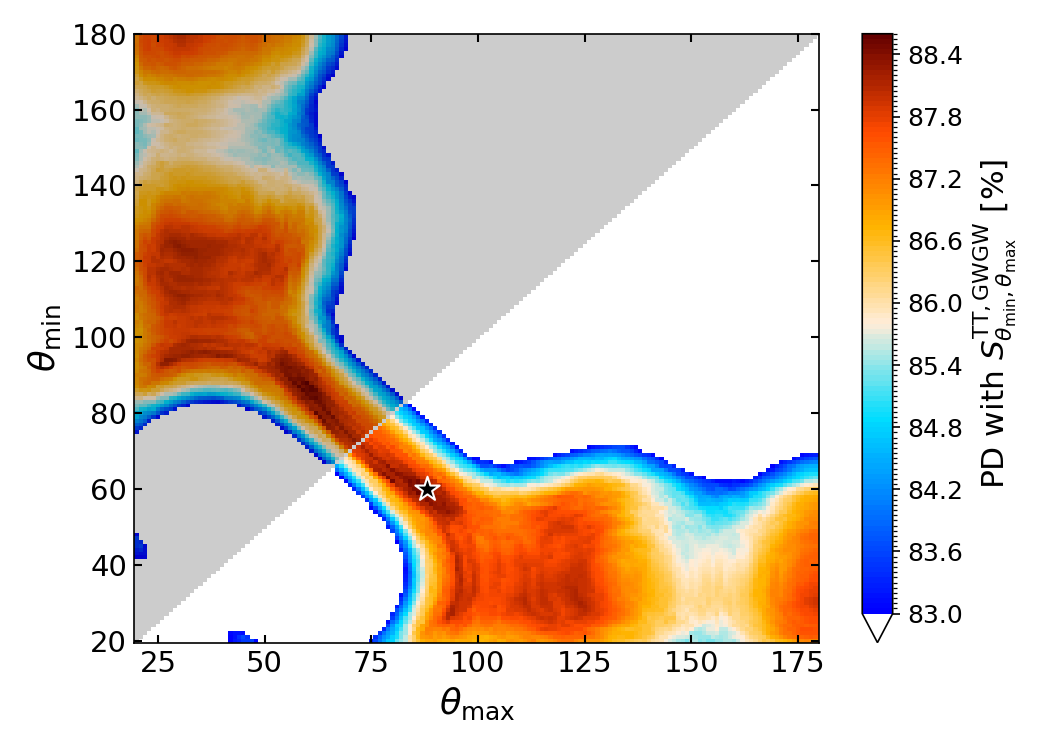}
    \includegraphics[width = 0.49\textwidth]{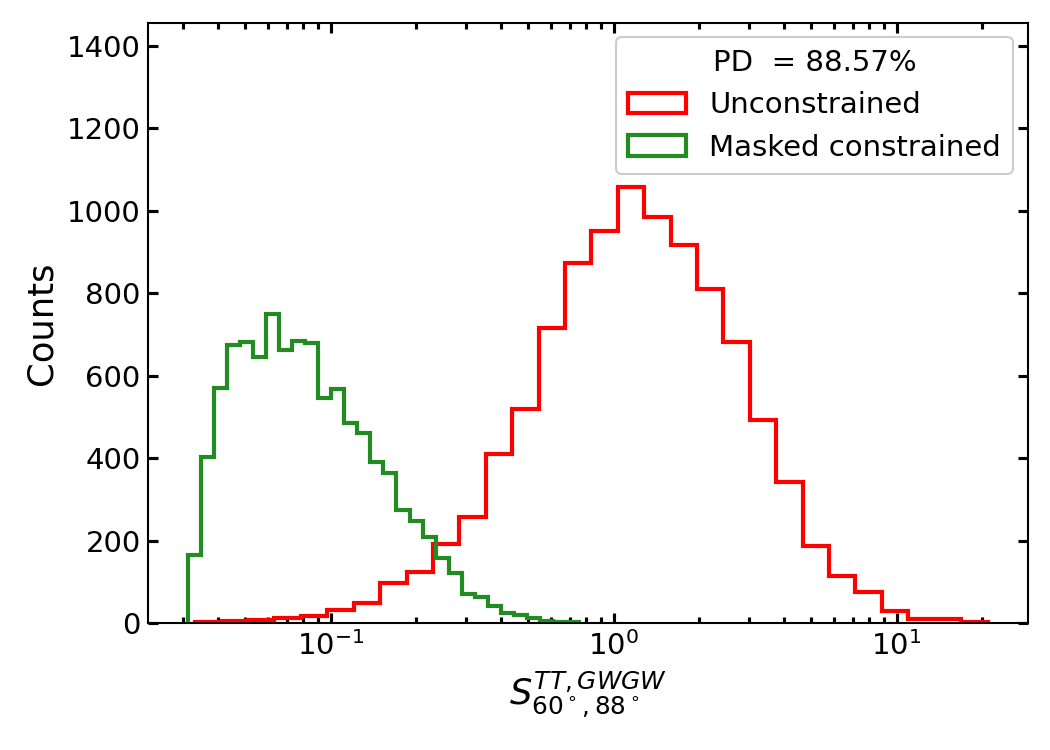}
    \caption{Optimal angles for the combination of TT and GWGW. The left and right panels show respectively optimal angles analysis and the results at the optimal range. Here we assume $\ell_{\rm max}= 10$.}
    \label{fig: masked_TTGW_lmax10}
\end{figure}

\begin{figure}[t]
    \centering
    \includegraphics[width = 0.49\textwidth]{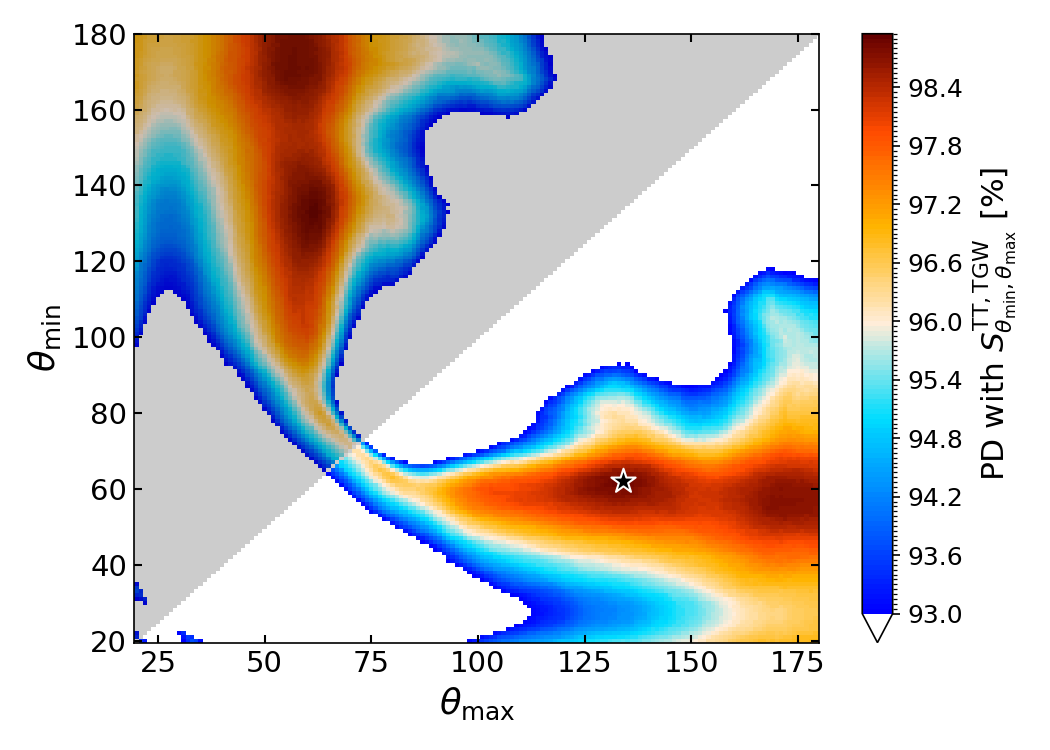}
    \includegraphics[width = 0.49\textwidth]{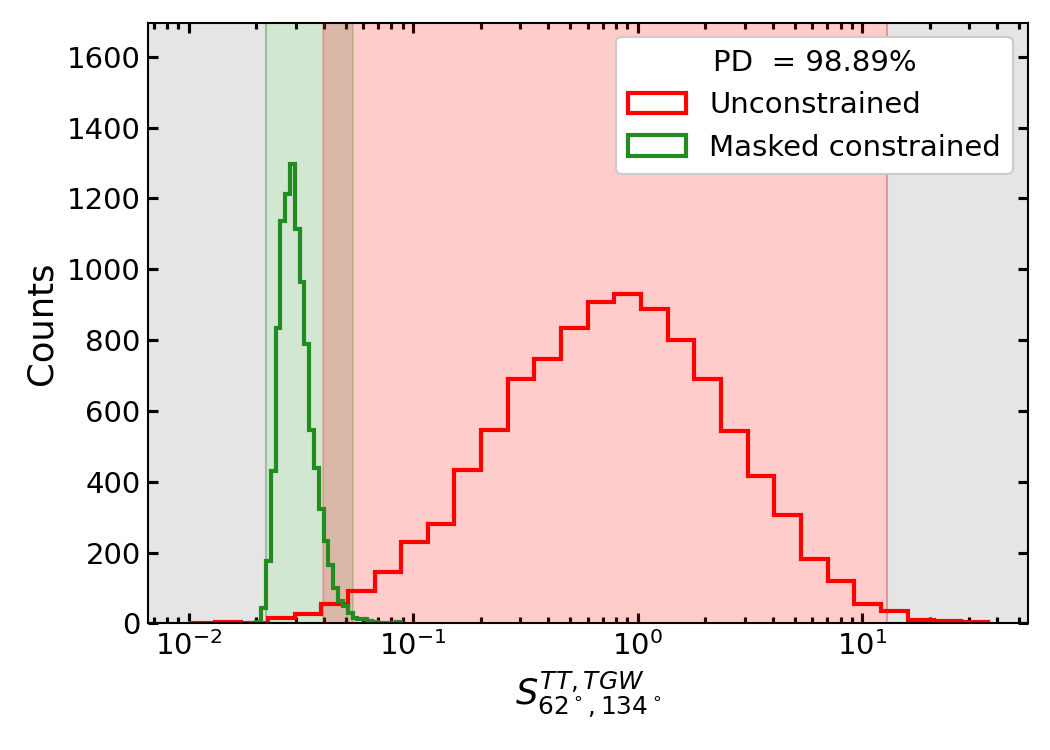}
    \caption{Optimal angles for the combination of TT and TGW. The left and right panels show respectively optimal angles analysis and the results at the optimal range. Here we assume $\ell_{\rm max}= 10$.}
    \label{fig: masked_TTX_lmax10}
\end{figure}

\begin{figure}[t]
    \centering
    \includegraphics[width = 0.49\textwidth]{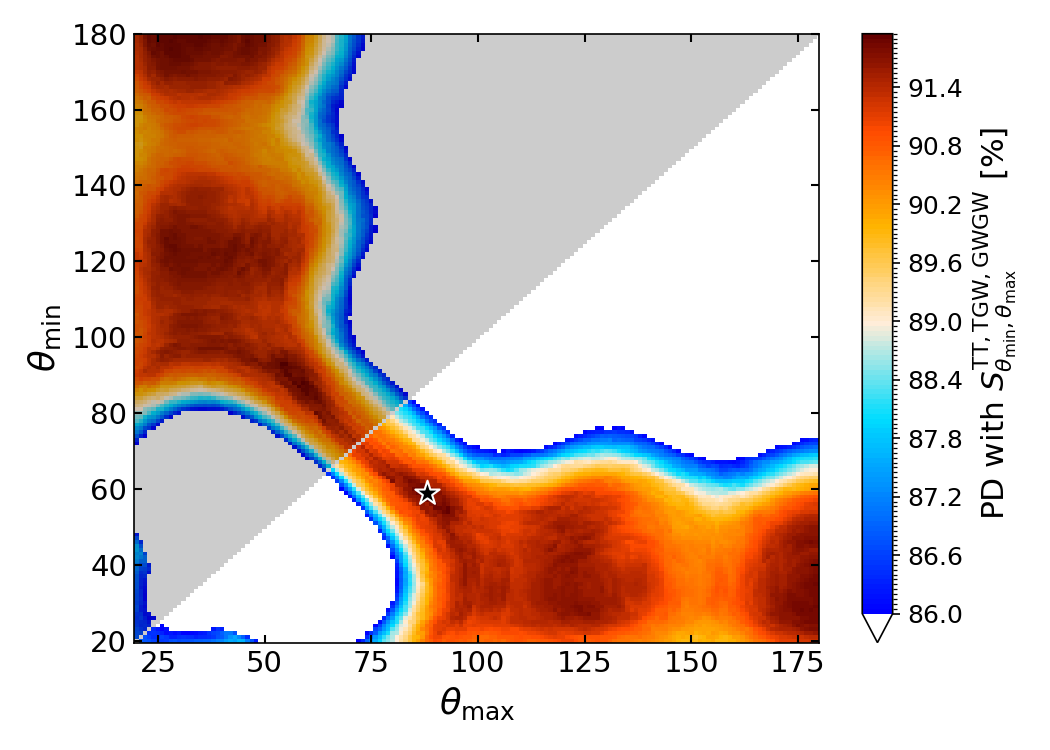}
    \includegraphics[width = 0.49\textwidth]{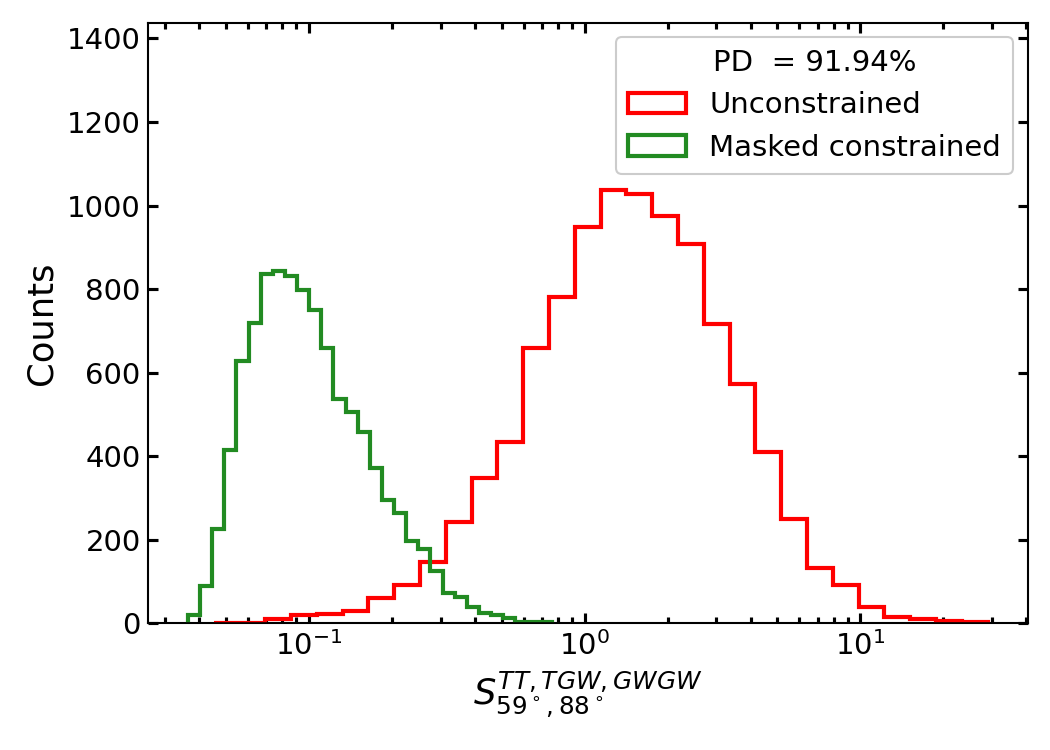}
    \caption{Optimal angles for the combination of TT, TGW, and GWGW. The left and right panels show respectively optimal angles analysis and the results at the optimal range. Here we assume $\ell_{\rm max}= 10$.}
    \label{fig: masked_TTXGW_lmax10}
\end{figure}

Instead, figures~\ref{fig: masked_TTGW_lmax10}-\ref{fig: masked_TTX_lmax10}-\ref{fig: masked_TTXGW_lmax10} show the same respectively for $S^{\rm TT, GWGW}_{\theta_{\rm min}, \theta_{\rm max}}$, $S^{\rm TT, TGW}_{\theta_{\rm min}, \theta_{\rm max}}$ and $S^{\rm TT, TGW, GWGW}_{\theta_{\rm min}, \theta_{\rm max}}$. Once again, there seems to be relatively high-PD region where the original $S_{1/2}$ estimator is defined.


\subsection{Significance with \texorpdfstring{$\ell_{\rm max} = 10$}{TEXT}}

\begin{figure}[t]
    \centering
    \includegraphics[width = 0.49\textwidth]{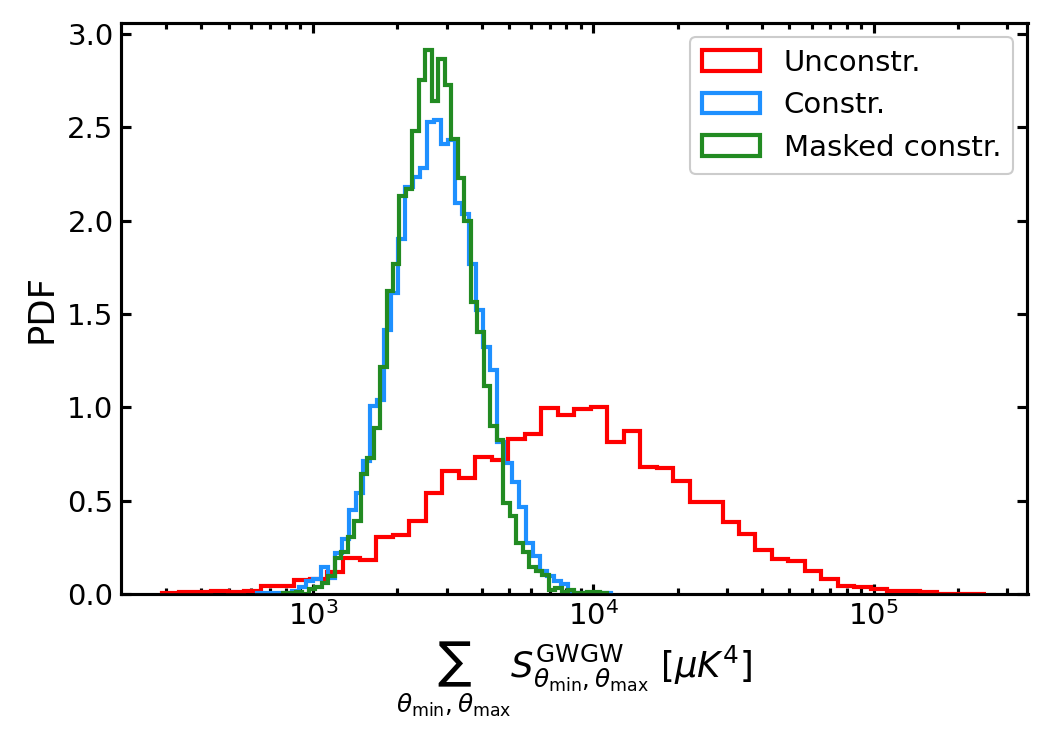}
    \includegraphics[width = 0.49\textwidth]{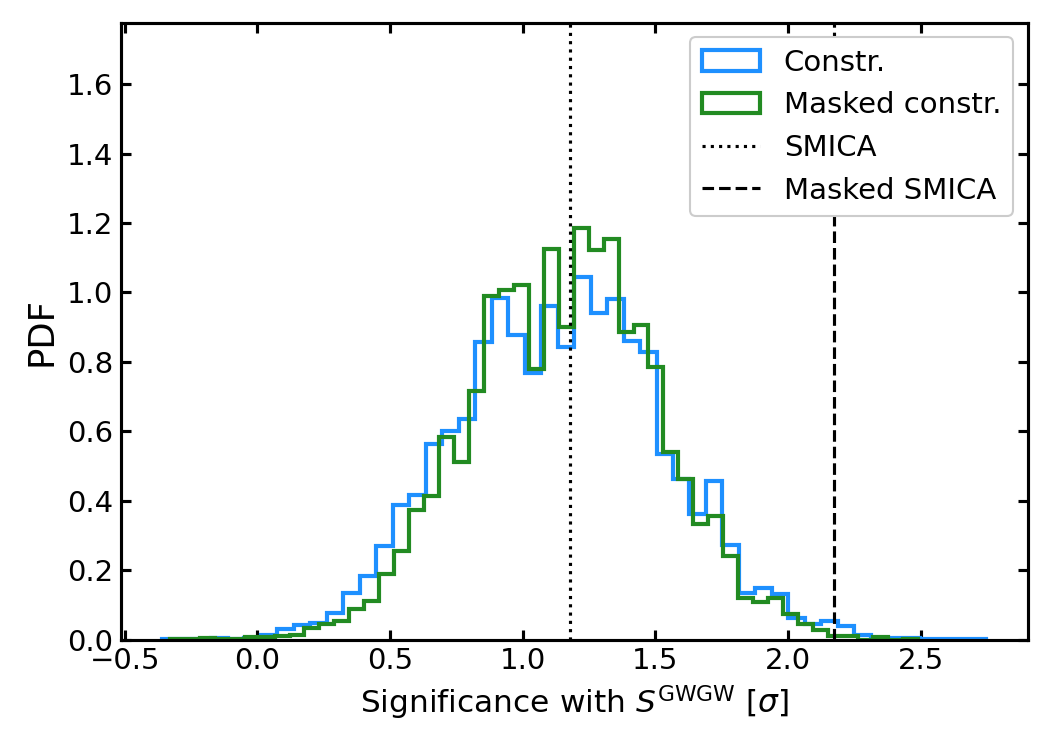}
    \caption{On the left panel, the value of $S^{\rm GWGW}$ for full-sky and masked constrained realizations of the CGWB and the $\Lambda$CDM realizations ($\ell_{\rm max}=10$). On the right panel, corresponding significance in terms of $\sigma$ w.r.t. the unconstrained realizations. The dotted and dashed vertical lines indicate the full-sky and masked SMICA-alone significance, respectively.}
    \label{fig:signi_GWGW_lmax10}
\end{figure}

\begin{figure}[t]
    \centering
    \includegraphics[width = 0.49\textwidth]{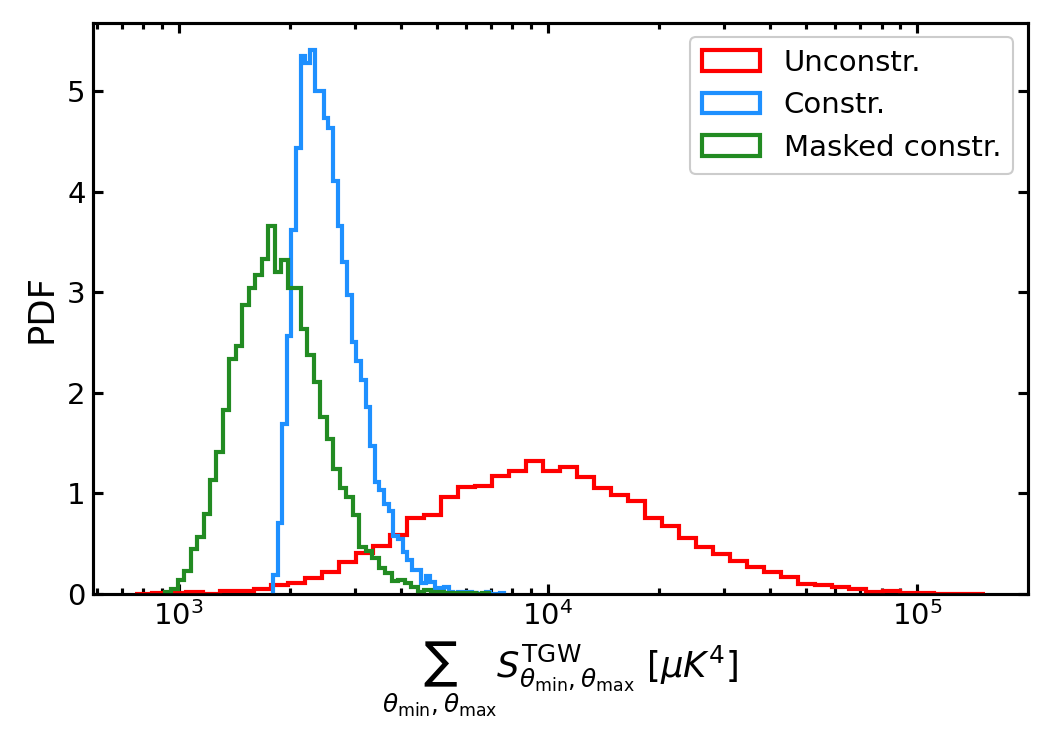}
    \includegraphics[width = 0.49\textwidth]{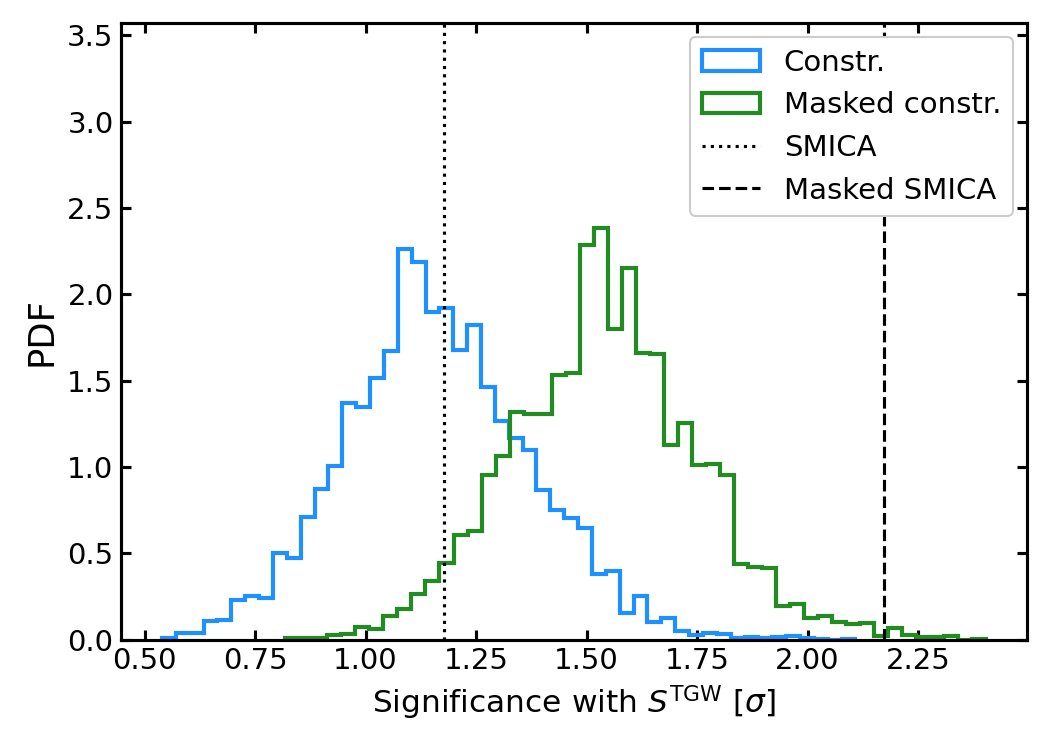}
    \caption{On the left panel, the value of $S^{\rm TGW}$ for full-sky and masked constrained realizations of the CGWB and the $\Lambda$CDM realizations ($\ell_{\rm max}=10$). On the right panel, corresponding significance in terms of $\sigma$ w.r.t. the unconstrained realizations. The dotted and dashed vertical lines indicate the full-sky and masked SMICA-alone significance, respectively.}
    \label{fig:signi_X_lmax10}
\end{figure}

As regards the significance, starting from $S^{\rm GWGW}$, we obtain figure~\ref{fig:signi_GWGW_lmax10}, while for $S^{\rm TGW}$ we get figure~\ref{fig:signi_X_lmax10}.


\begin{figure}[t]
    \centering
    \includegraphics[width = 0.49\textwidth]{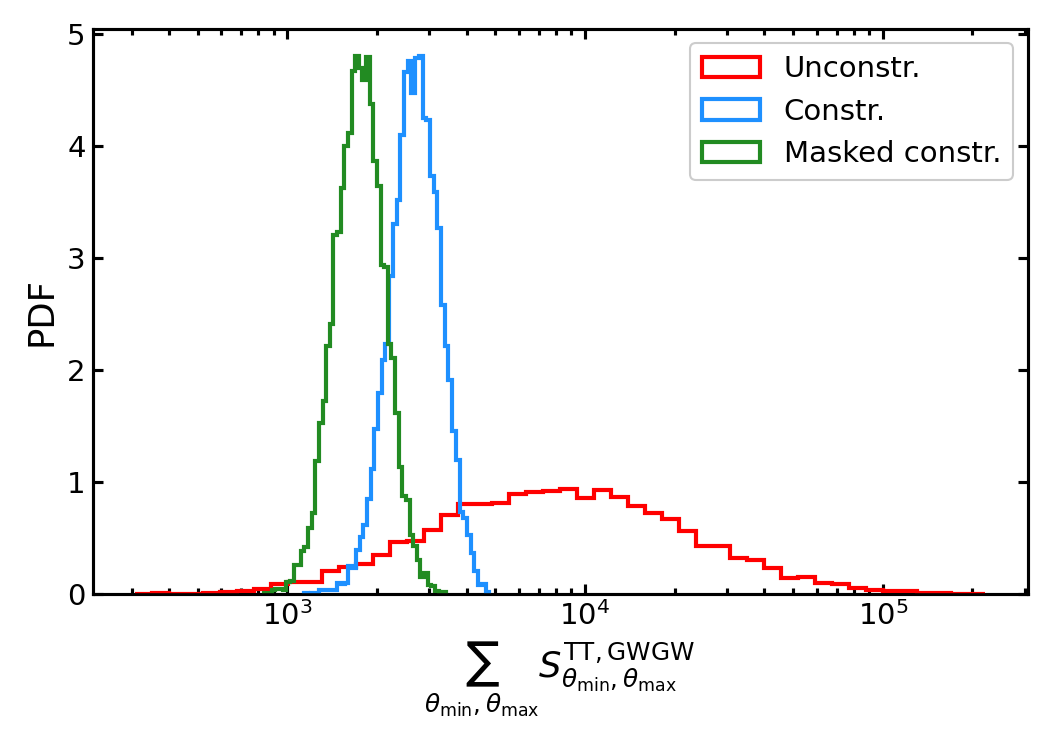}
    \includegraphics[width = 0.49\textwidth]{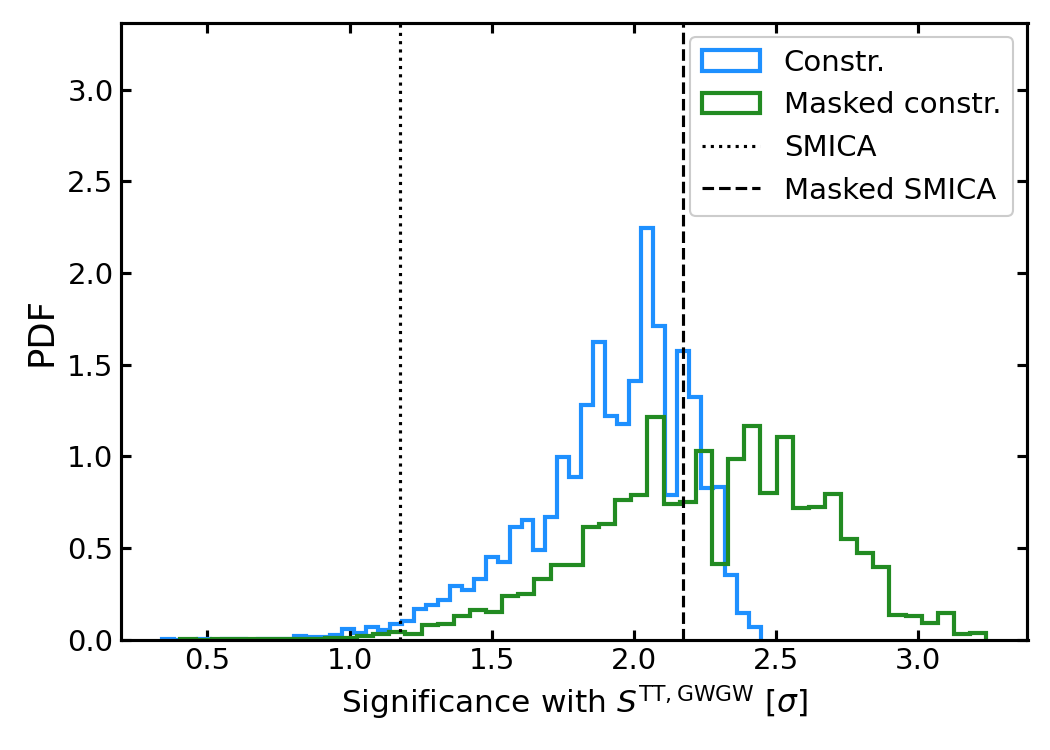}
    \caption{On the left panel, the value of $S^{\rm TT, GWGW}$ for full-sky and masked constrained realizations of the CGWB and the $\Lambda$CDM realizations ($\ell_{\rm max}=10$). On the right panel, corresponding significance in terms of $\sigma$ w.r.t. the unconstrained realizations. The dotted and dashed vertical lines indicate the full-sky and masked SMICA-alone significance, respectively.}
    \label{fig:signi_TTGW_lmax10}
\end{figure}

\begin{figure}[t]
    \centering
    \includegraphics[width = 0.49\textwidth]{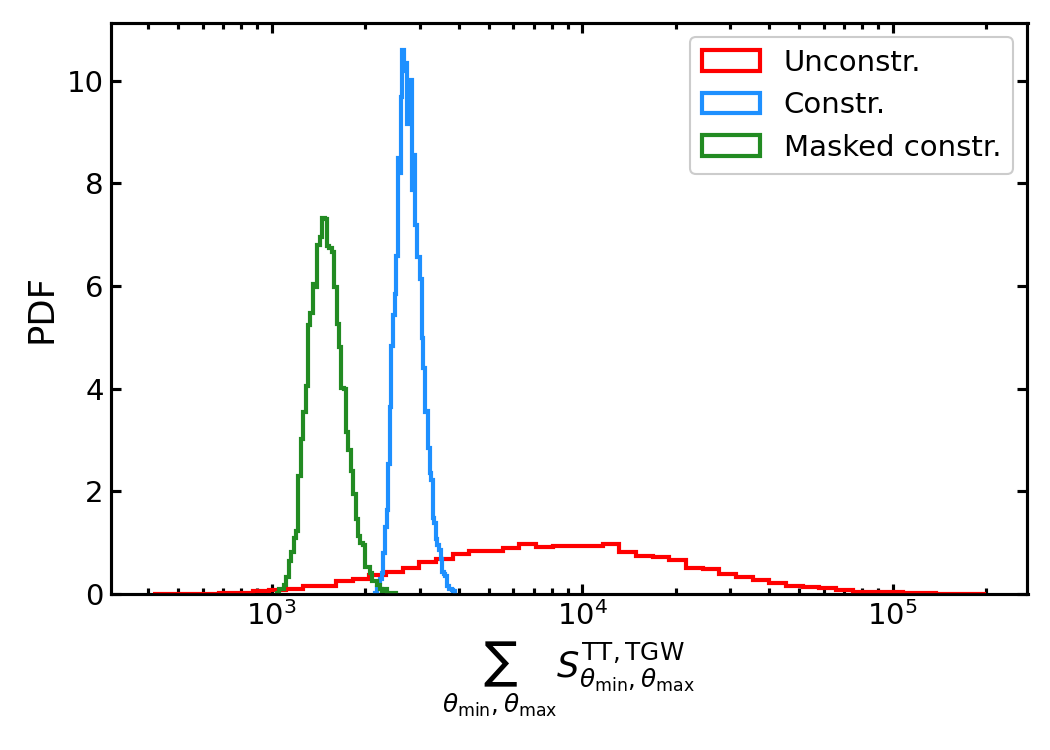}
    \includegraphics[width = 0.49\textwidth]{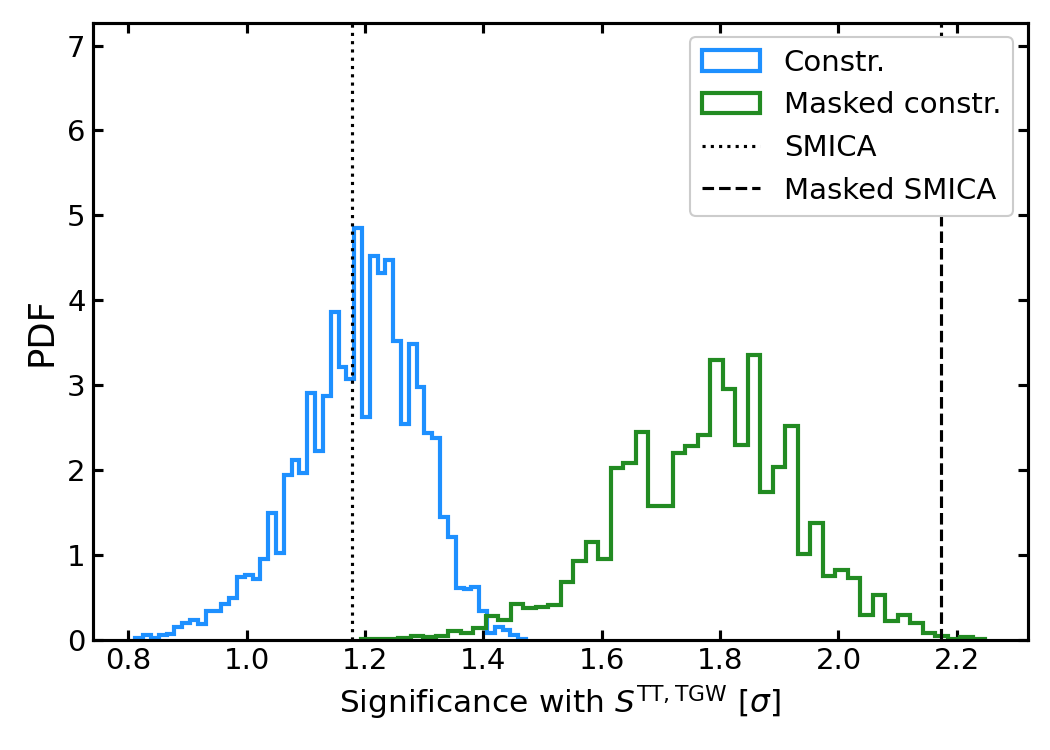}
    \caption{On the left panel, the value of $S^{\rm TT, TGW}$ for full-sky and masked constrained realizations of the CGWB and the $\Lambda$CDM realizations ($\ell_{\rm max}=10$). On the right panel, corresponding significance in terms of $\sigma$ w.r.t. the unconstrained realizations. The dotted and dashed vertical lines indicate the full-sky and masked SMICA-alone significance, respectively.}
    \label{fig:signi_TTX_lmax10}
\end{figure}

\begin{figure}[t]
    \centering
    \includegraphics[width = 0.49\textwidth]{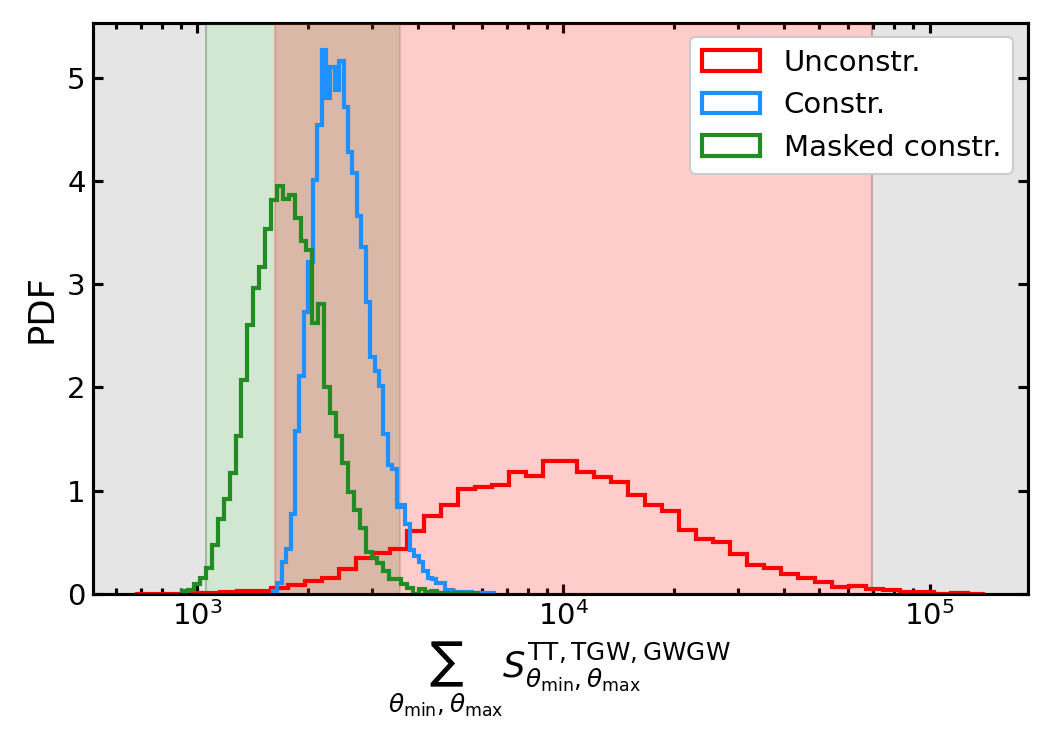}
    \includegraphics[width = 0.49\textwidth]{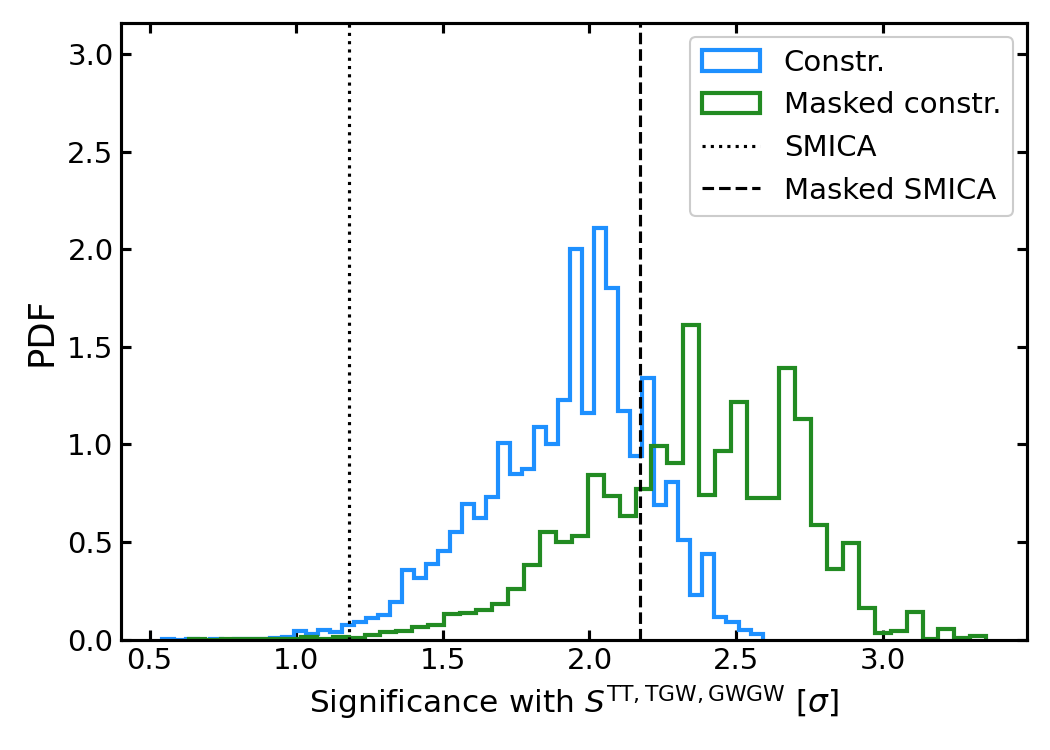}
    \caption{On the left panel, the value of $S^{\rm TT, TGW, GWGW}$ for full-sky and masked constrained realizations of the CGWB and the $\Lambda$CDM realizations ($\ell_{\rm max}=10$). On the right panel, corresponding significance in terms of $\sigma$ w.r.t. the unconstrained realizations. The dotted and dashed vertical lines indicate the full-sky and masked SMICA-alone significance, respectively.}
    \label{fig:signi_TTXGW_lmax10}
\end{figure}

Switching to the multi-field analysis, we obtain figures~\ref{fig:signi_TTGW_lmax10}-\ref{fig:signi_TTX_lmax10}-\ref{fig:signi_TTXGW_lmax10} for $S^{\rm TT, GWGW}$, $S^{\rm TT, TGW}$, and $S^{\rm TT, TGW, GWGW}$.


\end{appendix}

\end{document}